\documentclass[aps,prd,twocolumn,nofootinbib,superscriptaddress,preprintnumbers,floatfix]{revtex4-2}
\usepackage{dcolumn}
\usepackage{amsmath,amssymb}
\usepackage{bm}
\usepackage{graphicx}
\usepackage[hypertexnames=false]{hyperref}

\begin{document}

\preprint{\vbox{\hbox{MIT-CTP/5580} }}
\title{Lattice QCD study of $\pi\Sigma$--$\bar{K}N$ scattering and the $\Lambda(1405)$ resonance } 

\author{John Bulava}
\affiliation{Deutsches Elektronen-Synchrotron (DESY), Platanenallee 6, 15738 Zeuthen, Germany}

\author{B\'{a}rbara Cid-Mora}
\affiliation{GSI Helmholtz Centre for Heavy Ion Research, 64291 Darmstadt, Germany}

\author{Andrew D. Hanlon}
\affiliation{Physics Department, Brookhaven National Laboratory, Upton, New York 11973, USA}

\author{Ben H\"{o}rz}
\affiliation{Intel Deutschland GmbH, Dornacher Str. 1, 85622 Feldkirchen, Germany}

\author{Daniel Mohler}
\affiliation{Institut f\"ur Kernphysik, Technische Universit\"at Darmstadt,
Schlossgartenstrasse 2, 64289 Darmstadt, Germany}
\affiliation{GSI Helmholtz Centre for Heavy Ion Research, 64291 Darmstadt, Germany}

\author{Colin Morningstar}
\affiliation{Department of Physics, Carnegie Mellon University, Pittsburgh, Pennsylvania 15213, USA}

\author{Joseph Moscoso}
\affiliation{Department of Physics and Astronomy, University of North Carolina, Chapel Hill, NC 27516-3255, USA}

\author{Amy Nicholson}
\affiliation{Department of Physics and Astronomy, University of North Carolina, Chapel Hill, NC 27516-3255, USA}

\author{Fernando Romero-L\'{o}pez}
\affiliation{Center for Theoretical Physics, Massachusetts Institute of Technology, Cambridge, MA 02139, USA}

\author{Sarah Skinner}
\affiliation{Department of Physics, Carnegie Mellon University, Pittsburgh, Pennsylvania 15213, USA}

\author{Andr\'{e} Walker-Loud}
\affiliation{Nuclear Science Division, Lawrence Berkeley National Laboratory, Berkeley, CA 94720, USA}

\collaboration{for the Baryon Scattering (BaSc) Collaboration}

\date{\today}

\begin{abstract}
A lattice QCD computation of the coupled channel $\pi\Sigma$--$\bar{K}N$ scattering amplitudes in the 
$\Lambda(1405)$ region is detailed.  Results are obtained using a single ensemble of gauge field 
configurations with $N_{\rm f} = 2+1$ dynamical quark flavors and $m_{\pi} \approx 200~{\rm  MeV}$ and 
$m_K\approx487$~MeV.  Hermitian correlation matrices using both single baryon and meson-baryon 
interpolating operators for a variety of different total momenta and irreducible representations are used. 
Several parametrizations of the two-channel scattering $K$-matrix are utilized to obtain the 
scattering amplitudes from the finite-volume spectrum.  The amplitudes, continued to the complex
energy plane, exhibit a virtual bound state below the $\pi\Sigma$ threshold and a resonance pole 
just below the $\bar{K}N$ threshold.
\end{abstract}

\maketitle

\section{Introduction}

In meson-baryon scattering with strangeness $S=-1$ and isospin $I=0$, the Particle Data 
Group\cite{ParticleDataGroup:2022pth} currently recognizes a 4-star resonance of spin $J=1/2$, negative 
parity, and mass near 1405~MeV, known as the $\Lambda(1405)$.  A possible nearby resonance of the same 
quantum numbers, referred to as the $\Lambda(1380)$, is listed with only 2-star status. The issue of 
whether or not this lower-lying resonance actually exists is of great interest in hadron physics.  
In fact, Ref.~\cite{ParticleDataGroup:2022pth} includes an entire Section 83 dedicated to discussing 
the pole structure of the $\Lambda(1405)$ region.

The $\Lambda(1405)$ resonance first appeared when low-energy $K^-p$ amplitudes measured in bubble 
chamber experiments~\cite{Dalitz:1959dn,Dalitz:1960du} implied a resonance in the $\pi^{-} \Sigma^{+}$ 
spectrum just below the $K^{-}p$ threshold.  For a review of experimental progress in this system, see 
Refs.~\cite{Hyodo:2011ur, Hyodo:2020czb}.  Recent precise measurements of the energy shift and width 
of kaonic hydrogen by the SIDDHARTHA collaboration at DA$\Phi$NE~\cite{SIDDHARTA:2011dsy} have led to
improved determinations of the $K^-p$ scattering length, as discussed in Ref.~\cite{Meissner:2004jr}.  
The CLAS collaboration at JLab investigated the angular dependence of the reaction 
$\gamma + p \rightarrow K^{+} + \Sigma + \pi$, determining the line shapes~\cite{CLAS:2013rjt} and 
confirming~\cite{CLAS:2014tbc} that the $\Lambda(1405)$ resonance has spin and parity $J^P=1/2^-$.  
Using a chiral unitary framework, Refs.~\cite{Mai:2014xna,Roca:2013cca} found the CLAS data to be 
consistent with a two-pole picture. Recent studies by the BGOOD collaboration~\cite{BGOOD:2021sog} 
and the ALICE collaboration~\cite{ALICE:2022yyh} also support a two-pole scenario.  A preliminary 
analysis by the GlueX collaboration~\cite{Wickramaarachchi:2022mhi} favors two isoscalar poles, while
J-PARC~\cite{J-PARCE31:2022plu} claim a single pole describes their data. An overarching analysis in 
Ref.~\cite{Anisovich:2020lec} favors a single resonance, but does not rule out the two-pole picture.

Scattering in the $\Lambda(1405)$ region also poses a challenge for theory. Accommodating the
low mass and quantum numbers of the $\Lambda(1405)$ resonance in constituent quark models, such as 
Ref.~\cite{Isgur:1978xj}, is problematic.  The presence of two poles in this region was first 
suggested in Ref.~\cite{Oller:2000fj}. Early applications of chiral effective theory 
to scattering in this energy region were presented in Refs.~\cite{Kaiser:1995eg,Oset:1997it}.
Nearly all approaches based on $SU(3)$ chiral effective theory, 
which are reviewed in Refs.~\cite{Mai:2018rjx,Mai:2020ltx}, predict the presence of two poles in the 
scattering matrix analytically continued to complex center-of-mass energies, but disagree about the 
position of the lower pole. See Refs.~\cite{Ezoe:2020piq,Myint:2018ypc,Miyahara:2018onh,Miyahara:2018lud,
Azizi:2023tmw,Xie:2023cej,Lu:2022hwm,Hyodo:2022xhp,MartinezTorres:2012yi} for some other recent 
theoretical studies of the $\Lambda(1405)$ resonances. 
    
The above considerations suggest that a first-principles investigation of the pole structure in
the region of the $\Lambda(1405)$ resonance is warranted.  All previous lattice QCD computations of the 
$\Lambda(1405)$ have not computed scattering amplitudes and instead aimed only to isolate the lowest 
finite-volume energy eigenstate using single-baryon three-quark interpolating fields~\cite{Gubler:2016viv,
Menadue:2011pd,Engel:2012qp,Engel:2013ig,Nemoto:2003ft,Burch:2006cc,Takahashi:2009bu,Meinel:2021grq,
Hall:2014uca}.  However, the use of local single-hadron interpolating fields alone is insufficient to 
correctly determine the finite-volume spectrum above two-hadron thresholds~\cite{Lang:2012db,Mohler:2012na,
Wilson:2015dqa}. The $\bar{K}N$ scattering length for $I=0$ has also been computed long ago using the 
quenched approximation in Ref.~\cite{Fukugita:1994ve}, but neglecting the mixing with the kinematically-open
$\pi\Sigma$ channel and ignoring unitarity violation due to the quenched approximation, which invalidates 
the relation between the finite-volume spectrum and scattering amplitudes~\cite{Bernard:1995ez}.
The $\pi\Sigma$ and $\bar{K}N$ scattering lengths in other (non-singlet) flavor and isospin combinations
not directly relevant for the $\Lambda(1405)$ have also been computed in 
Refs.~\cite{Detmold:2015qwf,Torok:2009dg,Meng:2003gm}. 

A recent study~\cite{Bulava:2022vpq} of nucleon-pion
scattering in the region of the $\Delta$-resonance demonstrates that current lattice QCD techniques
are sufficiently efficacious for studying simple baryon resonances.  In this work, we apply these
techniques to the isospin $I=0$ and strangeness $S=-1$ coupled-channel $\pi\Sigma-\bar{K}N$ scattering 
amplitudes below the $\pi\pi\Lambda$ threshold for the first time.  Hermitian correlation matrices 
using both single baryon and meson-baryon interpolating operators for a variety of different total 
momenta and irreducible representations are used to obtain the finite-volume stationary-state energies.  
A set of parametrizations of the scattering amplitudes are then constrained by fits to the finite-volume 
energy spectrum using a well-known quantization condition.  The amplitudes, continued to the complex
energy plane, exhibit a virtual bound state below the $\pi\Sigma$ threshold and a resonance pole 
just below the $\bar{K}N$ threshold, the positions of which vary little with differing fit forms and 
are broadly consistent with predictions from chiral effective theory. 
This work constitutes the first coupled-channel scattering study in lattice QCD to include baryons; 
only coupled-meson systems have been previously studied~\cite{Wilson:2015dqa,Moir:2016srx,Briceno:2017qmb,
Woss:2019hse,Dudek:2014qha,Prelovsek:2020eiw}.  Highlights of this study were previously
presented in Ref.~\cite{Bulava:2023rmn}. Further technical details of the investigation are reported here.

This work is organized as follows.  The determination of the finite-volume stationary-state energies
is presented in Sec.~\ref{sec:comp}, including the ensemble details, method of evaluating the
correlation functions, and the extraction of the energies.  Details on the determinations of the scattering 
amplitudes are then presented in Sec.~\ref{sec:scatamp}. The quantization condition that relates the
amplitudes to the finite-volume spectrum is reviewed, the parametrizations of the $K$-matrix that we use
are described, and the fits using these parametrizations and the quantization condition are detailed.
Analytic continuation of the scattering transition amplitudes to complex energies is then used
to determine nearby $S$-matrix poles.

\section{Spectrum computation}
\label{sec:comp}

This section describes the determination of the finite-volume stationary-state energies.
The procedure used follows an approach similar to Ref.~\cite{Bulava:2022vpq},
but for the convenience of the reader, some of the main details of the method are 
repeated here, along with a summary of our results.

\subsection{Ensemble details}
\label{sec:ens}

A single ensemble of QCD gauge configurations is employed with dynamical mass-degenerate $u$- and 
$d$-quarks which are heavier than physical, and an $s$-quark lighter than physical, resulting in a 
pion mass $m_{\pi} \approx 200$~MeV and a kaon mass $m_{\rm K}\approx487$~MeV, which differ slightly 
from their physical values $m_\pi^{\rm phys}\approx 140$~MeV and $m_{\rm K}^{\rm phys}\approx 495$~MeV. 
The key properties of this so-called D200 ensemble are summarized in Table~\ref{tab:comp_deets}.  The 
configurations were generated by the Coordinated Lattice Simulations (CLS) consortium~\cite{Bruno:2014jqa} 
using the tree-level improved L\"uscher-Weisz gauge action~\cite{Luscher:1984xn} and a non-perturbatively 
$O(a)$-improved Wilson fermion action~\cite{Bulava:2013cta}. Open temporal boundary 
conditions~\cite{Luscher:2011kk} were employed to reduce the autocorrelation of the global topological 
charge, but this then requires that all interpolating fields are sufficiently far from the boundaries, 
limiting the maximum temporal separation in correlation functions to $t_{\rm max} = 25a$. 
The algorithm used to generate the D200 ensemble is described in Ref.~\cite{Bruno:2014jqa}. 
Low-lying eigenvalues of the Dirac matrix of either the light quark doublet or the strange 
quark can cause instabilities in the Hybrid Monte Carlo (HMC) or Rational Hybrid Monte Carlo 
(RHMC) algorithm\cite{Clark:2006fx} for generating the gauge fields, an issue which is 
sufficiently ameliorated for the D200 ensemble~\cite{Mohler:2020txx} by light- and 
strange-quark re-weighting~\cite{Luscher:2008tw}. Re-weighting factors must be included
in the analysis to convert the simulated action to the desired one, and we use the factors 
computed in \cite{Kuberski:2023zky}. The lattice spacing is determined in 
Ref.~\cite{Bruno:2016plf} and updated in Ref.~\cite{Strassberger:2021tsu}. 

\begin{table}[t]
\caption{
Parameters of the D200 ensemble~\cite{Bruno:2014jqa} with spatial extent 
$L=64a$ and temporal extent $T=128a$. The lattice spacing is given, as well as the masses 
$m_{\pi}, m_{\rm K}$ of the pion and kaon, respectively, in units of the lattice spacing $a$.
\label{tab:comp_deets}}
\begin{ruledtabular}
\begin{tabular}{c@{\hskip 12pt}c@{\hskip 12pt}c@{\hskip 12pt}c@{\hskip 12pt}}
 $a [\textup{fm}]$& $am_{\pi}$ & $am_{\rm K}$ & $m_\pi L$ \\ \hline
    0.0633(4)(6) & 0.06533(25) &  0.15602(16) & 4.181(16)
\end{tabular}
\end{ruledtabular}
\end{table}

\begin{figure}[t]
\centering
\includegraphics[width=\linewidth]{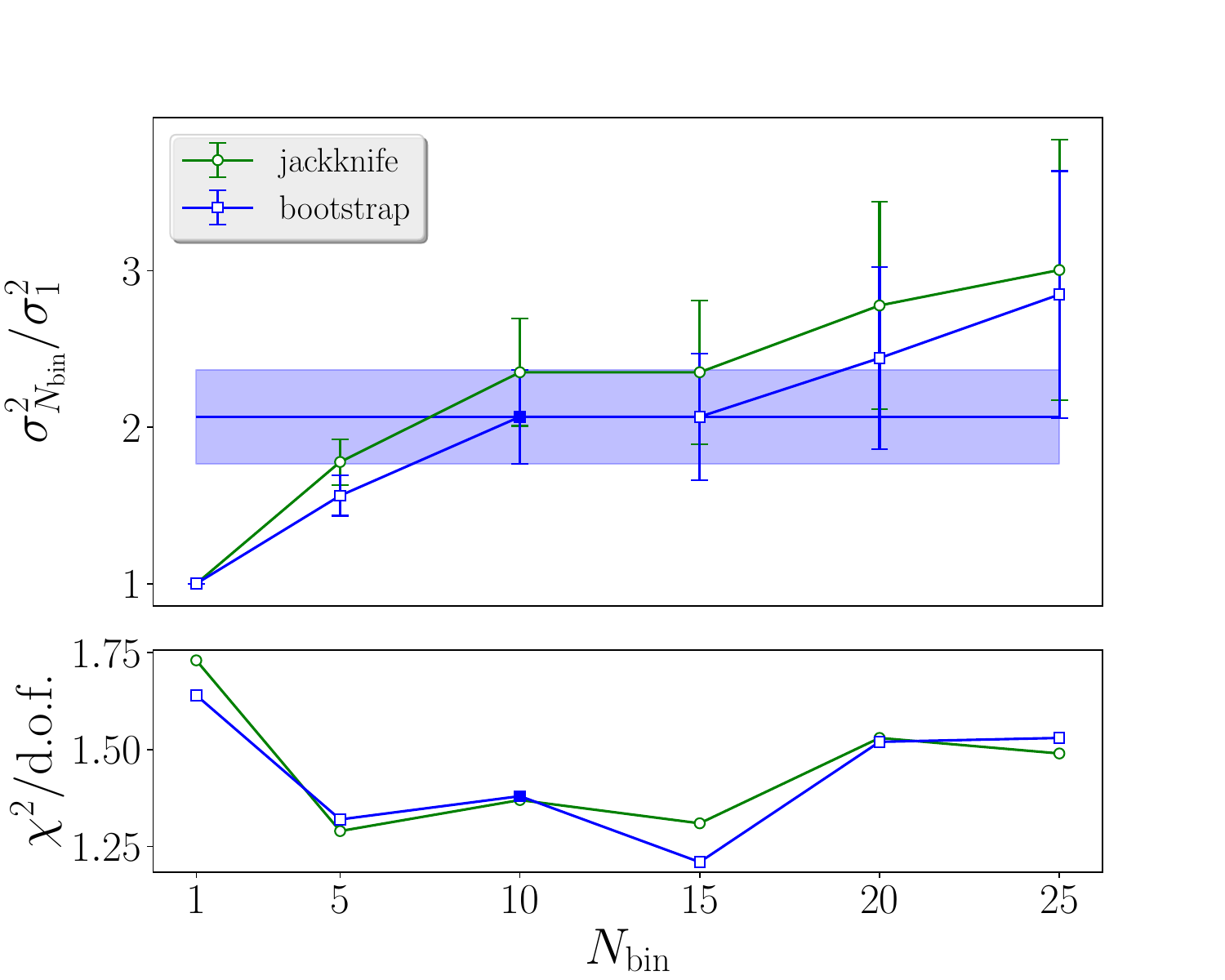}
 \caption{
 (Top) Ratios of variances for fits to determine $am_{\pi}$ for 
 various bin sizes $N_\textup{bin}$ over the variance for $N_{\rm bin}=1$. Both the 
 jackknife and bootstrap procedure are employed. (Bottom) The correlated-$\chi^2$ for 
 single-exponential fits to the pion correlator over a range in temporal separation 
 $[t_{\rm min}, t_{\rm max}] = [15a,25a]$ for various rebinning factors. Jackknife
 and bootstrap ($N_B = 800$) resampling methods are compared, and error bars are 
calculated using statistical estimation outlined in appendix F.2 of Ref.~\cite{RQCD:2022xux}.
\label{fig:rebin_analysis}}
\end{figure}

Correlation function evaluations are separated by four molecular dynamics units (MDU's) in the 
Monte Carlo Markov chain.  Our study of autocorrelations is summarized in Fig.~\ref{fig:rebin_analysis}.
The top panel of this figure shows how the variance in the pion mass determination
increases as the original measurements are rebinned.  The variance initially increases
with increasing $N_{\rm bin}$, until little difference is observed in going from 
$N_{\rm bin} = 10 $ to $N_\textup{bin}=20$.  The bottom panel displaying the correlated 
$\chi^2/\mathrm{dof}$ shows the expected reduction due to the larger variance with 
increasing $N_{\rm bin}$. For $N_{\rm bin} > 15$ the correlated-$\chi^2$ of the pion fit increases 
again, likely due to degrading estimates of the covariance matrix. Fits to determine the nucleon 
mass were also found to have similar behavior.  Hence, in this work, all primary quantities are 
first binned by averaging over $N_\textup{bin}=10$ consecutive gauge configurations. 

\subsection{Correlator determinations}
\label{sec:laph}

Our operator construction is described in Ref.~\cite{Morningstar:2013bda} and our method 
of evaluating the temporal correlators is detailed in Ref.~\cite{Morningstar:2011ka}. 
We use multi-hadron operators comprised of individual constituent hadrons, 
each corresponding to a definite momentum.  The single hadron operators are appropriate 
assemblages of gauge-covariantly smeared quark fields.  The quark fields are smeared
using the Laplacian Heaviside (LapH) procedure described in Ref.~\cite{Peardon:2009gh}.
This smearing employs a projection onto the subspace spanned by the $N_\textup{ev}$ lowest 
eigenmodes of the gauge-covariant three-dimensional Laplacian operator, expressed in terms
of link variables which are stout smeared~\cite{Morningstar:2003gk}.  The stout smearing
parameters are denoted by $(\rho, n_{\rho})$.  The multi-hadron operators used in 
this study are presented in Tables~\ref{tab:opdefs0}-\ref{tab:opdefs3} of 
Appendix~\ref{sec:operators}.

Evaluating the temporal correlations of our operator sets requires time-slice to 
time-slice quark propagators, which we estimate using the stochastic LapH 
method~\cite{Morningstar:2011ka}.  This method employs variance reduction using
noise dilution projectors.  Each projector is a product of time (`T'), spin (`S'), and 
Laplacian eigenvector index (`L') projectors.  For each, the different schemes used
are denoted by `F' for full dilution and `I$n$' for some number $n$ of 
uniformly interlaced projectors. Different dilution schemes are used for fixed-time 
quark lines, denoted `fix', which propagate from the source time-slice to the sink time-slice,
and relative-time lines (`rel') which start and end at the same time.  In this work, the 
relative-time quark lines are only used at the sink time, while the fixed-time lines 
are used for quark propagation starting and ending at the source time. The smearing 
parameters and the dilution schemes used are specified in Table~\ref{tab:laph_deets}. 
Source times $t_0=35a, 64a$ are used for correlations going forwards in time, and 
$t_0=64a, 92a$ are used for correlations going backwards in time.  Correlators
for the different rows of the little group irreducible representations (irreps) and for 
total momenta in all directions of the same magnitudes squared are averaged to increase 
statistics. An advantageous feature of the stochastic LapH method is source-sink 
factorization of the correlators which greatly facilitates the evaluation of
large Hermitian correlation matrices containing single-baryon, $\pi\Sigma$, and $\bar{K}N$ 
interpolating operators via optimized tensor contractions~\cite{Horz:2019rrn}.
 
\begin{table}[t]
\caption{Parameters of the stochastic LapH implementation used to compute temporal 
correlators in this work. The stout smearing parameters for the spatial links in 
the gauge-covariant Laplace operator are $(\rho,n_\rho)$, and $N_\textup{ev}$ denotes the 
dimension of the LapH subspace. Notation used to specify the dilution scheme for each line 
type is explained in the text.
\label{tab:laph_deets}}
\begin{ruledtabular}
\begin{tabular}{c @{\hskip 12pt} c @{\hskip 12pt} c @{\hskip 12pt} c }
 $(\rho,n_\rho)$ & $N_\textup{ev}$ & Noise dilution \\ \hline
 (0.1,\ 36) & 448 & (TF,\ SF,\ LI16$)_\textup{fix}$\quad (TI8,\ SF,\ LI16$)_\textup{rel}$ 
\end{tabular}
\end{ruledtabular}
\end{table}

\subsection{Finite-volume energies}
\label{sec:energies}

Once Markov-chain Monte Carlo estimates of the correlation functions are obtained, the 
determination of finite-volume energies can then be achieved.  Single-hadron energies
corresponding to the lowest-lying mesons and baryons are obtained from the single diagonal 
correlators of the relevant single-hadron operators.  For the spectra of stationary-state 
energies, the entire correlation matrices involving all operators in a given symmetry 
channel must be used.  The symmetry channels are labeled by their total isospin $I$,
total strangeness $S$, and the irreducible representation (irrep) $\Lambda(\bm{d}^2)$ of 
the little group for the total momentum squared $\bm{P}^2=(2\pi/L)^2\bm{d}^2$, where
$\bm{d}$ is a three-vector of integers and $L$ is the extent of the $L^3$ spatial
lattice.  Both single-baryon and meson-baryon operators are used in the correlation
matrices evaluated in this work.  The correlator matrices are then diagonalized, as described
in Ref.~\cite{Bulava:2022vpq} and summarized below, using solutions of a judiciously-formed 
generalized eigenvalue problem (GEVP).  Each finite-volume stationary-state energy is then 
extracted from fits to each of the resulting diagonal correlators.

Energies are determined from correlated-$\chi^2$ fits to both single- and two-exponential 
fit forms over various fit ranges $[t_{\rm min}, t_{\rm max}]$, which are additionally 
compared to a ``geometric-exp series'' form~\cite{Bulava:2022vpq}
\begin{gather}\label{e:geom}
    C(t) = \frac{A{\rm e}^{-Et}}{1 - B{\rm e}^{-Mt}},
\end{gather}
which consists of four free parameters. The geometric exponential series 
is sometimes found to be successful at estimating excited-state contamination of a correlator by
using an infinite tower of evenly-separated states.
The optimal fit for an energy determination is chosen so that the statistical error on the 
energy encompasses any variation between fit forms and is reasonably insensitive to $t_{\rm min}$.
However, the open temporal boundary conditions limit the range of the correlation functions. For our 
choice of source time-slices this results in $t_{\rm max}=25a$. At the same time, the energy gap 
between the desired ground state and unwanted excited states decreases as the pion mass is lowered 
to the physical point.  The limited time range and small energy gap often result in an insufficient 
description of the data using the single-exponential fit form.  The two-exponential form and the 
`geometric-exp' form, however, do result in suitable descriptions of the data and provide consistent 
energy determinations. 

\begin{figure*}[t]
\centering
\includegraphics[width=0.9\linewidth]{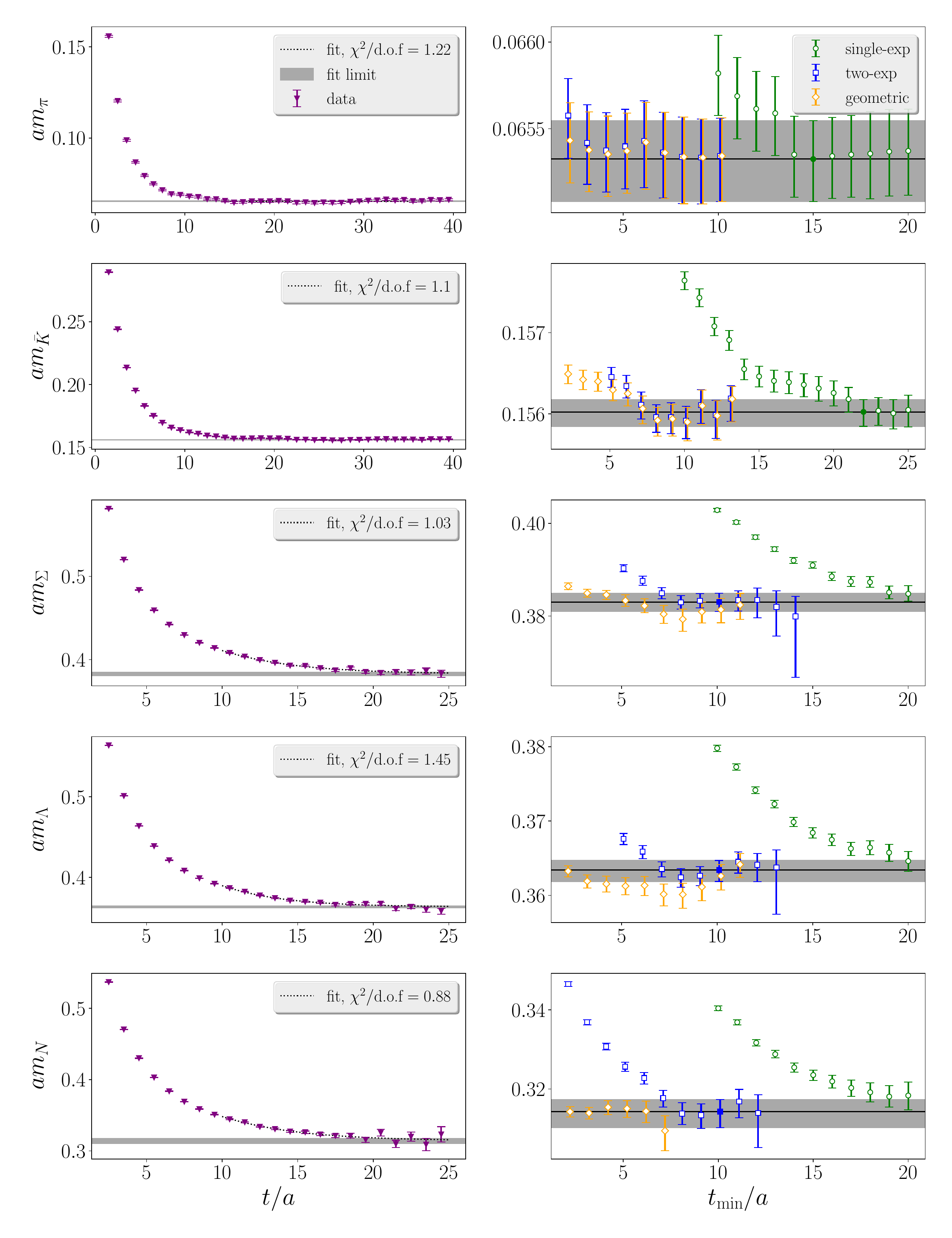}
\caption{\label{fig:rest_masses} Determination of single-hadron masses. Each row corresponds to 
 a particular hadron denoted by the label on the vertical axis. The left panel shows the 
 effective energy $\ln(C(t)/C(t+a))$ with two-exponential fits overlayed (dashed lines), except 
 for the pion and kaon for which a single-exponential fit is used. The 
 horizontal band corresponds to the 
 quoted mass and statistical error. The right panel compares different fit forms and different 
 $t_{\rm min}$ choices for fixed $t_{\rm max}=35a$ for the pion and kaon and $t_{\rm max}=25a$ 
 for the other hadrons, with the chosen fit denoted by the 
 horizontal band.}
\end{figure*}

\begin{table}
\caption{\label{tab:masses}   A summary of various hadron masses and decay constants 
(with normalization $f_\pi^{\rm phys}\approx130$~MeV) for the CLS D200 gauge ensemble used in this 
work. The $\eta$ mass is taken from 
Ref.~\cite{Bali:2021qem}, and the pion and kaon decay constants are taken from Ref.~\cite{Ce:2022kxy}. 
All other masses are determined in Fig.~\ref{fig:rest_masses}.}
\begin{ruledtabular}
\begin{tabular}{ll@{\qquad\quad}|ll}
 $am_\pi$     &   0.06533(25)    &     $af_\pi$       &   0.04226(13) \\
 $am_{ K}$    &   0.15602(16)    &     $af_{K}$       &  0.04910(11) \\
 $am_{\eta}$  &   0.1768(22)     &     $am_{\Lambda}$ & 0.3634(14) \\
 $am_{ N}$    &   0.3143(37)     &     $am_{\Sigma}$  &  0.3830(19)\\
\end{tabular}
\end{ruledtabular}
\end{table}

Determinations of the pion, kaon, nucleon, $\Sigma$, and $\Lambda$ masses are shown in 
Fig.~\ref{fig:rest_masses} and their values are presented in Table~\ref{tab:masses}.  This table 
also includes the $\eta$ mass, as well as the pion and kaon decay constants.  The hadrons listed 
in this table are all stable in the absence of electroweak interactions and in the isospin limit 
with $m_u=m_d$.

Before extracting a spectrum of stationary-state energies, the correlation matrix $C_{ij}(t)$ 
in a given symmetry channel is transformed\cite{Michael:1982gb} to a diagonal form $D_{ab}(t)$ using the
eigenvectors $v_n(t_0, t_{\rm d})$ of the GEVP
\begin{equation}
\label{eq:gevp}
    C(t_{\rm d})v_n(t_0, t_{\rm d}) = \lambda_n(t_0, t_{\rm d})\ C(t_0)\ v_n(t_0, t_{\rm d}),
\end{equation}
where $t_0$ is referred to as the \textit{metric time}, $t_{\rm d}$ is called 
the \textit{diagonalization time}, and $\lambda_n$ denote the eigenvalues.
The diagonal elements of the resulting matrix $D(t)$ are referred to here as the  \textit{rotated} 
correlators. In this work, two different ways of carrying out the above transformation are used.

In the simplest approach, known as a \textit{single pivot}, a single judicious choice of $t_0$ 
and $t_{\rm d}$ is used to transform the correlation matrix for all times $t$.  We typically
choose $t_0$ to be about half of $t_{\rm d}$ to minimize contamination from higher-lying
states~\cite{Luscher:1990ck,Blossier:2009kd} and choose $t_{\rm d}$ as small as
possible to minimize statistical errors but ensuring that the rotated correlation
matrix remains diagonal for all $t>t_{\rm d}$ within the statistical precision
of the calculations.  The eigenvectors $v_n$ utilized are computed using the mean values
of the correlation matrix.  In this approach, it is important to check insensitivity of 
the final spectrum results to a reasonable range of $(t_0, t_{\rm d})$ choices.

In the second approach, known as a \textit{rolling pivot},
a single value of $t_0$ is used, but $t_{\rm d}=t$ is used for rotating the correlation matrix
at time $t$.  In other words, the correlation matrix is separately rotated at every
time, keeping the metric time fixed.  The eigenvectors employed for each time are determined
using the mean values of the correlators.  This procedure is much more complicated than
the single pivot as the
order of the eigenvectors with changing time must be carefully considered.  The simplest
method of ordering the eigenvectors according to their eigenvalues at time $t$ can lead to
diagonal correlators which tend asymptotically towards different stationary-state
decay rates for different times $t$.  Hence, some method of eigenvector ``pinning'' is 
needed so that a given diagonal correlator is always tending towards the same
stationary-state behavior.  Diagonalizations at larger times can lead to increased
statistical errors, but this method ensures the correlation matrix remains diagonal 
at all times.

To check that uncertainties determined in the rolling pivot are not underestimated, a variant 
of the second approach is also used in which the variance in the diagonalized correlators include 
uncertainties from the GEVP. Instead of using the same
eigenvectors from the mean values of the correlators when bootstrapping, the
eigenvectors themselves are re-evaluated using the bootstrap resamplings of the
correlation matrix. 

\begin{figure}[t]
\begin{center}
\includegraphics[width=\linewidth]{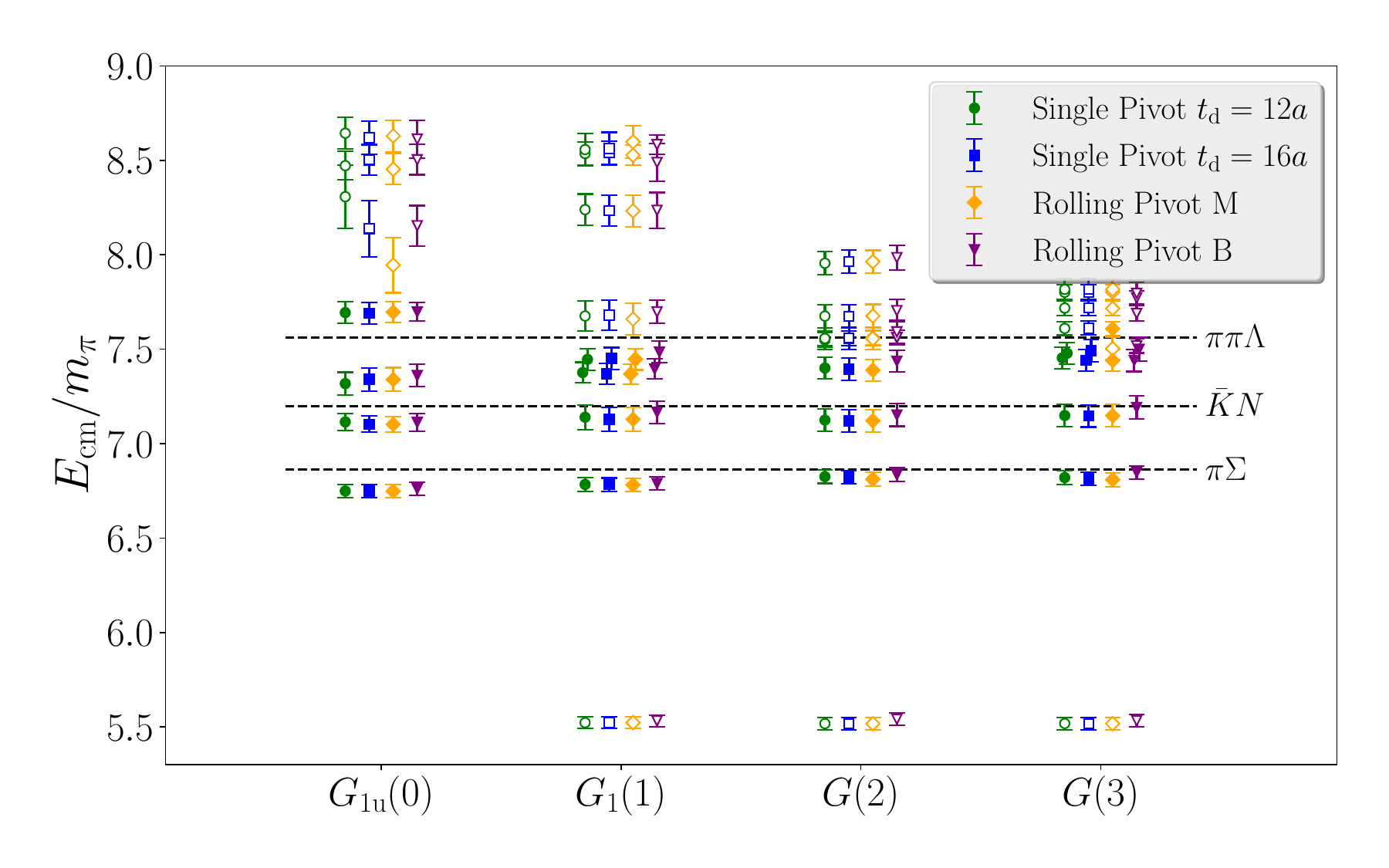}
\end{center}
\caption{\label{fig:gevp_stability} Stability of the finite-volume spectra under variation 
of the correlation matrix rotation method using the GEVP for four different symmetry sectors. 
The metric time is set to $t_0=4a$. ``Rolling Pivot M'' refers to the second approach that
uses eigenvectors determined using the mean values of the correlation matrix, and ``Rolling 
Pivot B'' refers to the bootstrapped variant of the rolling pivot.}
\end{figure}

Fig.~\ref{fig:gevp_stability} shows center-of-mass frame energy determinations in four 
symmetry channels using the above different approaches.  Two different choices of 
$t_{\rm d}$ in the single pivot method are also shown.  The same fitting strategy to extract
the energies from the diagonalized correlators was used as for the single hadron
energies.  One sees that the simplest single-pivot method produces nearly 
identical results to the other two more complicated methods as long as $(t_0,t_{\rm d})$ 
are chosen appropriately.  Given this stability of the results, our final results
used the single-pivot method with $t_0=4a$ and $t_{\rm d}=16a$.

In addition to using the one-exponential, two-exponential, and geometric-exp series
fit forms to directly determine the lab-frame stationary-state energies from the
diagonal elements of the rotated correlation matrix, a fourth method, already used in 
Ref.~\cite{CP-PACS:2007wro}, is also used to determine the energies.
After forming the rotated correlators $D_n(t)$, the following ratio of correlators
is taken 
\begin{equation}
\label{eq:ratio}
    R_n(t) = \frac{D_n(t)}{C_{A}(\boldsymbol{d}_{A}^2, t)\,C_{B}(\boldsymbol{d}_{B}^2, t) },
\end{equation}
where $(A,B)$ is either $(\pi,\Sigma)$ or $(\bar{K},N)$,
with $\boldsymbol{d}_{A}^2$ and $\boldsymbol{d}_{B}^2$  chosen so that 
\begin{gather}
    E_n^{\rm non.\, int.} = \sqrt{m_{A}^2 + \left(\frac{2\pi \boldsymbol{d}_{A}}{L}
    \right)^2} + \sqrt{m_{B}^2 + \left(\frac{2\pi \boldsymbol{d}_{B}}{L} \right)^2}
\end{gather}
corresponds to a non-interacting energy sum close to the energy expected for the
stationary state. The ratio $R_n(t)$ is then fit to a one-exponential ansatz to determine 
the energy interaction shift $a\Delta E_n$, from which the lab-frame energy can be 
reconstructed $aE_n^\textup{lab} = a\Delta E_n + aE_n^\textup{non.int.}$.  This method
hopes to take advantage of noise cancellation in the ratio of correlators to more
precisely determine the interaction shifts.  Due to the presence of both $\pi\Sigma$ 
and $\bar{K}N$ states, there are often nearby non-interacting sums of each type.  Since 
the reconstructed lab-frame energy should be independent of which product of non-interacting
correlators is used in the denominator, we use both types to check for consistency.

To select a final result for use in determining the scattering
amplitudes, several general rules of thumb are employed.  First, the selected fit must
have a $p$-value greater than 0.1 and/or $\chi^2/\textup{dof}$ less than 1.5.  Agreement of the fit
result with those from nearby $t_{\rm min}$ values is also considered. When choosing a final
fit, we also look for consistency with the other fit methods.  From these considerations,
a single-exp ratio fit is often selected, and we require that the final fit is within 
$2\sigma$ of the other ratio fit of the same $t_{\rm min}$. Consistency with the plateau 
regions of the two-exponential and geometric-exp fits is also required.
The final fit is chosen as that with the smallest statistical errors which is also stable between 
nearby $t_{\rm min}$ values and maintains consistency between methods.

\begin{figure}
\begin{center}
\includegraphics[width=\linewidth]{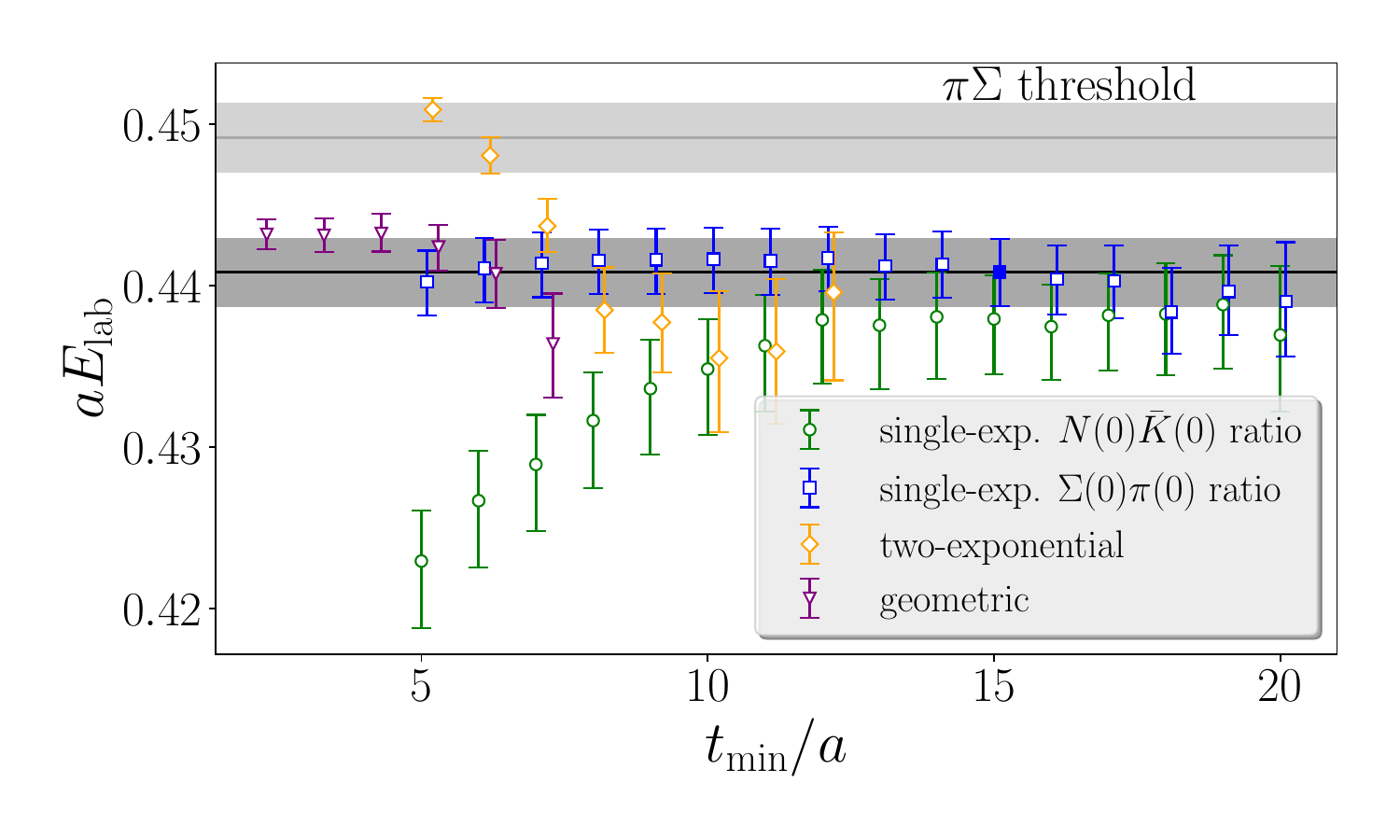}
\end{center}
\caption{\label{fig:g1u_example}
  Determination of the finite-volume stationary-state energy corresponding to the lowest 
  level of the $G_{\rm 1u}(0)$ irrep.  Each point shows the lab-frame energy from a particular 
  fit using temporal range $t_{\rm min}$, shown on the horizontal axis, to $t_{\rm max}=25a$.
  Four different fit methods are shown: two-exponential and geometric-exp fits to the
  rotated correlator, and single-exponential fits to the ratio of the rotated correlator
  over the product of single-hadron correlators for either $\bar{K}(0)N(0)$ or 
  $\pi(0)\Sigma(0)$. The dark horizontal band and filled symbol denote the final chosen fit
  selected as described in the text.}
\end{figure}

An example energy determination is shown in Fig.~\ref{fig:g1u_example}.  The energy
determined in this example corresponds to the lowest level of the $G_{\rm 1u}(0)$ irrep.  
Results from four different fit methods are shown: two-exponential and geometric-exp fits to the
rotated correlator, and single-exponential fits to the ratio of the rotated correlator
over the product of single-hadron correlators for either $\bar{K}(0)N(0)$ or $\pi(0)\Sigma(0)$,
the zeroes in parentheses referring to $\bm{d}^2$ of each hadron.
The single-exp ratio fit for $t_{\rm min}=15a$ is selected as the final estimate.
Analogous plots for other levels are given in Appendix~\ref{sec:appendix_tmin}.
     
\begin{figure}[t]
\begin{center}
\includegraphics[width=\linewidth]{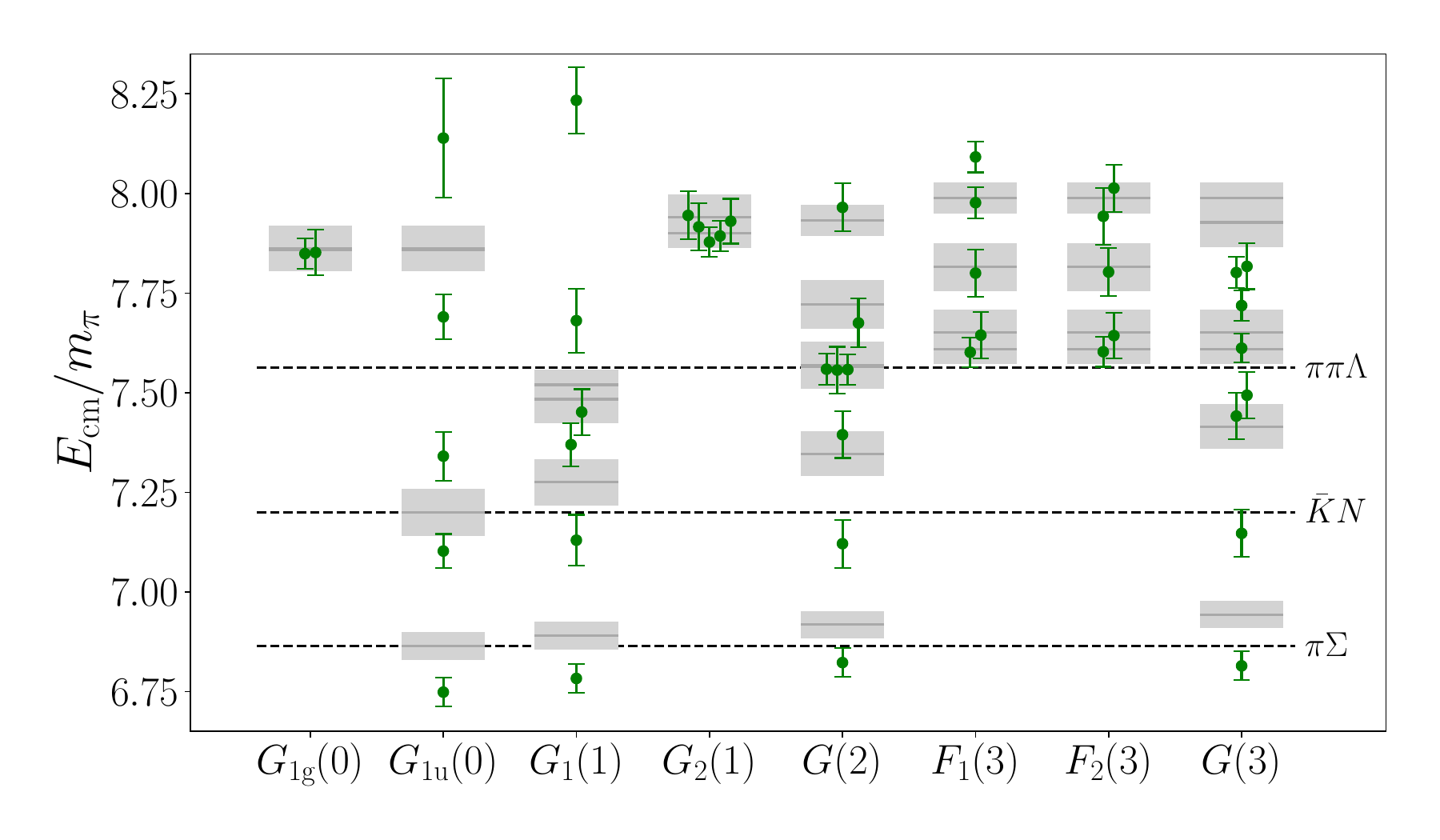}
\end{center}
\caption{Finite-volume stationary-state energy spectrum, shown as green points, in the 
center-of-mass frame for total isospin $I=0$, strangeness $S=-1$, and various symmetry channels 
indicated along the horizontal axis.  The gray bands show the locations of energy sums
for non-interacting two-particle combinations.  Various two and three particle thresholds
are shown as dashed horizontal lines. Energies are shown as ratios over the pion mass $m_\pi$.
\label{fig:energy_spectrum}}
\end{figure}

A summary of the total isospin $I=0$ and strangeness $S=-1$ spectrum in the center-of-mass frame 
for various symmetry channels is presented in Fig.~\ref{fig:energy_spectrum}.  
Each center-of-mass energy $E_{\rm cm}$ is obtained from the corresponding lab-frame
energy $E_{\rm lab}$ using
\begin{equation}
  E_{\rm cm}=\sqrt{E_{\rm lab}^2-\bm{P}^2}.
\end{equation}
The results are compared to the energy sums of non-interacting two-particle combinations.

\section{Scattering amplitudes}
\label{sec:scatamp}

Our determinations of the scattering amplitudes are presented in this section.
First, the quantization condition that relates the amplitudes to the finite-volume
spectrum is discussed.  Then the parametrizations of the $K$-matrix that we use
are described, and fits using these parametrizations and the quantization
condition are presented. 

\subsection{Quantization condition}
\label{sec:quant}

In lattice QCD, scattering amplitudes are obtained by solving a quantization 
condition~\cite{Luscher:1990ux,Rummukainen:1995vs, Kim:2005gf, He:2005ey, Bernard:2010fp, 
Gockeler:2012yj,Briceno:2012yi, Briceno:2014oea} which relates the amplitudes to the finite-volume 
spectrum of center-of-mass energies $E_{\rm cm}$.  Generally, the lowest partial wave amplitudes
must be somehow parametrized, then best-fit values of the parameters are determined by 
matching the finite-volume spectrum produced by the quantization condition to that
evaluated in lattice QCD.

The form of the quantization condition used here is given by
\begin{equation}
\label{eq:det}
{\rm det}[ \widetilde{K}^{-1}(E_{\rm cm}) - B^{\boldsymbol{P}}(E_{\rm cm})] 
  + {\rm O}({\rm e}^{-M L}) = 0 \ ,
\end{equation}
where $\widetilde{K}$ is related to the usual scattering $K$-matrix as described below, and
$B^{\boldsymbol{P}}(E_{\rm cm})$ for a particular total momentum 
$\boldsymbol{P} = (2\pi/L) \boldsymbol{d}$, with $\boldsymbol{d} \in \mathbb{Z}^3$, is the 
so-called box matrix, using the notation of Ref.~\cite{Morningstar:2017spu}. 
In Eq.~(\ref{eq:det}), the neglected correction terms are suppressed exponentially with the 
spatial extent $L$ and some relevant energy scale $M$, typically the pion mass.
Eq.~(\ref{eq:det}) applies only for energies below all thresholds of states containing
three or more particles.  The determinant can be taken over all unit normalized two-hadron
states $\vert Jm_J\ell Sa\rangle$
specified by total angular momentum $J$, the projection of $J$ along the $z$-axis $m_J$, 
the orbital angular momentum $\ell$, the total intrinsic spin $S$, and particle species $a$.  
Here, $a=0,1$, where species channel 0 is $\pi\Sigma$ and species channel 1 is $\bar{K}N$, 
and total spin $S=1/2$ is fixed, and therefore, need not be explicitly indicated.

The box matrix $\langle J'm'_J\ell' S'a'\vert\ B^{\boldsymbol{P}}(E_{\rm cm})\ \vert Jm_J\ell Sa\rangle$ 
encodes the effects of inserting the partial waves into the cubic box so as to maintain
the periodic boundary conditions.  This matrix is diagonal 
in the indices corresponding to total intrinsic spin and particle species, but not to any of the
other indices.  In particular, states of different total angular momentum can mix.
One can show that, under any symmetry transformation $Q$ of the cubic box which is an element 
of the little group of $\bm{P}$, the box matrix transforms as 
$QB^{\boldsymbol{P}}Q^\dagger=B^{\boldsymbol{P}}$.  This implies that the box matrix
can be block diagonalized by projecting onto the superpositions of states that transform
according to the irreps of the little group.  The $\widetilde{K}$ matrix similarly
block diagonalizes in such a basis, except for the total intrinsic spin and particle
species indices.  Hence, the determinant in Eq.~(\ref{eq:det}) can be dealt with separately
block by block.  A particular block can be denoted by the finite-volume irrep 
$\Lambda(\boldsymbol{d}^2)$ and a row of this irrep $\lambda$.  Since the spectrum is independent
of the row $\lambda$, this index is henceforth omitted. For a particular block, the 
block-diagonalized box-matrix is denoted $B^{\Lambda(\boldsymbol{d}^2)}_{J'\ell' n'; J\ell n}$,
where $n,n'$ are irrep occurrence numbers.  The expressions for all elements of $B^{\Lambda}(\bm{d}^2)$ 
relevant for this work are given in Ref.~\cite{Morningstar:2017spu}.  After transforming to the 
block diagonal matrix, the $\widetilde{K}$ matrix has the form given by Eq.~(35) in 
Ref.~\cite{Morningstar:2017spu}.  A truncation $\ell\leq\ell_{\rm max}$ in each block then makes 
the determinant condition manageable.

We use the same definition of the $K$-matrix as described in Ref.~\cite{Morningstar:2017spu}.
This matrix is real and symmetric and diagonal in total angular momentum and its projection:
\begin{equation}
\langle J'm'_J\ell' S'a'\vert K \vert Jm_J\ell Sa\rangle
 = \delta_{J'J}\delta_{m'_J m_J} K^{(J)}_{\ell'S'a'; \ell Sa}(E_{\rm cm}).
\end{equation}
The matrix $\widetilde{K}$ here is defined by
\begin{equation}
 \widetilde{K}^{(J)-1}_{\ell'S'a';\ \ell Sa}(E_{\rm cm})=\left(\frac{k_{a'}}{m_\pi}\right)^{\ell'+\frac{1}{2}}
 \!\!\!\!\!\!\!\! K^{(J)-1}_{\ell'S'a';\ \ell Sa}(E_{\rm cm})
  \left(\frac{k_a}{m_\pi}\right)^{\ell+\frac{1}{2}}\!\!\!,
\end{equation}
where
\begin{eqnarray}
    k_{0}^{2}&=& k_{\pi\Sigma}^2=\frac{1}{4E^{2}_{\rm cm }}\lambda_K(E^{2}_{\rm cm}, m_{\pi}^{2}, m_{\Sigma}^{2}),\\
    k_{1}^{2}&=& k_{\bar{K}N}^2=\frac{1}{4E^{2}_{\rm cm }}\lambda_K(E^{2}_{\rm cm}, m_{\bar{K}}^{2}, m_{\rm N}^{2}),
\end{eqnarray}
and $\lambda_K$ is the K\"all\'en function~\cite{KallenBook}  
\begin{equation}
    \lambda(x,y,z) = x^2+y^2+z^2-2xy-2xz- 2yz.
\end{equation}
Note that this definition of $\widetilde{K}$ differs
very slightly from that given in Ref.~\cite{Morningstar:2017spu}, and hence, the box matrix
here also is slightly different.  Simple factors related to the spatial extent $L$ have
been removed here.

In this work, the total intrinsic spin $S=S'=1/2$ is fixed, so these indices can be omitted.
For elastic scattering, $\widetilde{K}$ is diagonal in the orbital angular momentum, so
we take $\ell=\ell'$.  We restrict our attention to $\ell\leq\ell_{\rm max}$ where
$\ell_{\rm max}=0,1$.  Since parity for $\pi\Sigma$ and $\bar{K}N$ states is $P=(-1)^{\ell+1}$,
where the product of the intrinsics parities is $-1$, and we include only $\ell=0,1$
waves, we can change notation to remove $\ell$ in favor of $J^P$.  Thus, we use the
notation $\widetilde{K}^{(J^P)}_{a'a}(E_{\rm cm})$ in what follows, with $J^P=1/2^-$ for
$\ell_{\rm max}=0$ and $J^P=1/2^-,\, 1/2^+,\, 3/2^+$ for $\ell_{\rm max}=1$.

\subsection{Parametrizations and fits for $\ell_{\rm max}=0$}
\label{sec:lmaxzero}

For $\ell_{\rm max} = 0$, the finite-volume spectrum shown in Fig.~\ref{fig:energy_spectrum} constrains 
the coupled-channel scattering amplitude via Eq.~(\ref{eq:det}) at center-of-mass energies near the 
$\pi\Sigma$ and $\bar{K}N$ thresholds. All irreps in Table~1 of 
Ref.~\cite{Bulava:2022vpq} to which the $J^P=1/2^-$ partial wave contributes are employed.

Six different types of parametrizations of $\widetilde{K}$ are studied here 
using $\ell_{\rm max}=0$. In the expressions below, the subscripts $i$ and $j$ denote either of 
the two scattering channels (channel 0 is $\pi\Sigma$ and channel 1  is $\bar KN$), and the quantity
\begin{equation}
\Delta_{\pi \Sigma}(E_{\rm cm}) = \frac{E_{\rm cm}^{2} - (m_{\pi} + m_{\Sigma})^{2}}{(m_{\pi} 
 + m_{\Sigma})^{2}},
\end{equation}
is related to the center-of-mass energy gap above $\pi\Sigma$ threshold.  
The matrices $A$, $B$, $\widehat{A}$, $\widehat{B}$,
$\widetilde{A}$, $\widetilde{B}$, $\widehat C$, $A'$ and $B'$ below are 
real and symmetric.  The six forms of parametrizations are as follows:
\begin{enumerate}
\item An effective range expansion (ERE) of the form
    \begin{equation}
    \widetilde{K}_{ij} = \frac{m_\pi}{E_{\rm cm}}\Bigl(A_{ij} + B_{ij} \Delta_{\pi \Sigma}(E_{\rm cm})\Bigr).
     \label{eq:Kfitone}
  \end{equation}
   \item A variation of the first parametrization without the factor of $m_\pi/E_{\rm cm}$:
   \begin{equation}
   \widetilde{K}_{ij} = \widehat A_{ij} + \widehat B_{ij} \Delta_{\pi \Sigma}(E_{\rm cm}).
     \label{eq:Kfittwo}
  \end{equation}
   \item An ERE of $\widetilde{K}^{-1}$ of the form
    \begin{equation}
    \widetilde{K}^{-1}_{ij} = \frac{E_{\rm cm}}{m_\pi}
   \left(\widetilde A_{ij} + \widetilde B_{ij} \Delta_{\pi \Sigma}(E_{\rm cm}) \right).
     \label{eq:Kfitthree}
  \end{equation}
  \item A Blatt-Biederharn~\cite{Blatt:zz1952a} parametrization:
   \begin{equation}
     \widetilde{K} = C\ F\ C^{-1},
     \label{eq:Kfitfour}
   \end{equation}
   where
   \begin{eqnarray}
     C &=& \begin{pmatrix}
       \cos \epsilon & \sin \epsilon \\
       -\sin \epsilon & \cos \epsilon
   \end{pmatrix},\\
    F &=&
   \begin{pmatrix}
       f_0(E_{\rm cm}) & 0 \\ 0 &  f_1(E_{\rm cm})
   \end{pmatrix},
   \end{eqnarray}
  and
  \begin{equation}
      f_i(E_{\rm cm}) = \frac{m_\pi}{E_{\rm cm }} \frac{a_i + b_i \Delta_{\pi \Sigma}(E_{\rm cm})
         }{ 1 + c_i \Delta_{\pi \Sigma}(E_{\rm cm}) }.
  \end{equation}
  \item A parametrization based on the leading-order Weinberg-Tomozawa term~\cite{Oset:1997it}:
  \begin{equation}
       \widetilde{K}_{ij} = \frac{\widehat C_{ij}}{m_\pi} \left( 2 E_{\rm cm} - M_i - M_j \right),  
        \label{eq:Kfitfive}
  \end{equation}
  where $M_0 = m_\Sigma$ and  $M_1 = m_N$.
  \item An expansion that is linear in the energy around the $\pi \Sigma$ threshold:
  \begin{equation}
      \widetilde{K}_{ij} = A'_{ij} + \frac{B'_{ij}}{m_\pi} \left(E_{\rm cm } - m_\pi - m_\Sigma  \right).
     \label{eq:Kfitsix}
  \end{equation}
\end{enumerate}

Making use of these forms for $\widetilde{K}$, fits are carried out to determine the
best-fit values of the above parameters using the spectrum method~\cite{Guo:2012hv}. 
The correlated-$\chi^2$ minimized in these fits is defined similar to Eq.~(14) in 
Ref.~\cite{Bulava:2022vpq}, except that the residuals are formed in terms of
differences between the center-of-mass energy interaction shifts $\Delta E_{\rm cm}$ obtained 
from the quantization condition using the $\widetilde{K}$-matrix parametrization
and the interaction shifts determined from the Monte Carlo computations.
For each of the six parametrizations above, several fits were carried out,
setting various parameters to zero to check sensitivity to these parameters.
The results are presented in Tables~\ref{tab:fam1} to \ref{tab:fam6} of
Appendix~\ref{sec:fitres}.  In these tables, statistical uncertainties are estimated 
using a simple derivative method, as described in Eq.~(4.2) of Ref.~\cite{Draper:2023boj}. 
This method is faster than other methods, but it often yields slightly underestimated 
values.  However, this method is sufficient for the purposes of illustrating model 
dependence.  Each fit in these tables also shows the value of the Akaike Information 
Criterion (AIC)~\cite{Jay:2020jkz} defined by
\begin{equation}
    \text{AIC} = \chi^2 - 2\, n_{\rm dof},
\end{equation}
where $n_{\rm dof}$ denotes the number of degrees of freedom.

The best-fit values for the $\widetilde{K}$-matrix parameters in Tables~\ref{tab:fam1} to 
\ref{tab:fam6} show large variations from fit to fit.  However, only more physically-relevant
quantities, such as the scattering amplitudes or the $S$-matrix pole positions, 
are important.  We define a quantity $t_{ij}^{(J^P)}(E_{\rm cm})$ which is proportional 
to the scattering transition amplitude and is related to $\widetilde{K}$ by
\begin{equation}
t^{-1} = \widetilde{K}^{-1} -i \widehat k, \quad 
\label{eq:amplitude}
\end{equation}
where $m_\pi\widehat k = \text{diag }(k_{\pi \Sigma}, k_{\bar{K}N})$. 
The different para\-metri\-zations discussed above constrain the energy dependence of the 
amplitudes near the finite-volume energies, even if they do not accommodate left-hand 
(cross-channel) cuts. Knowledge over this limited range enables the analytic 
continuation of $t_{ij}(E_{\rm cm})$ to complex $E_{\rm cm}$ 
and the identification of poles close to the real axis on sheets adjacent to the 
physical one. Analytic continuation of the coupled channel $\pi\Sigma-\bar{K}N$ scattering 
amplitude involves four different Riemann sheets, each labelled by the sign of the 
imaginary parts of $(k_{\pi\Sigma},k_{\bar{K}N})$, with $(+,+)$ denoting the physical sheet. 
Complex poles in the scattering amplitudes correspond to vanishing eigenvalues in the 
right-hand side of Eq.~(\ref{eq:amplitude}) and are determined numerically. In the vicinity 
of a pole, the divergent part of the amplitude is
\begin{equation}
     t =  \frac{m_\pi}{  E_{\rm cm} -  E_{\rm pole}} \begin{pmatrix}
        c_{\pi\Sigma}^2 & c_{\pi\Sigma} \, c_{\bar{K}N} \\c_{\pi\Sigma} \, c_{\bar{K}N}  &  c_{\bar{K}N}^2
     \end{pmatrix}\,+\,\dots,
\end{equation}
where the (complex) residues $c_{\pi\Sigma}$, $c_{\bar{K}N}$ represent the coupling of the 
resonance pole to each channel.

\begin{figure}
\includegraphics[width=\linewidth]{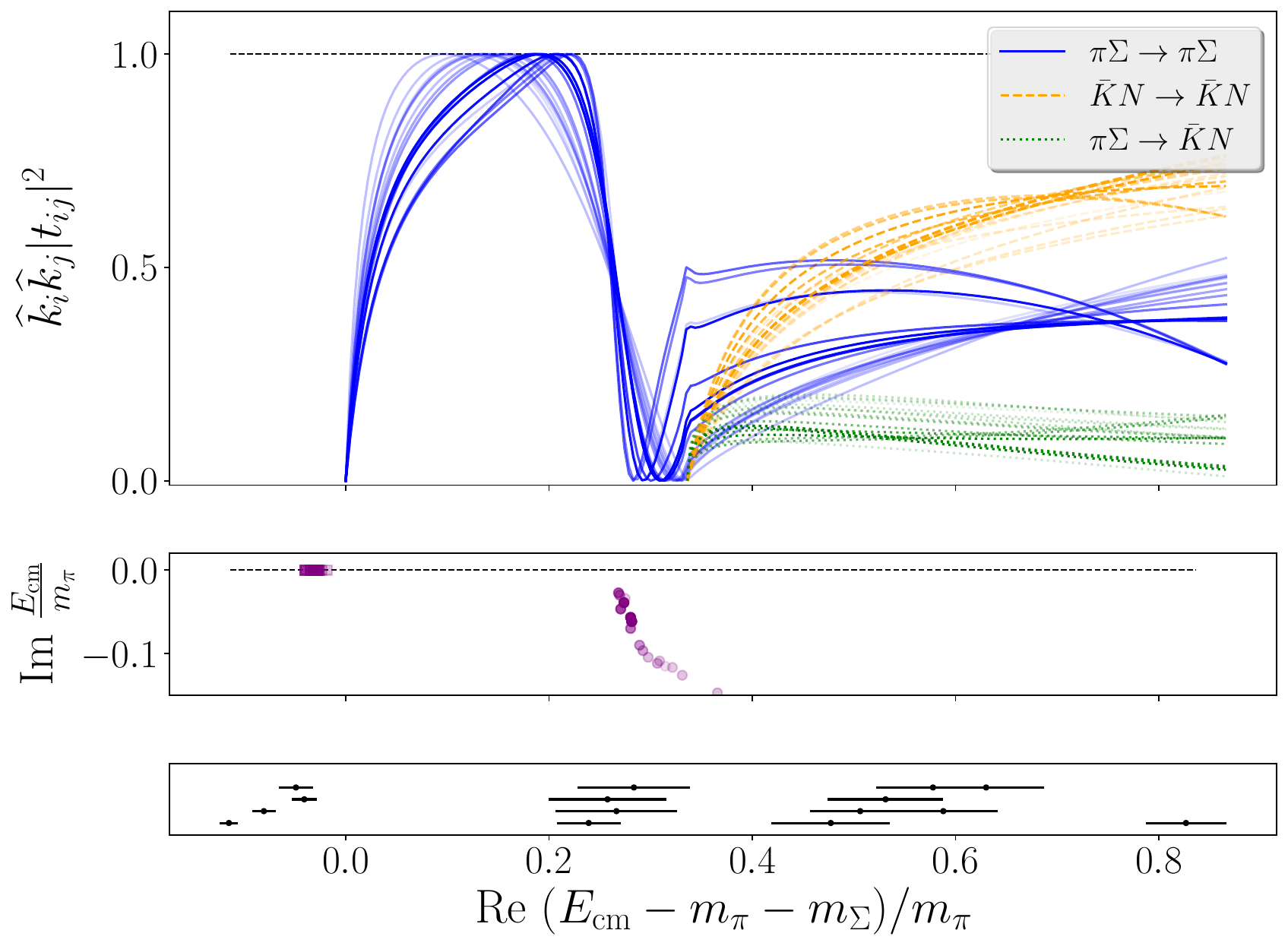}
\caption{ Scattering amplitudes and $S$-matrix pole locations against center-of-mass 
 energy difference to the $\pi\Sigma$ threshold for various fits using the six
 different $\widetilde{K}$-matrix parametrizations.  The quantities $t$ and $\hat{k}$ are 
 defined in Eq.~(\ref{eq:amplitude}), and the subscripts $i,j$ refer to the flavor channels.
 The different fits in Tables~\ref{tab:fam1} to \ref{tab:fam6} are shown as different lines in 
 the upper panel, with corresponding pairs of points in the middle panel which show the positions 
 of the $S$-matrix poles in the complex center-of-mass energy plane. The transparency parameter in 
 matplotlib~\cite{Hunter:2007} of each line and corresponding pair of points is set 
 to be  ${\texttt{alpha}=\exp (- \left(\text{AIC} - \text{AIC}_\text{min}\right)/2)}$. 
 The bottom panel shows the finite-volume spectrum used to
 constrain the fits involving the transition amplitudes.
 \label{fig:syst}}
\end{figure}

Results for the scattering transition amplitudes and the pole locations for each of
the fits in Tables~\ref{tab:fam1} to \ref{tab:fam6} are shown in Fig.~\ref{fig:syst}. 
The transition amplitudes are shown in the upper panel, and the resulting poles
from the analytic continuations are shown in the middle panel of 
Fig.~\ref{fig:syst}.  Each line in the upper panel corresponds to a different fit
in Tables~\ref{tab:fam1} to \ref{tab:fam6}, with corresponding pairs of points
in the middle panel.  The transparency of the lines and points is directly related to
the value of the AIC, darker implying a lower AIC.  One sees that the variations between
the results of the transition amplitudes and the $S$-matrix poles from the different 
fits are no longer large.

\begin{figure}
\includegraphics[width=\linewidth]{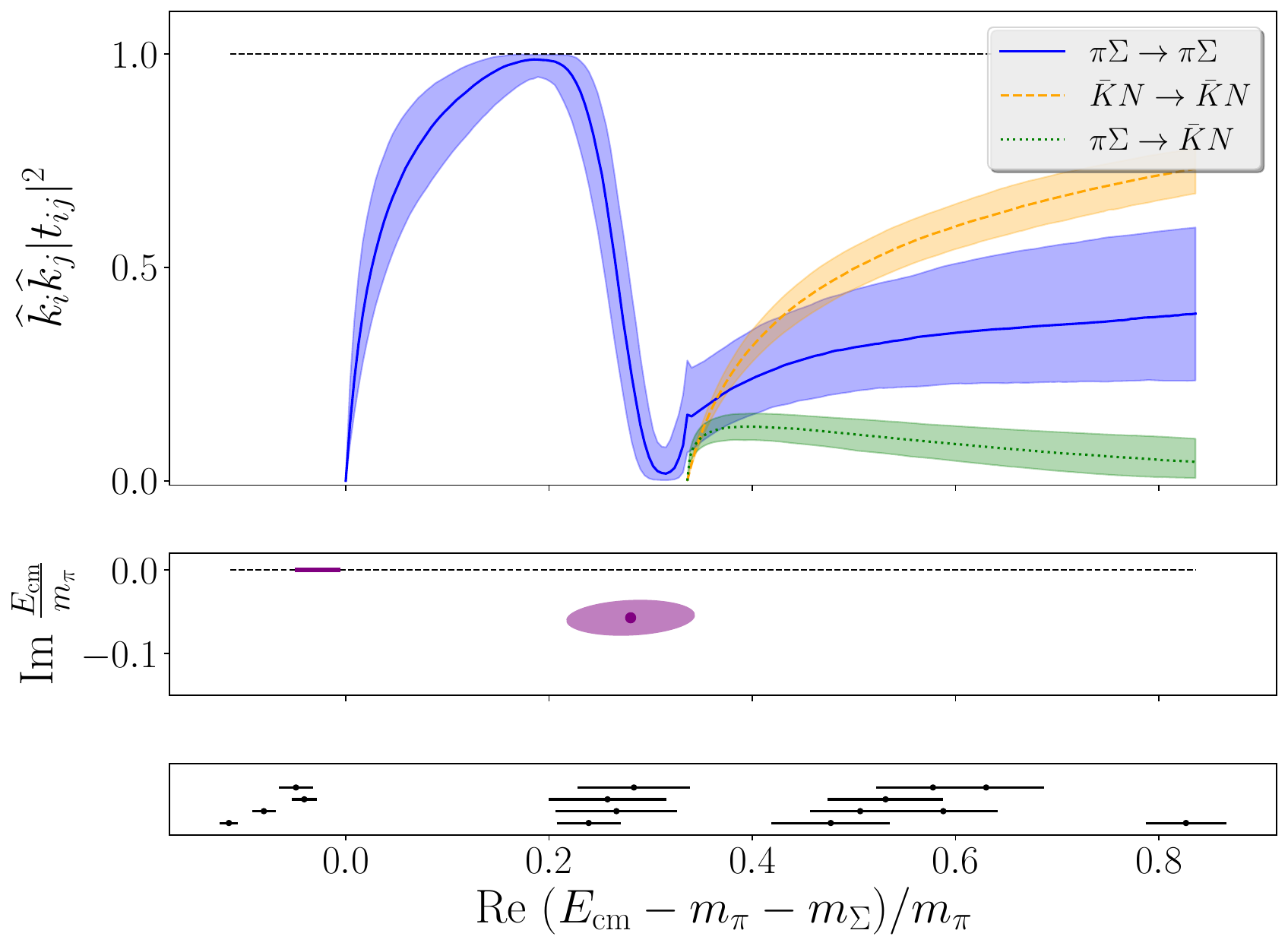}
\caption{The isospin $I=0$ and strangeness $S=-1$ coupled-channel $\pi\Sigma-\bar{K}N$ transition
 amplitudes computed on a single lattice QCD gauge-field ensemble with $m_{\pi} \approx 200\,{\rm MeV}$ 
 as a function of the center-of-mass energy difference to the $\pi\Sigma$ threshold.  
 Results are obtained using the best fit specified by Eqs.~(\ref{eq:bestfit}) and (\ref{eq:Kfitone}), 
 with uncertainties estimated by bootstrap resampling. The quantities $t$ and $\hat{k}$ are 
 defined in Eq.~(\ref{eq:amplitude}), and the subscripts $i,j$ refer to the flavor channels. 
 The middle panel shows the positions of the $S$-matrix poles in the complex center-of-mass energy 
 plane on the sheets closest to the physical one.  The bottom panel shows the finite-volume 
 spectrum used to constrain the fits involving the transition amplitudes.
 \label{fig:main}}
\end{figure}

\begin{figure}
 \centering
 \includegraphics[width=\linewidth]{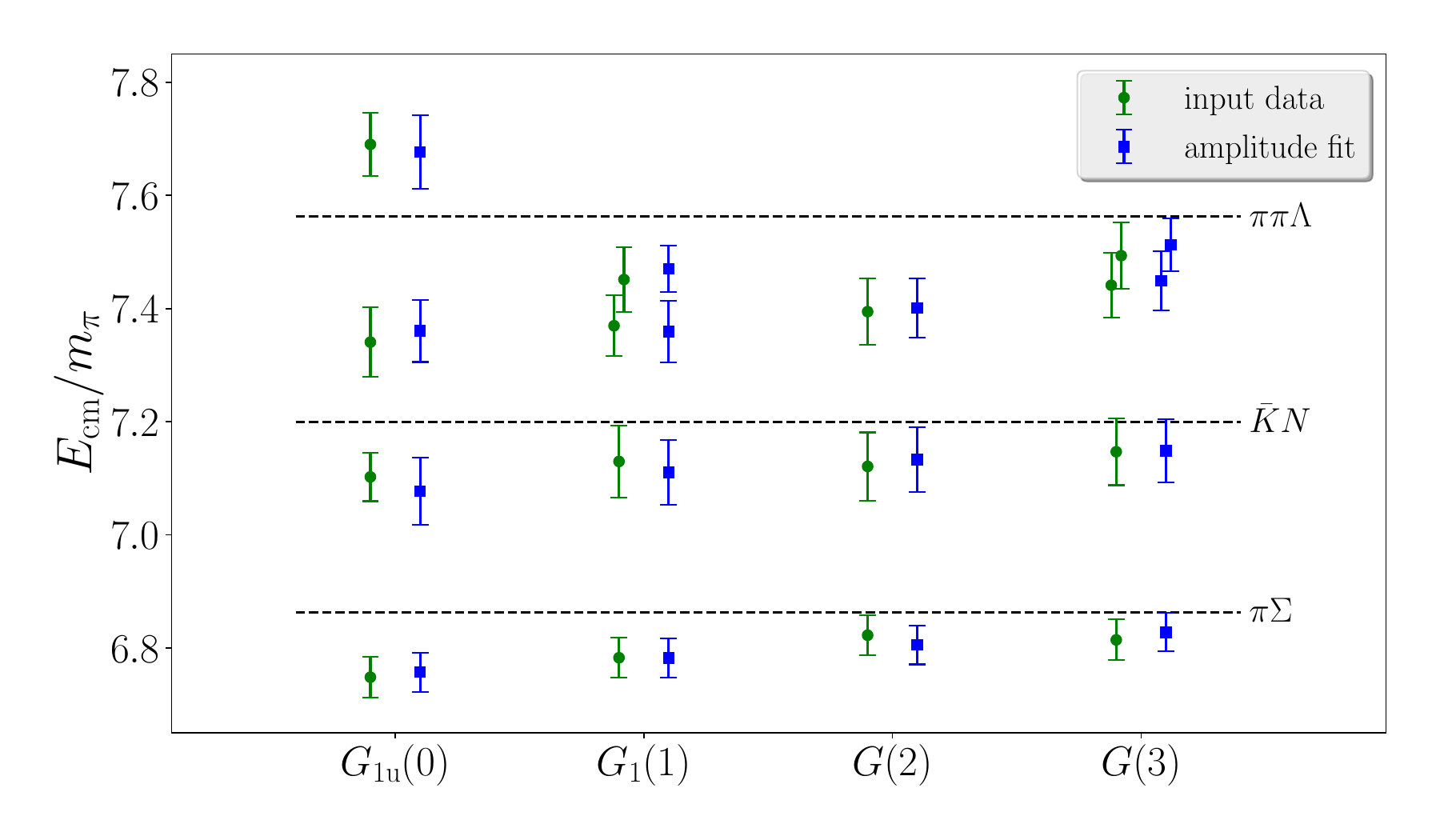}
 \caption{Finite-volume spectrum in the center-of-mass frame used as input data to constrain 
  parametrizations of the coupled-channel $\pi\Sigma$-$\bar{K}N$ scattering amplitude (green circles).
  Each column corresponds to a particular irrep $\Lambda(\boldsymbol{d}^2)$ of the little group of 
  total momentum $\boldsymbol{P}^2=(2\pi/L)^2\boldsymbol{d}^2$. 
 Only irreps where the $\ell=0$ partial wave contributes are included.  
 Dashed lines indicate various thresholds, as labeled. Model energies from the resultant fit are 
 shown as blue squares.
 \label{fig:spectrum_fit}}
\end{figure}

The first parametrization, given by Eq.~(\ref{eq:Kfitone}), leads to the lowest AIC, i.e.~AIC$_\text{min}$.  
Given this, a fit is carried out using this parametrization and the errors are more accurately 
determined using a bootstrap procedure with 800 samples.  This analysis also accounts for 
fluctuations in the single hadron masses, and thus the errors are larger than the ones estimated using 
the derivative method in Table~\ref{tab:fam1}.  The fit with the lowest AIC value is a 
four-parameter fit of the form of Eq.~(\ref{eq:Kfitone}), and the best-fit parameters values are
\begin{align}
\begin{split}
    &A_{00} = 4.1(1.8), \quad \ \, A_{11}=-10.5(1.1), \\ &A_{01}=10.3(1.5),
  \quad B_{01}=-29(18), 
\end{split}
\label{eq:bestfit}
\end{align}
with fixed $B_{00}=B_{11}=0$ and $\chi^{2}=10.52$ for 11 degrees of freedom. This fit 
is shown in Fig.~\ref{fig:main}.  
The finite-volume energies that are used in this fit are shown in Fig.~\ref{fig:spectrum_fit}.
The green points show the center-of-mass energies obtained from the Monte Carlo lattice
QCD computations, and the blue points show the energies obtained from the fit using the
quantization condition and the $\widetilde{K}$-matrix parametrization of Eq.~(\ref{eq:Kfitone})
with best-fit values given in Eq.~(\ref{eq:bestfit}).

For this preferred fit, two poles are found on the  $(-,+)$ sheet,  which is the one 
closest to physical scattering in the region between the two thresholds, with energies
\begin{align}\label{eq:poles}
\begin{split}
E_1/m_\pi &= 6.856(45)_\text{st}(06)_\text{md}, \\
E_2/m_\pi &= 7.144(63)_\text{st}(07)_\text{md} - i \, 0.057(22)_\text{st}(17)_\text{md},
\\
\\
E_1 &= 1392(9)_\text{st} (2)_\text{md} (16)_a \text{ MeV}
\\
E_2 &=  [1455(13)_\text{st}(2)_\text{md}(17)_a \\&\phantom{=}- i 
\times 11.5(4.4)_\text{st}(4)_\text{md}(0.1)_a] \text{ MeV},
\end{split}
\end{align}
while their couplings are
\begin{align}
\begin{split}
c^{(1)}_{\pi\Sigma} &= i \, 0.52(10)_\text{st}(2)_\text{md} , 
\\
c^{(1)}_{\bar{K}N} &= i \, 0.28(8)_\text{st}(6)_\text{md} ,
\\
c^{(2)}_{\pi\Sigma} &=  0.26(9)_\text{st}(5)_\text{md} - i \,0.13(3)_\text{st}(3)_\text{md} ,
\\
c^{(2)}_{\bar{K}N} &=  0.12(6)_\text{st}(4)_\text{md} - i\, 0.53(4)_\text{st}(2)_\text{md} .
\end{split}
\end{align}
The ratios of these couplings (with correlated uncertainties) show that the lower pole is more 
strongly coupled to the $\pi\Sigma$ channel, while the pole at a larger real energy is more 
strongly coupled to the $\bar{K}N$ channel:
\begin{align}
\begin{split}
\left|\frac{c^{(1)}_{\pi\Sigma}}{c^{(1)}_{\bar{K}N}}\right| &= 1.9(4)_\text{st}(6)_\text{md}, 
\\ \left| \frac{c^{(2)}_{\pi\Sigma}}{c^{(2)}_{\bar{K}N}}\right| &= 0.53(9)_\text{st}(10)_\text{md}.
\end{split}
\label{eq:couplings}
\end{align}
In the above results, the
first uncertainty is statistical, the second accounts for the model parametrization dependence, 
and when the pole positions are quoted in physical units, the third comes from the scale setting 
uncertainty in Table~\ref{tab:comp_deets}.  For this work, the systematic uncertainty due to the 
parametrization dependence is estimated by considering all models with 
$\text{AIC} - \text{AIC}_\text{min} < 1$, and taking half the difference in the maximal spread of 
values as the model uncertainty. In future work, when a full extrapolation to the physical point 
is performed, a more thorough determination of the model-average and systematic uncertainty will 
be performed.

As shown in Fig.~\ref{fig:syst}, the existence of two poles remains robust under variations 
of the underlying parametrization of the $K$-matrix.  Our $\widetilde{K}$-matrix parametrizations
make no assumptions about the number or locations of the $S$-matrix poles, and can accommodate
zero, one, or two poles.  The pole at $E_1$ is most likely a virtual bound state 
(except in 0.5\% of the bootstrap samples where the pole is in the physical sheet and is thus a 
bound state), while the one at $E_2$ is a resonance. The first pole has a stronger coupling 
to $\pi \Sigma$, while the second couples more strongly to $\bar{K}N$. 

\begin{figure}
\includegraphics[width=\linewidth]{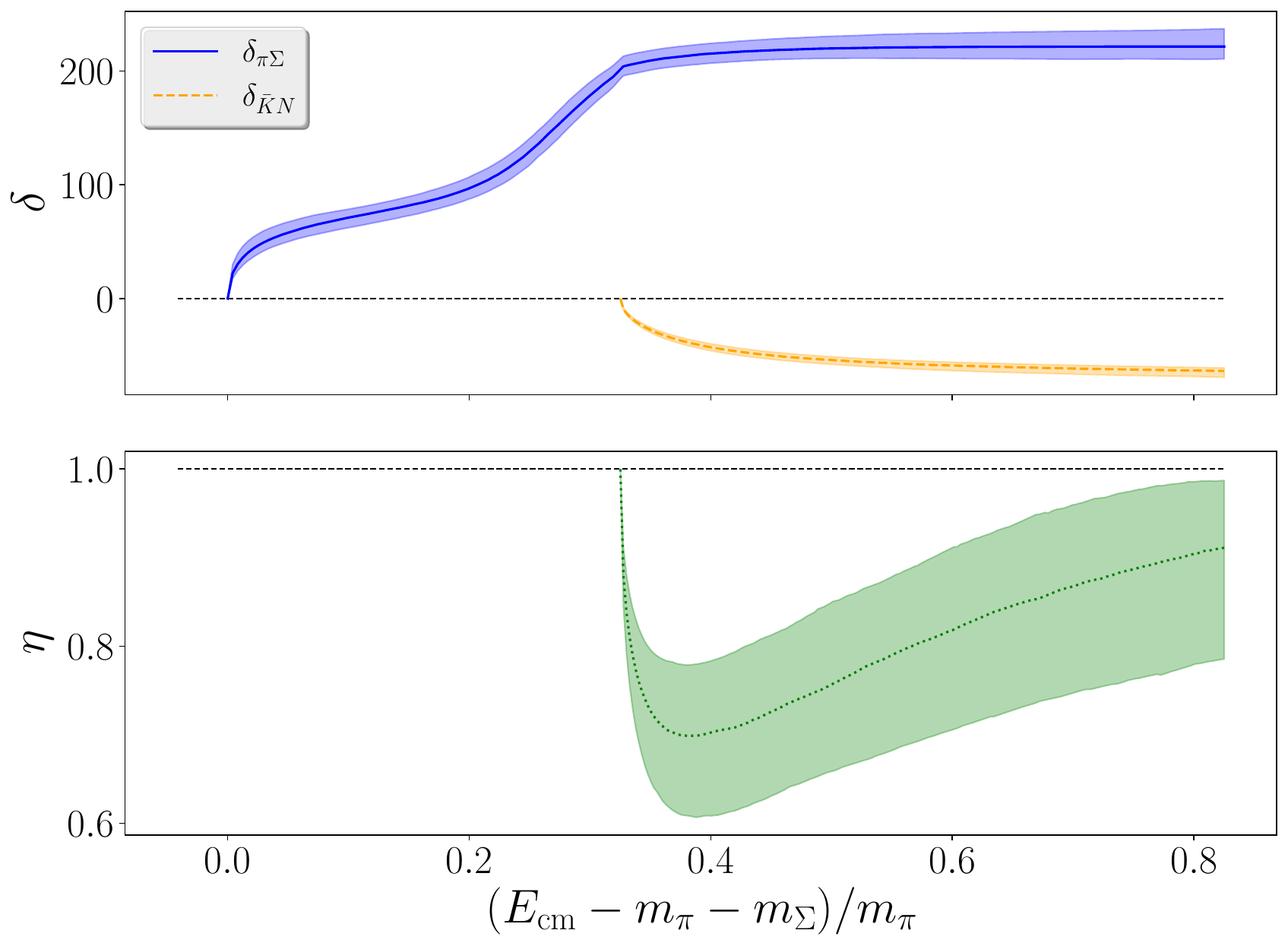}
\caption{ Inelasticity $\eta$ and phase shifts $\delta_{\pi\Sigma}$ and $\delta_{\bar{K}N}$ 
 against center-of-mass energy difference to the $\pi\Sigma$ threshold.
 These quantities are defined in Eq.~(\ref{eq:inelas}).  Results are obtained using the 
 best fit specified by Eqs.~(\ref{eq:bestfit}) and (\ref{eq:Kfitone}), with 
 uncertainties estimated by bootstrap resampling.
 \label{fig:inelas}}
\end{figure}

Another way of presenting the results for the amplitudes is to show the scattering phase 
shifts $\delta_i$ and the inelasticity $\eta$, which are related to $t$ by
\begin{equation}
\begin{aligned}
t_{00}  &=\frac{m_\pi(\eta e^{2 i \delta_{\pi\Sigma}}-1)}{2 i  k_{\pi\Sigma}}, \\
t_{11}  &=\frac{m_\pi(\eta e^{2 i \delta_{\bar{K}N}}-1)}{2 i  k_{\bar{K}N}} , \\
t_{01}  &=\frac{m_\pi\sqrt{1-\eta^2} e^{i\left(\delta_{\pi\Sigma}+\delta_{\bar{K}N}\right)}}{2 
  \sqrt{ k_{\pi\Sigma} k_{\bar{K}N}}},
\end{aligned}
\label{eq:inelas}
\end{equation}
where the indices again indicate the flavor channel: $0$ for $\pi\Sigma$ and 1 for $\bar{K}N$.
The results are shown in Fig.~\ref{fig:inelas}. This figure illustrates two features that can
be related to the two-pole structure. First, a rapid increase of the phase at the $\Sigma \pi$ 
threshold is related to the presence of a close virtual bound state. Second, the phase 
crosses $90$ degrees near the position of the real part of the second pole. 

\subsection{Single-channel $\pi\Sigma$ scattering}
\label{sec:pisigmascat}

In the vicinity of the $\pi \Sigma$ threshold, and well below the $\bar K N$ threshold, the 
standard single-channel L\"uscher formalism can be used to study the elastic $\pi \Sigma$ 
scattering amplitude. Compared to the multi-channel approach, the 
single-channel approach is equivalent up to effects that are exponentially-suppressed with 
respect to the distance to the $\bar K N$ threshold. Such an analysis is used here to show
further evidence of the existence of the lower pole and its nature as a virtual bound state.

\begin{figure}[ht]
 \includegraphics[width=\linewidth]{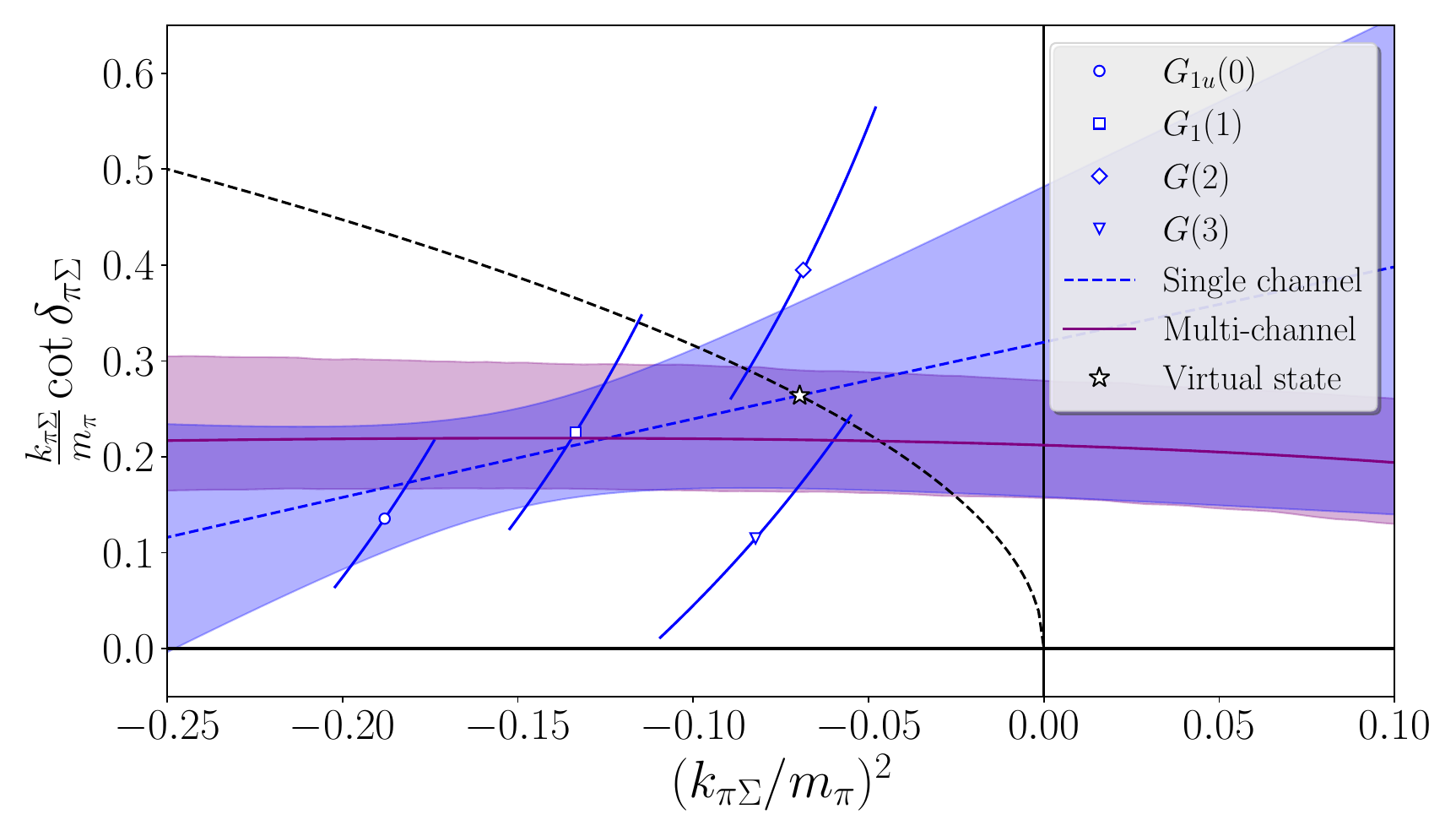}
 \caption{The $\pi\Sigma$ elastic phase shift as a function of the center-of-mass momentum squared,
 determined in a single-channel analysis and compared against the multi-channel result.
 The dashed black line corresponds to the virtual bound state condition in Eq.~(\ref{e:vbs}). The blue 
 dashed line and corresponding band 
 show the fit to an effective range expansion with statistical errors. The star labels the position of 
 the virtual bound state in the single-channel analysis. The solid purple line and the associated band 
 show the result from the multi-channel fit with lowest AIC value. The scattering phase shifts from 
 each energy level are shown by the hollow blue  symbols. 
\label{fig:1channel}}
\end{figure}

We perform this single-channel analysis using the lowest energies in each frame in the 
$G_{\rm 1u}(0), G_1(1)$, $G(2)$ and $G(3)$ irreps (4 in total), as shown in Fig.~\ref{fig:spectrum_fit}. 
The elastic scattering phase shift can be parametrized by
\begin{equation}
    \frac{k_{\pi\Sigma}}{m_\pi} \cot \delta_{\pi \Sigma} = \frac{E_{\rm cm}}{m_\pi}(a_{\pi\Sigma} 
+ b_{\pi\Sigma} \Delta_{\pi\Sigma}  ),
\end{equation}
where $a_{\pi\Sigma}$ and $b_{\pi\Sigma}$ are fit parameters.  The results of performing a 
fit using the spectrum method as in Ref.~\cite{Bulava:2022vpq} are
\begin{equation}
    a_{\pi\Sigma} = 0.047(14), \quad b_{\pi\Sigma} = 0.65(50), \quad \chi^2= 5.04,
\end{equation}
with 2 degrees of freedom.
A visualization of the single-channel phase shift from the fit is shown in Fig.~\ref{fig:1channel}, 
along with a comparison to the elastic phase shift from the multi-channel analysis in 
Eq.~(\ref{eq:bestfit}). One observes that the 
phase shift curve (blue band) intersects the virtual bound state condition (black dashed line). That is,
\begin{equation}\label{e:vbs}
   k_{\pi\Sigma} \cot \delta_{\pi \Sigma} - i   k_{\pi\Sigma} = 0,
\end{equation}
for purely imaginary and negative $k_{\pi\Sigma}$.
The position of the virtual bound state is found at
\begin{align}
\begin{split}
E_1/m_\pi &= 6.822(37),
\\
E_1 &= 1389(8)_\text{st} (16)_a \text{ MeV}.
\end{split}
\end{align}
Thus, the results are consistent with those obtained in the multichannel analysis, and 
confirm the existence of the lower pole in a model-independent way.  The larger $\chi^2$ per 
degree of freedom compared to the multi-channel analysis may arise from the proximity of the $\bar K N$
threshold, suggesting the need for the multi-channel approach to study this system. 

\subsection{Effect of higher partial waves}
\label{sec:higherpw}

The analysis above only includes $s$ waves and the $J^P=1/2^-$ amplitudes. 
For the energies in the rest frame in the $G_{\rm 1u}(0)$ irrep, this is a good approximation, 
since this irrep only receives contamination from $\ell=4$ partial waves. By contrast, energy 
levels in moving frames, such as the $G_{1}(1)$, $G(2)$, and $G(3)$ irreps, receive contamination 
from $p$-waves. In particular, the dominant contamination from higher partial waves is 
expected to be from the $J^P=1/2^+$ and $J^P=3/2^+$ channels. The scattering amplitude
for $J^P=1/2^+$ is the leading contribution to the finite-volume energy shifts in the 
$G_{\rm 1g}(0)$ irrep, while in the $F_1(3)$ and $F_2(3)$ irreps, the $J^P=3/2^+$ wave dominates. 
Since a few energy levels in these irreps are determined in this work, as shown
in Fig.~\ref{fig:energy_spectrum}, we can use these energies to estimate the strength of the 
interactions in these two waves and how they impact the main fit of this work.

To assess this, we perform two additional fits. In the first additional fit, sixteen energies
are used, which include the fifteen energies from the $s$-wave fits and one $G_{\rm 1g}(0)$ level, 
and we incorporate a $J^P=1/2^+$ wave in the $\widetilde{K}$-matrix parametrization.
In the second additional fit, seventeen energies are used, which include the fifteen energies 
from the $s$-wave fits and one level each in the $F_1(3), F_2(3)$ irreps, and we add
a $J^P=3/2^+$ wave in the $\widetilde{K}$-matrix parametrization. For each of these additional 
partial waves, we use simple parametrizations of the form
\begin{equation}
   \tilde{K}^{J^P} = \text{diag }\left( A^{J^P}_{00},  A^{J^P}_{11} \right).
   \label{eq:parametrizationB}
\end{equation}
The results of these additional fits are presented in Table~\ref{tab:highwaves} of 
Appendix~\ref{sec:fitres}.  One sees that the shifts in the parameters due to considering 
additional waves is about an order of magnitude lower than the statistical uncertainty 
when considering only the leading partial wave. Thus, it can be concluded that the effects of 
higher partial waves can be neglected to the 
present statistical uncertainty for energies below the $\pi\pi\Lambda$ threshold.

\section{Conclusion}
\label{sec:conc}

Hermitian correlation matrices using both single baryon and meson-baryon interpolating operators 
for a variety of different total momenta and irreducible representations were used 
to determine the finite-volume stationary-state energies in the isospin $I=0$ strangeness $S=-1$
sector in lattice QCD with the stochastic LapH method. Results were obtained using a single 
ensemble of gauge field configurations with $N_{\rm f} = 2+1$ dynamical quark flavors and 
$m_{\pi} \approx 200~{\rm  MeV}$ and $m_K\approx487$~MeV on a $64^3\times 128$ lattice with spacing 
$a=0.0633(4)(6)$~fm. Various $K$-matrix parametrizations restricted to $s$-waves were then employed 
in the quantization condition to match the spectrum obtained from lattice QCD.  The resulting best-fit
parameter values from the $K$-matrix parametrization yielding the lowest Akaike Information 
Criterion produced the $\pi\Sigma$-$\bar{K}N$ transition amplitudes shown in 
Fig.~\ref{fig:main}. Analytic continuation of the transition amplitudes into 
complex center-of-mass energies revealed two poles, suggesting a virtual bound state below 
the $\pi\Sigma$ threshold and a resonance just below the higher $\bar{K}N$ threshold. A single
channel fit of $\pi\Sigma$ scattering showed the robustness of the lower pole, and the
effects of including the leading $p$-wave contributions were examined and found to be negligible at 
the current statistical precision.

The results presented here demonstrate the feasibility of using current lattice QCD techniques 
to study coupled-channel dynamics for baryon resonances such as the controversial $\Lambda(1405)$.  We 
plan to obtain results at physical quark masses, as well as results for several lattice spacings to 
extrapolate to the continuum-limit.  The importance of studying lattice spacing errors has been 
underscored recently by the observation of large discretization effects in the $H$-dibaryon binding 
energy\cite{Green:2021qol} using the same lattice discretization as employed here. However, that work 
suggests that at the fine lattice spacing used here the result is unlikely to differ qualitatively 
from the continuum. Note that moving to physical quark masses requires the consideration of 
three-particle effects due to $\pi \pi \Lambda$ states. The lattice QCD determination of levels above 
this threshold should not present a major problem in the region relevant for the $\Lambda(1405)$, and 
the formalism for three-hadrons including particles with spin is progressing rapidly~\cite{Draper:2023xvu}.  Exploring the quark-mass trajectory toward the SU(3)-symmetric point will 
be useful for testing chiral effective theories.  This work opens the door to investigating 
other baryon resonances, such as the $N(1535)$, $\Lambda(1670)$, $\Sigma(1620)$, and $\Xi(1620)$, 
among others.

\begin{acknowledgments}
We acknowledge helpful discussions with R.J.~Hudspith, A.~Jackura, M.F.M.~Lutz and M.~Mai.  
We thank our colleagues within the CLS consortium for sharing ensembles.
Computations were carried out on Frontera~\cite{frontera} at the Texas Advanced Computing Center (TACC), 
and at the National Energy Research Scientific Computing Center (NERSC), a U.S.~Department of Energy 
Office of Science User Facility located at Lawrence Berkeley National Laboratory, operated under 
Contract No.~DE-AC02-05CH11231 using NERSC awards NP-ERCAP0005287, NP-ERCAP0010836 and NP-ERCAP0015497.
This work was supported in part by:
the U.S.~National Science Foundation under awards PHY-1913158 and PHY-2209167 (CJM, SS), 
the Faculty Early Career Development Program (CAREER) under award PHY-2047185 (AN)
and by the Graduate Research Fellowship Program under Grant No.~DGE-2040435 (JM);
the U.S.~Department of Energy, Office of Science, Office of Nuclear Physics, under grant contract 
numbers DE-SC0011090 and DE-SC0021006 (FRL), DE-SC0012704 (ADH), DE-AC02-05CH11231 (AWL) and within 
the framework of Scientific Discovery through Advanced Computing (SciDAC) award ``Fundamental Nuclear 
Physics at the Exascale and Beyond'' (ADH); the Mauricio and Carlota Botton Fellowship (FRL);
and the Heisenberg Programme of the Deutsche Forschungsgemeinschaft (DFG, German Research Foundation) 
project number 454605793 (DM). NumPy~\cite{harris2020array}, matplotlib~\cite{Hunter:2007}, and the 
CHROMA software suite~\cite{Edwards:2004sx} were used for analysis, plotting, and correlator evaluation.
\end{acknowledgments}

\appendix

\section{Single- and two-hadron operators}
\label{sec:operators}

The single- and two-hadron operators used in this study are specified in 
Tables~\ref{tab:opdefs0}-\ref{tab:opdefs3} in 
this section. We use multi-hadron operators comprised of individual constituent hadrons, 
each corresponding to a definite momentum. Each single-hadron operator is specified 
by its flavor structure, such as $\Lambda$, $\Sigma$, $N$, $\pi$, $\bar{K}$, then in
square brackets, the irrep of its little group, with the squared spatial momentum, in 
units of $(2\pi/L)^2$, shown in parentheses.  The subscript indicates a spatial 
identification number. The spin and orbital structure associated with each identification 
number can be obtained from the authors upon request. 

All operators used in this study 
are single-site operators. The notation for the irreps follows the conventions in 
Ref.~\cite{Morningstar:2013bda}. The subscripts $g/u$ denote even/odd parity, and the 
superscripts $+/-$ denote even/odd $G$-parity. 
Whenever there are more than one independent Clebsch-Gordan combinations, the
multiplicity is indicated by an integer in parentheses to the right of the operator
identification. The Clebsch-Gordan coefficients that 
fully define each operator are not given, but are available upon request.

\begin{table}[ht]
\caption{Single- and two-hadron operators used in each symmetry sector with
total momentum $\bm{d}^2=0$.  Operator notation is described in the text.
\label{tab:opdefs0}}
\begin{ruledtabular}
\begin{tabular}{l@{\hspace*{12mm}}l}
 $\Lambda(\bm{d}^2)$ & Operators \\
  \hline
    $H_{\rm u}(0)$ 
 &  $\pi[A_{1u}^{-}(0)]_0\ \Sigma[Hg(0)]_0$ \\
 &  $\pi[A_{2}^{-}(1)]_1\ \Sigma[G_1(1)]_0$\\
 &  $\bar{K}[A_{2}(1)]_1\ N[G_1(1)]_0$\\ \hline
   $G_{\rm 1g}(0)$
 & $\Lambda[G_{1g}(0)]_0$\\
 & $\Lambda[G_{1g}(0)]_1$\\
 & $\Lambda[G_{1g}(0)]_3$\\
 & ${\bar K}[A_{2}(1)]_1\ N[G_1(1)]_0 $\\
 & $\pi[A_{2}^{-}(1)]_1\ \Sigma[G_1(1)]_0 $\\ \hline
   $G_{\rm 1u}(0)$
 & $\Lambda[G_{1u}(0)]_0$\\
 & $\Lambda[G_{1u}(0)]_1$\\
 & $\Lambda[G_{1u}(0)]_2$\\
 & $\Lambda[G_{1u}(0)]_3$\\
 & $\bar{K}[A_{1u}(0)]_0\ N[G_{1g}(0)]_0$\\
 & $    \pi[A_{1u}^{-}(0)]_0\ \Sigma[G_{1g}(0)]_0$\\
 & $\bar{K}[A_{2}(1)]_1\ N[G_1(1)]_0$\\
 & $    \pi[A_{2}^{-}(1)]_1\ \Sigma[G_1(1)]_0$ 
\end{tabular}
\end{ruledtabular}
\end{table}

\begin{table}[ht]
\caption{ Same as Table~\ref{tab:opdefs0} with $\bm{d}^2=1$.
\label{tab:opdefs1}}
\begin{ruledtabular}
\begin{tabular}{l@{\hspace*{12mm}}l}
 $\Lambda(\bm{d}^2)$ & Operators \\
\hline
  $G_1(1)$
 & $\Lambda[G_1(1)]_0$\\
 & $\Lambda[G_1(1)]_1$\\
 & $\Lambda[G_1(1)]_2$\\
 & $\Lambda[G_1(1)]_4$\\
 & $\Lambda[G_1(1)]_6$\\
 & $\bar{K}[A_{1u}(0)]_0   \ N[G_1(1)]_0$\\
 & $\pi[A_{1u}^{-}(0)]_0\ \Sigma[G_1(1)]_0$\\
 & $\bar{K}[A_{2}(1)]_1 \ N[G_1g(0)]_0$\\
 & $\pi[A_{2}^{-}(1)]_1 \ \Sigma[G_1g(0)]_0$\\
\hline
  $G_2(1)$
 & $\Lambda[G_2(1)]_0$\\
 & $\Lambda[G_2(1)]_1$\\
 & $\pi[A_{1u}^-(0)]_0 \ \Sigma[G_2(1)]_0$\\
 & $\pi[A_2^-(1)]_1 \ \Sigma[G(2)]_0\quad  (2)$\\
 & $\bar{K}[A_2(1)]_1 \ N[G(2)]_0\quad  (2)$\\
\end{tabular}
\end{ruledtabular}
\end{table}

\begin{table}
\caption{Same as Table~\ref{tab:opdefs0} with $\bm{d}^2=2$.
\label{tab:opdefs2}}
\begin{ruledtabular}
\begin{tabular}{l@{\hspace*{12mm}}l}
 $\Lambda(\bm{d}^2)$ & Operators \\
\hline
  $G(2)$
 &  $\Lambda[G(2)]_0$\\
 &  $\Lambda[G(2)]_1$\\
 &  $\Lambda[G(2)]_2$\\
 &  $\Lambda[G(2)]_3$\\
 &  $\Lambda[G(2)]_5$\\
 &  $\Lambda[G(2)]_6$\\
 &  $ \bar{K}[A_{1u}(0)]_0 \ N[G(2)]_0$\\
 &  $ \pi[A_{1u}^{-}(0)]_0 \ \Sigma[G(2)]_0$\\
 &  $ \bar{K}[A_{2}(1)]_1  \ N[G_1(1)]_0\quad  (2)$\\
 &  $ \pi[A_{2}^{-}(1)]_1  \ \Sigma[G_1(1)]_0\quad  (2)$\\
 &  $ \bar{K}[A_{2}(2)]_0  \ N[G_{1g}(0)]_0$\\
 &  $ \pi[A_{1u}^{-}(0)]_0 \ \Sigma[G(2)]_1$\\
 &  $ \pi[A_{1u}^{-}(0)]_0 \ \Sigma[G(2)]_7$\\
\end{tabular}
\end{ruledtabular}
\end{table}

\begin{table}
\caption{Same as Table~\ref{tab:opdefs0} with $\bm{d}^2=3$.
\label{tab:opdefs3}}
\begin{ruledtabular}
\begin{tabular}{l@{\hspace*{12mm}}l}
 $\Lambda(\bm{d}^2)$ & Operators \\
\hline
  $F_1(3)$
 &  $\Lambda[F_1(3)]_0$\\
 &  $ \pi[A_{2}^{-}(1)]_1 \ \Sigma[G(2)]_0$\\
 &  $ \bar{K}[A_{2}(1)]_1 \ N[G(2)]_0$\\
 &  $ \pi[A_{2}^{-}(2)]_0 \ \Sigma[G_1(1)]_0$\\
 &  $ \bar{K}[A_{2}(2)]_0 \ N[G_1(1)]_0$\\
 &  $ \pi[A_{1u}^{-}(0)]_0 \ \Sigma[F_2(3)]_0$\\
\hline
  $F_2(3)$ 
 &  $\Lambda[F_2(3)]_0$\\
 &  $\pi[A_{2}^{-}(1)]_1 \ \Sigma[G(2)]_0$\\
 &  $\bar{K}[A_{2}(1)]_1 \ N[G(2)]_0$\\
 &  $\pi[A_{2}^{-}(2)]_0 \ \Sigma[G_1(1)]_0$\\
 &  $\bar{K}[A_{2}(2)]_0 \ N[G_1(1)]_0$\\
 &  $\pi[A_{1u}^{-}(0)]_0 \ \Sigma[F_1(3)]_0$\\
\hline
  $G(3)$
 &  $\Lambda[G(3)]_0$\\
 &  $\Lambda[G(3)]_1$\\
 &  $\Lambda[G(3)]_4$\\
 &  $\Lambda[G(3)]_5$\\
 &  $ \bar{K}[A_{1u}(0)]_0 \ N[G(3)]_0$\\
 &  $ \pi[A_{1u}^{-}(0)]_0 \ \Sigma[G(3)]_0$\\
 &  $ \bar{K}[A_{2}(1)]_1 \ N[G(2)]_0\quad  (2)$\\
 &  $ \pi[A_{2}^{-}(1)]_1 \ \Sigma[G(2)]_0\quad (2)$\\
 &  $ \bar{K}[A_{2}(2)]_0 \ N[G_1(1)]_0\quad (2)$\\
 &  $ \bar{K}[A_{2}(2)]_0 \ N[G_{1g}(0)]_0$\\
\end{tabular}
\end{ruledtabular}
\end{table}

\section{Energy extractions}
\label{sec:appendix_tmin}

\begin{figure*}[ht]
\centering
\includegraphics[width=\linewidth]{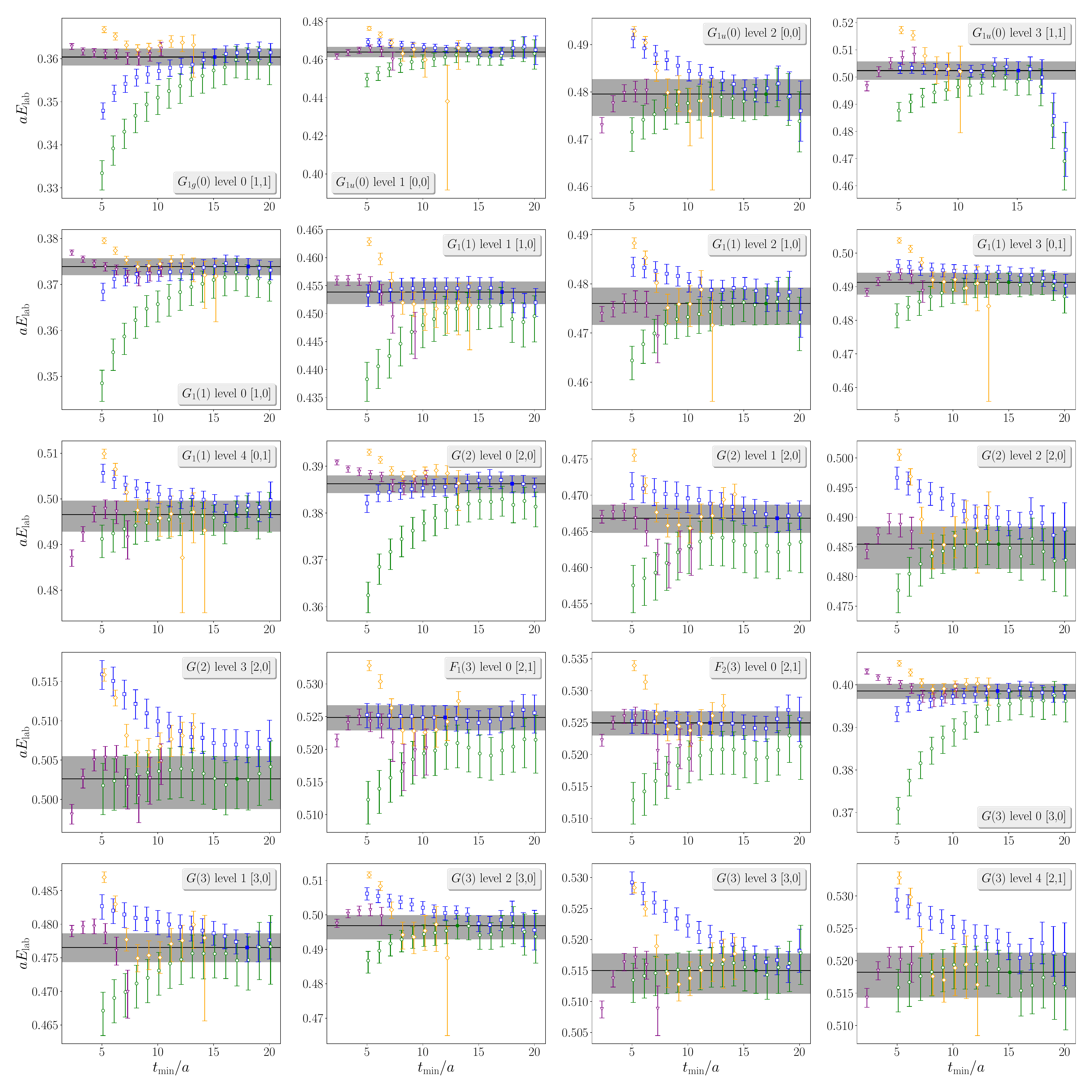}
\caption{\label{fig:fitsall}
Fit results for the stationary-state energies indicated. These plots are similar to that shown in 
Fig.~\ref{fig:g1u_example}.  Orange diamonds and purple triangles indicate results from fits using 
a two-exponential and a geometric-exp series, respectively. Green circles and blue squares indicate 
results from single-exponential fits to ratios of the rotated diagonal correlator over the product 
of $N(\bm{d}_1^2)\ \overline{K}(\bm{d}_2^2)$ and $\Sigma(\bm{d}_1^2)\ \pi(\bm{d}_2^2)$ single-hadron 
correlators, respectively, as described in Sec.~\ref{sec:energies}. In each plot, the integers in 
square brackets in the legend show the values $[\bm{d}_1^2, \bm{d}_2^2]$. The dark horizontal band and
filled symbol denote the final chosen fit selected as described
in Sec.~\ref{sec:energies}.}
\end{figure*}

Energy determinations are shown in Fig.~\ref{fig:fitsall} in this section.   
Results from four different fit methods are shown: two-exponential and geometric-exp series fits to the 
rotated correlator, and single-exponential fits to the ratio of the rotated diagonal correlator
over the product of single-hadron correlators for either $\bar{K}(\bm{d}_{\bar K}^2)N(\bm{d}_N^2)$ 
or $\pi(\bm{d}_\pi^2)\Sigma(\bm{d}_\Sigma^2)$.
The dark horizontal band and filled symbol denote the final chosen fit for each level
selected as described in the Sec.~\ref{sec:energies}.
Bootstrap samples for extracted energies are available at Ref.~\cite{zenododata}.

\section{Amplitude parameter fit results}
\label{sec:fitres}

Best-fit results for the parameters in the various $\widetilde{K}$-matrix parametrizations
are presented here.  Results are obtained by fitting the spectrum obtained from the 
$\widetilde{K}$-matrix parametrizations and the quantization condition to the spectrum
determined in the lattice QCD computations. Tables~\ref{tab:fam1}-\ref{tab:fam6} contain
results from fits using $\ell_{\rm max}=0$, and  Table~\ref{tab:highwaves}
using $\ell_{\rm max}=1$.

\begin{table*}
\caption{Fit results for $\widetilde{K}$ parametrization class 1 shown in Eq.~(\ref{eq:Kfitone}). 
 Errors are propagated through the derivative method. Empty entries indicate parameters set to zero in a fit.  
 AIC refers to Akaike Information Criterion.
\label{tab:fam1}}
\begin{ruledtabular}
\begin{tabular}{lddd@{\hspace*{7mm}}d@{\hspace*{-5mm}}d@{\hspace*{-5mm}}d@{\hspace*{-17mm}}d@{\hspace*{6mm}}d}
Fit &  \multicolumn{1}{c}{$\qquad A_{00}$} & \multicolumn{1}{c}{$\qquad A_{11}$} 
    & \multicolumn{1}{c}{$\quad A_{01}$}  & \multicolumn{1}{l}{$\quad B_{00}$} 
    & \multicolumn{1}{l}{$\quad B_{11}$}  & \multicolumn{1}{l}{$\qquad B_{01}$}  
    & \multicolumn{1}{c}{$\quad \chi^2/\text{dof}$} & \multicolumn{1}{c}{AIC}  \\ \hline
a &  1.5(1.4)&  -8.78(72)&  8.30(65)&         &        &        & 15.68/(15-3)& -8.32 \\
b &  4.1(1.2)& -10.5(1.1)& 10.3(1.3)&         &        & -29(15)& 10.52/(15-4)& -11.48 \\
c &  2.3(1.3)&  -8.62(58)&  7.60(80)&         & -18(11)&        & 12.29/(15-4)& -9.71 \\
d & 15.1(5.3)& -11.8(1.3)&  7.6(1.3)&  -56(19)&        &        & 11.48/(15-4)& -10.52 \\
e &  9.6(6.2)& -12.7(3.4)& 11.1(2.8)&  -23(26)&  18(31)& -37(29)& 9.70/(15-6)& -8.30 \\
\end{tabular}
\end{ruledtabular}
\end{table*}
  
\begin{table*}
\caption{Fit results for $\widetilde{K}$ parametrization class 2 shown in Eq.~(\ref{eq:Kfittwo}). 
 Errors are propagated through the derivative method. Empty entries indicate parameters set to zero in a fit.  
 AIC refers to Akaike Information Criterion.
\label{tab:fam2}}
\begin{ruledtabular}
\begin{tabular}{lddddddd@{\hspace*{4mm}}d}
Fit &  \multicolumn{1}{c}{$\qquad \widehat A_{00}$} 
 & \multicolumn{1}{c}{$\qquad \widehat A_{11}$}  
 & \multicolumn{1}{c}{$\qquad \widehat A_{01}$}   
& \multicolumn{1}{c}{$\qquad \widehat B_{00}$} 
&  \multicolumn{1}{c}{$\qquad \widehat B_{11}$}  
& \multicolumn{1}{c}{$\qquad\quad \widehat B_{01}$}
  & \multicolumn{1}{c}{$\qquad \chi^2/\text{dof}$} & \multicolumn{1}{c}{AIC}  \\ \hline
a &  0.16(19)& -1.229(91)& 1.140(88)&         &        &        & 15.44/(15-3)& -8.56 \\
b &  0.52(18)&  -1.45(15)&  1.42(18)&         &        & -3.9(2.0)& 10.73/(15-4)& -11.27 \\
\end{tabular}
\end{ruledtabular}
\end{table*}

\begin{table*}
\caption{Fit results for $\widetilde{K}$ parametrization class 3 shown in Eq.~(\ref{eq:Kfitthree}). 
 Errors are propagated through the derivative method. Empty entries indicate parameters set to zero in a fit.  
 AIC refers to Akaike Information Criterion.
\label{tab:fam3}}
\begin{ruledtabular}
\begin{tabular}{l@{\hspace*{-5mm}}d@{\hspace*{-5mm}}d@{\hspace*{-5mm}}d@{\hspace*{-5mm}}d@{\hspace*{-5mm}}d@{\hspace*{-5mm}}d@{\hspace*{-5mm}}d@{\hspace*{5mm}}d}
Fit & \multicolumn{1}{c}{$\qquad\qquad\widetilde A_{00}$}
& \multicolumn{1}{c}{$\qquad\qquad\widetilde A_{11}$}
& \multicolumn{1}{c}{$\qquad\qquad\widetilde A_{01}$}  
& \multicolumn{1}{c}{$\qquad\qquad\widetilde B_{00}$}
& \multicolumn{1}{c}{$\qquad\qquad\widetilde B_{11}$} 
& \multicolumn{1}{c}{$\qquad\qquad\widetilde B_{01}$} 
& \multicolumn{1}{c}{$\qquad\chi^2/\text{dof}$} & \multicolumn{1}{c}{AIC}  \\ \hline
a & 0.092(21)& -0.036(15)& 0.082(20)&  0.28(15)&        &        & 11.73/(15-4)& -10.27 \\
b & 0.114(25)& -0.041(24)& 0.096(19)&         & 0.19(16)&        & 14.57/(15-4)& -7.43 \\
c & 0.137(33)& -0.019(14)& 0.119(21)&         &        & -0.142(85)& 13.10/(15-4)& -8.90 \\
\end{tabular}
\end{ruledtabular}
\end{table*}

\begin{table*}
\caption{Fit results for $\widetilde{K}$ parametrization class 4 shown in Eq.~(\ref{eq:Kfitfour}). 
 Errors are propagated through the derivative method. Empty entries indicate parameters set 
 to zero in a fit. AIC refers to Akaike Information Criterion.
\label{tab:fam4}}
\begin{ruledtabular}
\begin{tabular}{l@{\hspace*{-4mm}}d@{\hspace*{0mm}}d@{\hspace*{4mm}}d@{\hspace*{-7mm}}d@{\hspace*{-12mm}}d@{\hspace*{-2mm}}d@{\hspace*{-5mm}}d@{\hspace*{-5mm}}d@{\hspace*{5mm}}d}
Fit & \multicolumn{1}{c}{$\qquad a_{0}$}
& \multicolumn{1}{c}{$a_1$}
& \multicolumn{1}{c}{$\!\!\!b_0$} 
& \multicolumn{1}{c}{$\ b_1$}
& \multicolumn{1}{c}{$\qquad c_0$} 
& \multicolumn{1}{c}{$\qquad\quad c_1$} 
& \multicolumn{1}{c}{$\qquad\qquad \epsilon$}  
& \multicolumn{1}{c}{$\qquad \chi^2/\text{dof}$} & \multicolumn{1}{c}{AIC}  \\ \hline
a &  5.7(1.2)& -11.4(1.2)&        & -27(15)&        &        &  0.451(56)& 13.27/(15-4)& -8.73 \\
b & 13.7(4.1)& -14.06(86)& -37(17)&        &        &        &  0.349(75)& 10.63/(15-4)& -11.37 \\
c &  5.8(1.2)& -11.8(1.1)&        &        &        & -1.62(95)&  0.468(48)& 13.54/(15-4)& -8.46 \\
d & 12.2(3.4)& -14.06(87)&        &        & 5.8(3.2)&        &  0.360(82)& 11.13/(15-4)& -10.87 \\

\end{tabular}
\end{ruledtabular}
\end{table*}

\begin{table*}
\caption{Fit results for $\widetilde{K}$ parametrization class 5 shown in Eq.~(\ref{eq:Kfitfive}). 
 Errors are propagated through the derivative method. Empty entries indicate parameters set 
 to zero in a fit. AIC refers to Akaike Information Criterion.
\label{tab:fam5}}
\begin{ruledtabular}
\begin{tabular}{l@{\hspace*{-4mm}}ddddd}
Fit & \multicolumn{1}{c}{$\qquad \widehat C_{00}$}
& \multicolumn{1}{c}{$\qquad\widehat C_{11}$}
& \multicolumn{1}{c}{$\qquad\widehat C_{01}$} 
& \multicolumn{1}{c}{$\qquad\qquad\chi^2/\text{dof}$} & \multicolumn{1}{c}{AIC}  \\ \hline
a &  0.005(58)& -0.270(12)&   -0.295(22) & 15.28/(15-3)& -8.72 \\
\end{tabular}
\end{ruledtabular}
\end{table*}

\begin{table*}
\caption{Fit results for $\widetilde{K}$ parametrization class 6 shown in Eq.~(\ref{eq:Kfitsix}). 
 Errors are propagated through the derivative method. Empty entries indicate parameters set to zero in a fit.  
 AIC refers to Akaike Information Criterion.
\label{tab:fam6}}
\begin{ruledtabular}
\begin{tabular}{l@{\hspace*{-4mm}}dddddd@{\hspace*{-7mm}}d@{\hspace*{4mm}}d}
Fit &  \multicolumn{1}{c}{$\qquad A'_{00}$} & \multicolumn{1}{c}{$\qquad A'_{11}$} 
    & \multicolumn{1}{c}{$\qquad A'_{01}$}  & \multicolumn{1}{c}{$\qquad B'_{00}$} 
    & \multicolumn{1}{c}{$\qquad B'_{11}$}  & \multicolumn{1}{c}{$\qquad\qquad B_{01}$}  
    & \multicolumn{1}{c}{$\qquad\quad \chi^2/\text{dof}$} & \multicolumn{1}{c}{AIC}  \\ \hline
a & 2.11(77) &  -1.64(19) & 1.08(17)  &  -2.51(88)        &        &        & 11.88/(15-4)& -10.12 \\
b & 0.27(18) &  -1.222(75) &  1.05(11) &         &  -0.69(44)      &   & 12.63/(15-4)& -9.37 \\
c &0.51(18)  & -1.46(15) & 1.43(18) &         &&   -1.18(62)     & 10.76/(15-4)& -11.24 \\
\end{tabular}
\end{ruledtabular}
\end{table*}

\clearpage
\begin{table*}[t]
\caption{Fit results for $\widetilde{K}$ parametrization class 1 shown in Eq.~(\ref{eq:Kfitone})
 for the $J^P=1/2^-$ wave, and Eq.~(\ref{eq:parametrizationB}) for the $J^P=1/2^+, 3/2^+$ waves using $\ell_{\rm max}=1$.
 Errors are propagated through the derivative method. Empty entries indicate parameters set to zero in a fit.  
 AIC refers to Akaike Information Criterion.
\label{tab:highwaves}}
\begin{ruledtabular}
\begin{tabular}{ccccccccccc}
$J^P$ partial waves & $A_{00}$ & $A_{11}$  & $A_{01}$   & $B_{01}$ & $A^{1/2^+}_{00}$ & $A^{1/2^+}_{11}$ & $A^{3/2^+}_{00}$ & $A^{3/2^+}_{11}$  & $\chi^2/\text{dof}$ & AIC  \\ \hline
 $1/2^-$ &       4.1(1.2) & $-10.5(1.1)$ &10.3(1.3) & $-29(15)$  &   &  &  &  & 10.52/(15-4) & $-11.48$ \\ 
$1/2^-$ and $1/2^+$ &  4.1(1.2)& $-10.5(1.1)$ & 10.3(1.3)& $-30(15)$ & 0.0088(39)& 0.031(15)& & & 10.52/(17-6)& $-11.48$ \\
$1/2^-$ and $3/2^+$ &  4.1(1.1)& $-10.9(1.1)$ & 10.4(1.3)& $-32(15)$ &         &         & 0.0172(48)& 0.0218(48)& 14.10/(21-6)& $-15.90$ \\\end{tabular}
\end{ruledtabular}
\end{table*}

\bibliography{cited_refs}   

\begin{thebibliography}{98}%
\makeatletter
\providecommand \@ifxundefined [1]{%
 \@ifx{#1\undefined}
}%
\providecommand \@ifnum [1]{%
 \ifnum #1\expandafter \@firstoftwo
 \else \expandafter \@secondoftwo
 \fi
}%
\providecommand \@ifx [1]{%
 \ifx #1\expandafter \@firstoftwo
 \else \expandafter \@secondoftwo
 \fi
}%
\providecommand \natexlab [1]{#1}%
\providecommand \enquote  [1]{``#1''}%
\providecommand \bibnamefont  [1]{#1}%
\providecommand \bibfnamefont [1]{#1}%
\providecommand \citenamefont [1]{#1}%
\providecommand \href@noop [0]{\@secondoftwo}%
\providecommand \href [0]{\begingroup \@sanitize@url \@href}%
\providecommand \@href[1]{\@@startlink{#1}\@@href}%
\providecommand \@@href[1]{\endgroup#1\@@endlink}%
\providecommand \@sanitize@url [0]{\catcode `\\12\catcode `\$12\catcode
  `\&12\catcode `\#12\catcode `\^12\catcode `\_12\catcode `\%12\relax}%
\providecommand \@@startlink[1]{}%
\providecommand \@@endlink[0]{}%
\providecommand \url  [0]{\begingroup\@sanitize@url \@url }%
\providecommand \@url [1]{\endgroup\@href {#1}{\urlprefix }}%
\providecommand \urlprefix  [0]{URL }%
\providecommand \Eprint [0]{\href }%
\providecommand \doibase [0]{https://doi.org/}%
\providecommand \selectlanguage [0]{\@gobble}%
\providecommand \bibinfo  [0]{\@secondoftwo}%
\providecommand \bibfield  [0]{\@secondoftwo}%
\providecommand \translation [1]{[#1]}%
\providecommand \BibitemOpen [0]{}%
\providecommand \bibitemStop [0]{}%
\providecommand \bibitemNoStop [0]{.\EOS\space}%
\providecommand \EOS [0]{\spacefactor3000\relax}%
\providecommand \BibitemShut  [1]{\csname bibitem#1\endcsname}%
\let\auto@bib@innerbib\@empty
\bibitem [{\citenamefont {Workman}\ \emph {et~al.}(2022)\citenamefont {Workman}
  \emph {et~al.}}]{ParticleDataGroup:2022pth}%
  \BibitemOpen
  \bibfield  {author} {\bibinfo {author} {\bibfnamefont {R.~L.}\ \bibnamefont
  {Workman}} \emph {et~al.} (\bibinfo {collaboration} {Particle Data Group}),\
  }\bibfield  {title} {\bibinfo {title} {{Review of Particle Physics}},\ }\href
  {https://doi.org/10.1093/ptep/ptac097} {\bibfield  {journal} {\bibinfo
  {journal} {PTEP}\ }\textbf {\bibinfo {volume} {2022}},\ \bibinfo {pages}
  {083C01} (\bibinfo {year} {2022})}\BibitemShut {NoStop}%
\bibitem [{\citenamefont {Dalitz}\ and\ \citenamefont
  {Tuan}(1959)}]{Dalitz:1959dn}%
  \BibitemOpen
  \bibfield  {author} {\bibinfo {author} {\bibfnamefont {R.~H.}\ \bibnamefont
  {Dalitz}}\ and\ \bibinfo {author} {\bibfnamefont {S.~F.}\ \bibnamefont
  {Tuan}},\ }\bibfield  {title} {\bibinfo {title} {{Possible resonant state in
  pion-hyperon scattering}},\ }\href
  {https://doi.org/10.1103/PhysRevLett.2.425} {\bibfield  {journal} {\bibinfo
  {journal} {Phys. Rev. Lett.}\ }\textbf {\bibinfo {volume} {2}},\ \bibinfo
  {pages} {425} (\bibinfo {year} {1959})}\BibitemShut {NoStop}%
\bibitem [{\citenamefont {Dalitz}\ and\ \citenamefont
  {Tuan}(1960)}]{Dalitz:1960du}%
  \BibitemOpen
  \bibfield  {author} {\bibinfo {author} {\bibfnamefont {R.~H.}\ \bibnamefont
  {Dalitz}}\ and\ \bibinfo {author} {\bibfnamefont {S.~F.}\ \bibnamefont
  {Tuan}},\ }\bibfield  {title} {\bibinfo {title} {{The phenomenological
  representation of $\overline{K}$-nucleon scattering and reaction
  amplitudes}},\ }\href {https://doi.org/10.1016/0003-4916(60)90001-4}
  {\bibfield  {journal} {\bibinfo  {journal} {Annals Phys.}\ }\textbf {\bibinfo
  {volume} {10}},\ \bibinfo {pages} {307} (\bibinfo {year} {1960})}\BibitemShut
  {NoStop}%
\bibitem [{\citenamefont {Hyodo}\ and\ \citenamefont
  {Jido}(2012)}]{Hyodo:2011ur}%
  \BibitemOpen
  \bibfield  {author} {\bibinfo {author} {\bibfnamefont {T.}~\bibnamefont
  {Hyodo}}\ and\ \bibinfo {author} {\bibfnamefont {D.}~\bibnamefont {Jido}},\
  }\bibfield  {title} {\bibinfo {title} {{The nature of the $\Lambda(1405)$
  resonance in chiral dynamics}},\ }\href
  {https://doi.org/10.1016/j.ppnp.2011.07.002} {\bibfield  {journal} {\bibinfo
  {journal} {Prog. Part. Nucl. Phys.}\ }\textbf {\bibinfo {volume} {67}},\
  \bibinfo {pages} {55} (\bibinfo {year} {2012})},\ \Eprint
  {https://arxiv.org/abs/1104.4474} {arXiv:1104.4474 [nucl-th]} \BibitemShut
  {NoStop}%
\bibitem [{\citenamefont {Hyodo}\ and\ \citenamefont
  {Niiyama}(2021)}]{Hyodo:2020czb}%
  \BibitemOpen
  \bibfield  {author} {\bibinfo {author} {\bibfnamefont {T.}~\bibnamefont
  {Hyodo}}\ and\ \bibinfo {author} {\bibfnamefont {M.}~\bibnamefont
  {Niiyama}},\ }\bibfield  {title} {\bibinfo {title} {{QCD and the strange
  baryon spectrum}},\ }\href {https://doi.org/10.1016/j.ppnp.2021.103868}
  {\bibfield  {journal} {\bibinfo  {journal} {Prog. Part. Nucl. Phys.}\
  }\textbf {\bibinfo {volume} {120}},\ \bibinfo {pages} {103868} (\bibinfo
  {year} {2021})},\ \Eprint {https://arxiv.org/abs/2010.07592}
  {arXiv:2010.07592 [hep-ph]} \BibitemShut {NoStop}%
\bibitem [{\citenamefont {Bazzi}\ \emph {et~al.}(2011)\citenamefont {Bazzi}
  \emph {et~al.}}]{SIDDHARTA:2011dsy}%
  \BibitemOpen
  \bibfield  {author} {\bibinfo {author} {\bibfnamefont {M.}~\bibnamefont
  {Bazzi}} \emph {et~al.} (\bibinfo {collaboration} {SIDDHARTA}),\ }\bibfield
  {title} {\bibinfo {title} {{A New Measurement of Kaonic Hydrogen X-rays}},\
  }\href {https://doi.org/10.1016/j.physletb.2011.09.011} {\bibfield  {journal}
  {\bibinfo  {journal} {Phys. Lett. B}\ }\textbf {\bibinfo {volume} {704}},\
  \bibinfo {pages} {113} (\bibinfo {year} {2011})},\ \Eprint
  {https://arxiv.org/abs/1105.3090} {arXiv:1105.3090 [nucl-ex]} \BibitemShut
  {NoStop}%
\bibitem [{\citenamefont {Meissner}\ \emph {et~al.}(2004)\citenamefont
  {Meissner}, \citenamefont {Raha},\ and\ \citenamefont
  {Rusetsky}}]{Meissner:2004jr}%
  \BibitemOpen
  \bibfield  {author} {\bibinfo {author} {\bibfnamefont {U.~G.}\ \bibnamefont
  {Meissner}}, \bibinfo {author} {\bibfnamefont {U.}~\bibnamefont {Raha}},\
  and\ \bibinfo {author} {\bibfnamefont {A.}~\bibnamefont {Rusetsky}},\
  }\bibfield  {title} {\bibinfo {title} {{Spectrum and decays of kaonic
  hydrogen}},\ }\href {https://doi.org/10.1140/epjc/s2004-01859-4} {\bibfield
  {journal} {\bibinfo  {journal} {Eur. Phys. J. C}\ }\textbf {\bibinfo {volume}
  {35}},\ \bibinfo {pages} {349} (\bibinfo {year} {2004})},\ \Eprint
  {https://arxiv.org/abs/hep-ph/0402261} {arXiv:hep-ph/0402261} \BibitemShut
  {NoStop}%
\bibitem [{\citenamefont {Moriya}\ \emph {et~al.}(2013)\citenamefont {Moriya}
  \emph {et~al.}}]{CLAS:2013rjt}%
  \BibitemOpen
  \bibfield  {author} {\bibinfo {author} {\bibfnamefont {K.}~\bibnamefont
  {Moriya}} \emph {et~al.} (\bibinfo {collaboration} {CLAS}),\ }\bibfield
  {title} {\bibinfo {title} {{Measurement of the
  \ensuremath{\Sigma}\ensuremath{\pi} photoproduction line shapes near the
  \ensuremath{\Lambda}(1405)}},\ }\href
  {https://doi.org/10.1103/PhysRevC.87.035206} {\bibfield  {journal} {\bibinfo
  {journal} {Phys. Rev. C}\ }\textbf {\bibinfo {volume} {87}},\ \bibinfo
  {pages} {035206} (\bibinfo {year} {2013})},\ \Eprint
  {https://arxiv.org/abs/1301.5000} {arXiv:1301.5000 [nucl-ex]} \BibitemShut
  {NoStop}%
\bibitem [{\citenamefont {Moriya}\ \emph {et~al.}(2014)\citenamefont {Moriya}
  \emph {et~al.}}]{CLAS:2014tbc}%
  \BibitemOpen
  \bibfield  {author} {\bibinfo {author} {\bibfnamefont {K.}~\bibnamefont
  {Moriya}} \emph {et~al.} (\bibinfo {collaboration} {CLAS}),\ }\bibfield
  {title} {\bibinfo {title} {{Spin and parity measurement of the Lambda(1405)
  baryon}},\ }\href {https://doi.org/10.1103/PhysRevLett.112.082004} {\bibfield
   {journal} {\bibinfo  {journal} {Phys. Rev. Lett.}\ }\textbf {\bibinfo
  {volume} {112}},\ \bibinfo {pages} {082004} (\bibinfo {year} {2014})},\
  \Eprint {https://arxiv.org/abs/1402.2296} {arXiv:1402.2296 [hep-ex]}
  \BibitemShut {NoStop}%
\bibitem [{\citenamefont {Mai}\ and\ \citenamefont
  {Mei\ss{}ner}(2015)}]{Mai:2014xna}%
  \BibitemOpen
  \bibfield  {author} {\bibinfo {author} {\bibfnamefont {M.}~\bibnamefont
  {Mai}}\ and\ \bibinfo {author} {\bibfnamefont {U.-G.}\ \bibnamefont
  {Mei\ss{}ner}},\ }\bibfield  {title} {\bibinfo {title} {{Constraints on the
  chiral unitary $\bar KN$ amplitude from $\pi\Sigma K^+$ photoproduction
  data}},\ }\href {https://doi.org/10.1140/epja/i2015-15030-3} {\bibfield
  {journal} {\bibinfo  {journal} {Eur. Phys. J. A}\ }\textbf {\bibinfo {volume}
  {51}},\ \bibinfo {pages} {30} (\bibinfo {year} {2015})},\ \Eprint
  {https://arxiv.org/abs/1411.7884} {arXiv:1411.7884 [hep-ph]} \BibitemShut
  {NoStop}%
\bibitem [{\citenamefont {Roca}\ and\ \citenamefont
  {Oset}(2013)}]{Roca:2013cca}%
  \BibitemOpen
  \bibfield  {author} {\bibinfo {author} {\bibfnamefont {L.}~\bibnamefont
  {Roca}}\ and\ \bibinfo {author} {\bibfnamefont {E.}~\bibnamefont {Oset}},\
  }\bibfield  {title} {\bibinfo {title} {{Isospin 0 and 1 resonances from $\pi
  \Sigma$ photoproduction data}},\ }\href
  {https://doi.org/10.1103/PhysRevC.88.055206} {\bibfield  {journal} {\bibinfo
  {journal} {Phys. Rev. C}\ }\textbf {\bibinfo {volume} {88}},\ \bibinfo
  {pages} {055206} (\bibinfo {year} {2013})},\ \Eprint
  {https://arxiv.org/abs/1307.5752} {arXiv:1307.5752 [nucl-th]} \BibitemShut
  {NoStop}%
\bibitem [{\citenamefont {Scheluchin}\ \emph {et~al.}(2022)\citenamefont
  {Scheluchin} \emph {et~al.}}]{BGOOD:2021sog}%
  \BibitemOpen
  \bibfield  {author} {\bibinfo {author} {\bibfnamefont {G.}~\bibnamefont
  {Scheluchin}} \emph {et~al.} (\bibinfo {collaboration} {BGOOD}),\ }\bibfield
  {title} {\bibinfo {title} {{Photoproduction of $K^+\Lambda(1405)\rightarrow
  K^+\pi^0\Sigma^0$ extending to forward angles and low momentum transfer}},\
  }\href {https://doi.org/10.1016/j.physletb.2022.137375} {\bibfield  {journal}
  {\bibinfo  {journal} {Phys. Lett. B}\ }\textbf {\bibinfo {volume} {833}},\
  \bibinfo {pages} {137375} (\bibinfo {year} {2022})},\ \Eprint
  {https://arxiv.org/abs/2108.12235} {arXiv:2108.12235 [nucl-ex]} \BibitemShut
  {NoStop}%
\bibitem [{\citenamefont {Acharya}\ \emph {et~al.}(2023)\citenamefont {Acharya}
  \emph {et~al.}}]{ALICE:2022yyh}%
  \BibitemOpen
  \bibfield  {author} {\bibinfo {author} {\bibfnamefont {S.}~\bibnamefont
  {Acharya}} \emph {et~al.} (\bibinfo {collaboration} {ALICE}),\ }\bibfield
  {title} {\bibinfo {title} {{Constraining the $\overline{K}\,N$ coupled
  channel dynamics using femtoscopic correlations at the LHC}},\ }\href
  {https://doi.org/10.1140/epjc/s10052-023-11476-0} {\bibfield  {journal}
  {\bibinfo  {journal} {Eur. Phys. J. C}\ }\textbf {\bibinfo {volume} {83}},\
  \bibinfo {pages} {340} (\bibinfo {year} {2023})},\ \Eprint
  {https://arxiv.org/abs/2205.15176} {arXiv:2205.15176 [nucl-ex]} \BibitemShut
  {NoStop}%
\bibitem [{\citenamefont {Wickramaarachchi}\ \emph {et~al.}(2022)\citenamefont
  {Wickramaarachchi}, \citenamefont {Schumacher},\ and\ \citenamefont
  {Kalicy}}]{Wickramaarachchi:2022mhi}%
  \BibitemOpen
  \bibfield  {author} {\bibinfo {author} {\bibfnamefont {N.}~\bibnamefont
  {Wickramaarachchi}}, \bibinfo {author} {\bibfnamefont {R.~A.}\ \bibnamefont
  {Schumacher}},\ and\ \bibinfo {author} {\bibfnamefont {G.}~\bibnamefont
  {Kalicy}} (\bibinfo {collaboration} {GlueX}),\ }\bibfield  {title} {\bibinfo
  {title} {{Decay of the \ensuremath{\Lambda}(1405) hyperon to
  \ensuremath{\Sigma^0} \ensuremath{\pi^0} measured at GlueX}},\ }\href
  {https://doi.org/10.1051/epjconf/202227107005} {\bibfield  {journal}
  {\bibinfo  {journal} {EPJ Web Conf.}\ }\textbf {\bibinfo {volume} {271}},\
  \bibinfo {pages} {07005} (\bibinfo {year} {2022})},\ \Eprint
  {https://arxiv.org/abs/2209.06230} {arXiv:2209.06230 [nucl-ex]} \BibitemShut
  {NoStop}%
\bibitem [{\citenamefont {Aikawa}\ \emph {et~al.}(2023)\citenamefont {Aikawa}
  \emph {et~al.}}]{J-PARCE31:2022plu}%
  \BibitemOpen
  \bibfield  {author} {\bibinfo {author} {\bibfnamefont {S.}~\bibnamefont
  {Aikawa}} \emph {et~al.} (\bibinfo {collaboration} {J-PARC E31}),\ }\bibfield
   {title} {\bibinfo {title} {{Pole position of \ensuremath{\Lambda}(1405)
  measured in d(K\ensuremath{^-},n)\ensuremath{\pi} \ensuremath{\Sigma}
  reactions}},\ }\href {https://doi.org/10.1016/j.physletb.2022.137637}
  {\bibfield  {journal} {\bibinfo  {journal} {Phys. Lett. B}\ }\textbf
  {\bibinfo {volume} {837}},\ \bibinfo {pages} {137637} (\bibinfo {year}
  {2023})},\ \Eprint {https://arxiv.org/abs/2209.08254} {arXiv:2209.08254
  [nucl-ex]} \BibitemShut {NoStop}%
\bibitem [{\citenamefont {Anisovich}\ \emph {et~al.}(2020)\citenamefont
  {Anisovich}, \citenamefont {Sarantsev}, \citenamefont {Nikonov},
  \citenamefont {Burkert}, \citenamefont {Schumacher}, \citenamefont {Thoma},\
  and\ \citenamefont {Klempt}}]{Anisovich:2020lec}%
  \BibitemOpen
  \bibfield  {author} {\bibinfo {author} {\bibfnamefont {A.~V.}\ \bibnamefont
  {Anisovich}}, \bibinfo {author} {\bibfnamefont {A.~V.}\ \bibnamefont
  {Sarantsev}}, \bibinfo {author} {\bibfnamefont {V.~A.}\ \bibnamefont
  {Nikonov}}, \bibinfo {author} {\bibfnamefont {V.}~\bibnamefont {Burkert}},
  \bibinfo {author} {\bibfnamefont {R.~A.}\ \bibnamefont {Schumacher}},
  \bibinfo {author} {\bibfnamefont {U.}~\bibnamefont {Thoma}},\ and\ \bibinfo
  {author} {\bibfnamefont {E.}~\bibnamefont {Klempt}},\ }\bibfield  {title}
  {\bibinfo {title} {{Hyperon III: $K^{-}p - \pi \Sigma $ coupled-channel
  dynamics in the $\Lambda (1405)$ mass region}},\ }\href
  {https://doi.org/10.1140/epja/s10050-020-00142-8} {\bibfield  {journal}
  {\bibinfo  {journal} {Eur. Phys. J. A}\ }\textbf {\bibinfo {volume} {56}},\
  \bibinfo {pages} {139} (\bibinfo {year} {2020})}\BibitemShut {NoStop}%
\bibitem [{\citenamefont {Isgur}\ and\ \citenamefont
  {Karl}(1978)}]{Isgur:1978xj}%
  \BibitemOpen
  \bibfield  {author} {\bibinfo {author} {\bibfnamefont {N.}~\bibnamefont
  {Isgur}}\ and\ \bibinfo {author} {\bibfnamefont {G.}~\bibnamefont {Karl}},\
  }\bibfield  {title} {\bibinfo {title} {{P Wave Baryons in the Quark Model}},\
  }\href {https://doi.org/10.1103/PhysRevD.18.4187} {\bibfield  {journal}
  {\bibinfo  {journal} {Phys. Rev. D}\ }\textbf {\bibinfo {volume} {18}},\
  \bibinfo {pages} {4187} (\bibinfo {year} {1978})}\BibitemShut {NoStop}%
\bibitem [{\citenamefont {Oller}\ and\ \citenamefont
  {Meissner}(2001)}]{Oller:2000fj}%
  \BibitemOpen
  \bibfield  {author} {\bibinfo {author} {\bibfnamefont {J.~A.}\ \bibnamefont
  {Oller}}\ and\ \bibinfo {author} {\bibfnamefont {U.~G.}\ \bibnamefont
  {Meissner}},\ }\bibfield  {title} {\bibinfo {title} {{Chiral dynamics in the
  presence of bound states: Kaon nucleon interactions revisited}},\ }\href
  {https://doi.org/10.1016/S0370-2693(01)00078-8} {\bibfield  {journal}
  {\bibinfo  {journal} {Phys. Lett. B}\ }\textbf {\bibinfo {volume} {500}},\
  \bibinfo {pages} {263} (\bibinfo {year} {2001})},\ \Eprint
  {https://arxiv.org/abs/hep-ph/0011146} {arXiv:hep-ph/0011146} \BibitemShut
  {NoStop}%
\bibitem [{\citenamefont {Kaiser}\ \emph {et~al.}(1995)\citenamefont {Kaiser},
  \citenamefont {Siegel},\ and\ \citenamefont {Weise}}]{Kaiser:1995eg}%
  \BibitemOpen
  \bibfield  {author} {\bibinfo {author} {\bibfnamefont {N.}~\bibnamefont
  {Kaiser}}, \bibinfo {author} {\bibfnamefont {P.~B.}\ \bibnamefont {Siegel}},\
  and\ \bibinfo {author} {\bibfnamefont {W.}~\bibnamefont {Weise}},\ }\bibfield
   {title} {\bibinfo {title} {{Chiral dynamics and the low-energy kaon -
  nucleon interaction}},\ }\href {https://doi.org/10.1016/0375-9474(95)00362-5}
  {\bibfield  {journal} {\bibinfo  {journal} {Nucl. Phys. A}\ }\textbf
  {\bibinfo {volume} {594}},\ \bibinfo {pages} {325} (\bibinfo {year}
  {1995})},\ \Eprint {https://arxiv.org/abs/nucl-th/9505043}
  {arXiv:nucl-th/9505043} \BibitemShut {NoStop}%
\bibitem [{\citenamefont {Oset}\ and\ \citenamefont
  {Ramos}(1998)}]{Oset:1997it}%
  \BibitemOpen
  \bibfield  {author} {\bibinfo {author} {\bibfnamefont {E.}~\bibnamefont
  {Oset}}\ and\ \bibinfo {author} {\bibfnamefont {A.}~\bibnamefont {Ramos}},\
  }\bibfield  {title} {\bibinfo {title} {{Nonperturbative chiral approach to
  $s$ wave $\overline{K} N$ interactions}},\ }\href
  {https://doi.org/10.1016/S0375-9474(98)00170-5} {\bibfield  {journal}
  {\bibinfo  {journal} {Nucl. Phys. A}\ }\textbf {\bibinfo {volume} {635}},\
  \bibinfo {pages} {99} (\bibinfo {year} {1998})},\ \Eprint
  {https://arxiv.org/abs/nucl-th/9711022} {arXiv:nucl-th/9711022} \BibitemShut
  {NoStop}%
\bibitem [{\citenamefont {Mai}(2018)}]{Mai:2018rjx}%
  \BibitemOpen
  \bibfield  {author} {\bibinfo {author} {\bibfnamefont {M.}~\bibnamefont
  {Mai}},\ }\bibfield  {title} {\bibinfo {title} {{Status of the $\Lambda
  (1405)$}},\ }\href {https://doi.org/10.1007/s00601-018-1389-4} {\bibfield
  {journal} {\bibinfo  {journal} {Few Body Syst.}\ }\textbf {\bibinfo {volume}
  {59}},\ \bibinfo {pages} {61} (\bibinfo {year} {2018})}\BibitemShut {NoStop}%
\bibitem [{\citenamefont {Mai}(2021)}]{Mai:2020ltx}%
  \BibitemOpen
  \bibfield  {author} {\bibinfo {author} {\bibfnamefont {M.}~\bibnamefont
  {Mai}},\ }\bibfield  {title} {\bibinfo {title} {{Review of the ${\Lambda
  }$(1405) A curious case of a strangeness resonance}},\ }\href
  {https://doi.org/10.1140/epjs/s11734-021-00144-7} {\bibfield  {journal}
  {\bibinfo  {journal} {Eur. Phys. J. ST}\ }\textbf {\bibinfo {volume} {230}},\
  \bibinfo {pages} {1593} (\bibinfo {year} {2021})},\ \Eprint
  {https://arxiv.org/abs/2010.00056} {arXiv:2010.00056 [nucl-th]} \BibitemShut
  {NoStop}%
\bibitem [{\citenamefont {Ezoe}\ and\ \citenamefont
  {Hosaka}(2020)}]{Ezoe:2020piq}%
  \BibitemOpen
  \bibfield  {author} {\bibinfo {author} {\bibfnamefont {T.}~\bibnamefont
  {Ezoe}}\ and\ \bibinfo {author} {\bibfnamefont {A.}~\bibnamefont {Hosaka}},\
  }\bibfield  {title} {\bibinfo {title} {{$\Lambda$(1405) as a $\overline{K}N$
  Feshbach resonance in the Skyrme model}},\ }\href
  {https://doi.org/10.1103/PhysRevD.102.014046} {\bibfield  {journal} {\bibinfo
   {journal} {Phys. Rev. D}\ }\textbf {\bibinfo {volume} {102}},\ \bibinfo
  {pages} {014046} (\bibinfo {year} {2020})},\ \Eprint
  {https://arxiv.org/abs/2006.03788} {arXiv:2006.03788 [hep-ph]} \BibitemShut
  {NoStop}%
\bibitem [{\citenamefont {Myint}\ \emph {et~al.}(2018)\citenamefont {Myint},
  \citenamefont {Akaishi}, \citenamefont {Hassanvand},\ and\ \citenamefont
  {Yamazaki}}]{Myint:2018ypc}%
  \BibitemOpen
  \bibfield  {author} {\bibinfo {author} {\bibfnamefont {K.~S.}\ \bibnamefont
  {Myint}}, \bibinfo {author} {\bibfnamefont {Y.}~\bibnamefont {Akaishi}},
  \bibinfo {author} {\bibfnamefont {M.}~\bibnamefont {Hassanvand}},\ and\
  \bibinfo {author} {\bibfnamefont {T.}~\bibnamefont {Yamazaki}},\ }\bibfield
  {title} {\bibinfo {title} {{Single-pole Nature of the Detectable
  Lambda(1405)}},\ }\href {https://doi.org/10.1093/ptep/pty075} {\bibfield
  {journal} {\bibinfo  {journal} {PTEP}\ }\textbf {\bibinfo {volume} {2018}},\
  \bibinfo {pages} {073D01} (\bibinfo {year} {2018})},\ \Eprint
  {https://arxiv.org/abs/1804.08240} {arXiv:1804.08240 [nucl-th]} \BibitemShut
  {NoStop}%
\bibitem [{\citenamefont {Miyahara}\ \emph {et~al.}(2018)\citenamefont
  {Miyahara}, \citenamefont {Hyodo},\ and\ \citenamefont
  {Weise}}]{Miyahara:2018onh}%
  \BibitemOpen
  \bibfield  {author} {\bibinfo {author} {\bibfnamefont {K.}~\bibnamefont
  {Miyahara}}, \bibinfo {author} {\bibfnamefont {T.}~\bibnamefont {Hyodo}},\
  and\ \bibinfo {author} {\bibfnamefont {W.}~\bibnamefont {Weise}},\ }\bibfield
   {title} {\bibinfo {title} {{Construction of a local $\bar K N-\pi \Sigma-\pi
  \Lambda$ potential and composition of the $\Lambda(1405)$}},\ }\href
  {https://doi.org/10.1103/PhysRevC.98.025201} {\bibfield  {journal} {\bibinfo
  {journal} {Phys. Rev. C}\ }\textbf {\bibinfo {volume} {98}},\ \bibinfo
  {pages} {025201} (\bibinfo {year} {2018})},\ \Eprint
  {https://arxiv.org/abs/1804.08269} {arXiv:1804.08269 [nucl-th]} \BibitemShut
  {NoStop}%
\bibitem [{\citenamefont {Miyahara}\ and\ \citenamefont
  {Hyodo}(2018)}]{Miyahara:2018lud}%
  \BibitemOpen
  \bibfield  {author} {\bibinfo {author} {\bibfnamefont {K.}~\bibnamefont
  {Miyahara}}\ and\ \bibinfo {author} {\bibfnamefont {T.}~\bibnamefont
  {Hyodo}},\ }\bibfield  {title} {\bibinfo {title} {{Theoretical study of
  $\Lambda(1405)$ resonance in $\Xi_b^0 \to D^0 (\pi \Sigma)$ decay}},\ }\href
  {https://doi.org/10.1103/PhysRevC.98.025202} {\bibfield  {journal} {\bibinfo
  {journal} {Phys. Rev. C}\ }\textbf {\bibinfo {volume} {98}},\ \bibinfo
  {pages} {025202} (\bibinfo {year} {2018})},\ \Eprint
  {https://arxiv.org/abs/1803.05572} {arXiv:1803.05572 [nucl-th]} \BibitemShut
  {NoStop}%
\bibitem [{\citenamefont {Azizi}\ \emph {et~al.}(2023)\citenamefont {Azizi},
  \citenamefont {Sarac},\ and\ \citenamefont {Sundu}}]{Azizi:2023tmw}%
  \BibitemOpen
  \bibfield  {author} {\bibinfo {author} {\bibfnamefont {K.}~\bibnamefont
  {Azizi}}, \bibinfo {author} {\bibfnamefont {Y.}~\bibnamefont {Sarac}},\ and\
  \bibinfo {author} {\bibfnamefont {H.}~\bibnamefont {Sundu}},\ }\bibfield
  {title} {\bibinfo {title} {{Investigation of $\Lambda(1405)$ as a molecular
  pentaquark state}},\ }\href@noop {} {\  (\bibinfo {year} {2023})},\ \Eprint
  {https://arxiv.org/abs/2306.07393} {arXiv:2306.07393 [hep-ph]} \BibitemShut
  {NoStop}%
\bibitem [{\citenamefont {Xie}\ \emph {et~al.}(2023)\citenamefont {Xie},
  \citenamefont {Lu}, \citenamefont {Geng},\ and\ \citenamefont
  {Zou}}]{Xie:2023cej}%
  \BibitemOpen
  \bibfield  {author} {\bibinfo {author} {\bibfnamefont {J.-M.}\ \bibnamefont
  {Xie}}, \bibinfo {author} {\bibfnamefont {J.-X.}\ \bibnamefont {Lu}},
  \bibinfo {author} {\bibfnamefont {L.-S.}\ \bibnamefont {Geng}},\ and\
  \bibinfo {author} {\bibfnamefont {B.-S.}\ \bibnamefont {Zou}},\ }\bibfield
  {title} {\bibinfo {title} {{Two-pole structures demystified: chiral dynamics
  at work}},\ }\href@noop {} {\bibfield  {journal} {\bibinfo  {journal}
  {(unpublished)}\ } (\bibinfo {year} {2023})},\ \Eprint
  {https://arxiv.org/abs/2307.11631} {arXiv:2307.11631 [hep-ph]} \BibitemShut
  {NoStop}%
\bibitem [{\citenamefont {Lu}\ \emph {et~al.}(2023)\citenamefont {Lu},
  \citenamefont {Geng}, \citenamefont {Doering},\ and\ \citenamefont
  {Mai}}]{Lu:2022hwm}%
  \BibitemOpen
  \bibfield  {author} {\bibinfo {author} {\bibfnamefont {J.-X.}\ \bibnamefont
  {Lu}}, \bibinfo {author} {\bibfnamefont {L.-S.}\ \bibnamefont {Geng}},
  \bibinfo {author} {\bibfnamefont {M.}~\bibnamefont {Doering}},\ and\ \bibinfo
  {author} {\bibfnamefont {M.}~\bibnamefont {Mai}},\ }\bibfield  {title}
  {\bibinfo {title} {{Cross-Channel Constraints on Resonant Antikaon-Nucleon
  Scattering}},\ }\href {https://doi.org/10.1103/PhysRevLett.130.071902}
  {\bibfield  {journal} {\bibinfo  {journal} {Phys. Rev. Lett.}\ }\textbf
  {\bibinfo {volume} {130}},\ \bibinfo {pages} {071902} (\bibinfo {year}
  {2023})},\ \Eprint {https://arxiv.org/abs/2209.02471} {arXiv:2209.02471
  [hep-ph]} \BibitemShut {NoStop}%
\bibitem [{\citenamefont {Hyodo}\ and\ \citenamefont
  {Weise}(2022)}]{Hyodo:2022xhp}%
  \BibitemOpen
  \bibfield  {author} {\bibinfo {author} {\bibfnamefont {T.}~\bibnamefont
  {Hyodo}}\ and\ \bibinfo {author} {\bibfnamefont {W.}~\bibnamefont {Weise}},\
  }\bibinfo {title} {{Theory of Kaon-Nuclear Systems}},\ in\ \href
  {https://doi.org/10.1007/978-981-15-8818-1_38-1} {\emph {\bibinfo {booktitle}
  {{Handbook of Nuclear Physics}}}},\ \bibinfo {editor} {edited by\ \bibinfo
  {editor} {\bibfnamefont {I.}~\bibnamefont {Tanihata}}, \bibinfo {editor}
  {\bibfnamefont {H.}~\bibnamefont {Toki}},\ and\ \bibinfo {editor}
  {\bibfnamefont {T.}~\bibnamefont {Kajino}}}\ (\bibinfo {year} {2022})\ pp.\
  \bibinfo {pages} {1--34},\ \Eprint {https://arxiv.org/abs/2202.06181}
  {arXiv:2202.06181 [nucl-th]} \BibitemShut {NoStop}%
\bibitem [{\citenamefont {Martinez~Torres}\ \emph {et~al.}(2012)\citenamefont
  {Martinez~Torres}, \citenamefont {Bayar}, \citenamefont {Jido},\ and\
  \citenamefont {Oset}}]{MartinezTorres:2012yi}%
  \BibitemOpen
  \bibfield  {author} {\bibinfo {author} {\bibfnamefont {A.}~\bibnamefont
  {Martinez~Torres}}, \bibinfo {author} {\bibfnamefont {M.}~\bibnamefont
  {Bayar}}, \bibinfo {author} {\bibfnamefont {D.}~\bibnamefont {Jido}},\ and\
  \bibinfo {author} {\bibfnamefont {E.}~\bibnamefont {Oset}},\ }\bibfield
  {title} {\bibinfo {title} {{Strategy to find the two $\Lambda(1405)$ states
  from lattice QCD simulations}},\ }\href
  {https://doi.org/10.1103/PhysRevC.86.055201} {\bibfield  {journal} {\bibinfo
  {journal} {Phys. Rev. C}\ }\textbf {\bibinfo {volume} {86}},\ \bibinfo
  {pages} {055201} (\bibinfo {year} {2012})},\ \Eprint
  {https://arxiv.org/abs/1202.4297} {arXiv:1202.4297 [hep-lat]} \BibitemShut
  {NoStop}%
\bibitem [{\citenamefont {Gubler}\ \emph {et~al.}(2016)\citenamefont {Gubler},
  \citenamefont {Takahashi},\ and\ \citenamefont {Oka}}]{Gubler:2016viv}%
  \BibitemOpen
  \bibfield  {author} {\bibinfo {author} {\bibfnamefont {P.}~\bibnamefont
  {Gubler}}, \bibinfo {author} {\bibfnamefont {T.~T.}\ \bibnamefont
  {Takahashi}},\ and\ \bibinfo {author} {\bibfnamefont {M.}~\bibnamefont
  {Oka}},\ }\bibfield  {title} {\bibinfo {title} {{Flavor structure of
  $\Lambda$ baryons from lattice QCD: From strange to charm quarks}},\ }\href
  {https://doi.org/10.1103/PhysRevD.94.114518} {\bibfield  {journal} {\bibinfo
  {journal} {Phys. Rev. D}\ }\textbf {\bibinfo {volume} {94}},\ \bibinfo
  {pages} {114518} (\bibinfo {year} {2016})},\ \Eprint
  {https://arxiv.org/abs/1609.01889} {arXiv:1609.01889 [hep-lat]} \BibitemShut
  {NoStop}%
\bibitem [{\citenamefont {Menadue}\ \emph {et~al.}(2012)\citenamefont
  {Menadue}, \citenamefont {Kamleh}, \citenamefont {Leinweber},\ and\
  \citenamefont {Mahbub}}]{Menadue:2011pd}%
  \BibitemOpen
  \bibfield  {author} {\bibinfo {author} {\bibfnamefont {B.~J.}\ \bibnamefont
  {Menadue}}, \bibinfo {author} {\bibfnamefont {W.}~\bibnamefont {Kamleh}},
  \bibinfo {author} {\bibfnamefont {D.~B.}\ \bibnamefont {Leinweber}},\ and\
  \bibinfo {author} {\bibfnamefont {M.~S.}\ \bibnamefont {Mahbub}},\ }\bibfield
   {title} {\bibinfo {title} {{Isolating the $\Lambda(1405)$ in Lattice QCD}},\
  }\href {https://doi.org/10.1103/PhysRevLett.108.112001} {\bibfield  {journal}
  {\bibinfo  {journal} {Phys. Rev. Lett.}\ }\textbf {\bibinfo {volume} {108}},\
  \bibinfo {pages} {112001} (\bibinfo {year} {2012})},\ \Eprint
  {https://arxiv.org/abs/1109.6716} {arXiv:1109.6716 [hep-lat]} \BibitemShut
  {NoStop}%
\bibitem [{\citenamefont {Engel}\ \emph
  {et~al.}(2013{\natexlab{a}})\citenamefont {Engel}, \citenamefont {Lang},\
  and\ \citenamefont {Sch\"afer}}]{Engel:2012qp}%
  \BibitemOpen
  \bibfield  {author} {\bibinfo {author} {\bibfnamefont {G.~P.}\ \bibnamefont
  {Engel}}, \bibinfo {author} {\bibfnamefont {C.~B.}\ \bibnamefont {Lang}},\
  and\ \bibinfo {author} {\bibfnamefont {A.}~\bibnamefont {Sch\"afer}}
  (\bibinfo {collaboration} {BGR (Bern-Graz-Regensburg)}),\ }\bibfield  {title}
  {\bibinfo {title} {{Low-lying $\Lambda$ baryons from the lattice}},\ }\href
  {https://doi.org/10.1103/PhysRevD.87.034502} {\bibfield  {journal} {\bibinfo
  {journal} {Phys. Rev. D}\ }\textbf {\bibinfo {volume} {87}},\ \bibinfo
  {pages} {034502} (\bibinfo {year} {2013}{\natexlab{a}})},\ \Eprint
  {https://arxiv.org/abs/1212.2032} {arXiv:1212.2032 [hep-lat]} \BibitemShut
  {NoStop}%
\bibitem [{\citenamefont {Engel}\ \emph
  {et~al.}(2013{\natexlab{b}})\citenamefont {Engel}, \citenamefont {Lang},
  \citenamefont {Mohler},\ and\ \citenamefont {Sch\"afer}}]{Engel:2013ig}%
  \BibitemOpen
  \bibfield  {author} {\bibinfo {author} {\bibfnamefont {G.~P.}\ \bibnamefont
  {Engel}}, \bibinfo {author} {\bibfnamefont {C.~B.}\ \bibnamefont {Lang}},
  \bibinfo {author} {\bibfnamefont {D.}~\bibnamefont {Mohler}},\ and\ \bibinfo
  {author} {\bibfnamefont {A.}~\bibnamefont {Sch\"afer}} (\bibinfo
  {collaboration} {BGR}),\ }\bibfield  {title} {\bibinfo {title} {{QCD with Two
  Light Dynamical Chirally Improved Quarks: Baryons}},\ }\href
  {https://doi.org/10.1103/PhysRevD.87.074504} {\bibfield  {journal} {\bibinfo
  {journal} {Phys. Rev. D}\ }\textbf {\bibinfo {volume} {87}},\ \bibinfo
  {pages} {074504} (\bibinfo {year} {2013}{\natexlab{b}})},\ \Eprint
  {https://arxiv.org/abs/1301.4318} {arXiv:1301.4318 [hep-lat]} \BibitemShut
  {NoStop}%
\bibitem [{\citenamefont {Nemoto}\ \emph {et~al.}(2003)\citenamefont {Nemoto},
  \citenamefont {Nakajima}, \citenamefont {Matsufuru},\ and\ \citenamefont
  {Suganuma}}]{Nemoto:2003ft}%
  \BibitemOpen
  \bibfield  {author} {\bibinfo {author} {\bibfnamefont {Y.}~\bibnamefont
  {Nemoto}}, \bibinfo {author} {\bibfnamefont {N.}~\bibnamefont {Nakajima}},
  \bibinfo {author} {\bibfnamefont {H.}~\bibnamefont {Matsufuru}},\ and\
  \bibinfo {author} {\bibfnamefont {H.}~\bibnamefont {Suganuma}},\ }\bibfield
  {title} {\bibinfo {title} {{Negative parity baryons in quenched anisotropic
  lattice QCD}},\ }\href {https://doi.org/10.1103/PhysRevD.68.094505}
  {\bibfield  {journal} {\bibinfo  {journal} {Phys. Rev. D}\ }\textbf {\bibinfo
  {volume} {68}},\ \bibinfo {pages} {094505} (\bibinfo {year} {2003})},\
  \Eprint {https://arxiv.org/abs/hep-lat/0302013} {arXiv:hep-lat/0302013}
  \BibitemShut {NoStop}%
\bibitem [{\citenamefont {Burch}\ \emph {et~al.}(2006)\citenamefont {Burch},
  \citenamefont {Gattringer}, \citenamefont {Glozman}, \citenamefont {Hagen},
  \citenamefont {Hierl}, \citenamefont {Lang},\ and\ \citenamefont
  {Sch{\"a}fer}}]{Burch:2006cc}%
  \BibitemOpen
  \bibfield  {author} {\bibinfo {author} {\bibfnamefont {T.}~\bibnamefont
  {Burch}}, \bibinfo {author} {\bibfnamefont {C.}~\bibnamefont {Gattringer}},
  \bibinfo {author} {\bibfnamefont {L.~Y.}\ \bibnamefont {Glozman}}, \bibinfo
  {author} {\bibfnamefont {C.}~\bibnamefont {Hagen}}, \bibinfo {author}
  {\bibfnamefont {D.}~\bibnamefont {Hierl}}, \bibinfo {author} {\bibfnamefont
  {C.~B.}\ \bibnamefont {Lang}},\ and\ \bibinfo {author} {\bibfnamefont
  {A.}~\bibnamefont {Sch{\"a}fer}},\ }\bibfield  {title} {\bibinfo {title}
  {{Excited hadrons on the lattice: Baryons}},\ }\href
  {https://doi.org/10.1103/PhysRevD.74.014504} {\bibfield  {journal} {\bibinfo
  {journal} {Phys. Rev. D}\ }\textbf {\bibinfo {volume} {74}},\ \bibinfo
  {pages} {014504} (\bibinfo {year} {2006})},\ \Eprint
  {https://arxiv.org/abs/hep-lat/0604019} {arXiv:hep-lat/0604019} \BibitemShut
  {NoStop}%
\bibitem [{\citenamefont {Takahashi}\ and\ \citenamefont
  {Oka}(2010)}]{Takahashi:2009bu}%
  \BibitemOpen
  \bibfield  {author} {\bibinfo {author} {\bibfnamefont {T.~T.}\ \bibnamefont
  {Takahashi}}\ and\ \bibinfo {author} {\bibfnamefont {M.}~\bibnamefont
  {Oka}},\ }\bibfield  {title} {\bibinfo {title} {{Low-lying Lambda Baryons
  with spin 1/2 in Two-flavor Lattice QCD}},\ }\href
  {https://doi.org/10.1103/PhysRevD.81.034505} {\bibfield  {journal} {\bibinfo
  {journal} {Phys. Rev. D}\ }\textbf {\bibinfo {volume} {81}},\ \bibinfo
  {pages} {034505} (\bibinfo {year} {2010})},\ \Eprint
  {https://arxiv.org/abs/0910.0686} {arXiv:0910.0686 [hep-lat]} \BibitemShut
  {NoStop}%
\bibitem [{\citenamefont {Meinel}\ and\ \citenamefont
  {Rendon}(2022)}]{Meinel:2021grq}%
  \BibitemOpen
  \bibfield  {author} {\bibinfo {author} {\bibfnamefont {S.}~\bibnamefont
  {Meinel}}\ and\ \bibinfo {author} {\bibfnamefont {G.}~\bibnamefont
  {Rendon}},\ }\bibfield  {title} {\bibinfo {title} {{Charm-baryon semileptonic
  decays and the strange $\Lambda^*$ resonances: New insights from lattice
  QCD}},\ }\href {https://doi.org/10.1103/PhysRevD.105.L051505} {\bibfield
  {journal} {\bibinfo  {journal} {Phys. Rev. D}\ }\textbf {\bibinfo {volume}
  {105}},\ \bibinfo {pages} {L051505} (\bibinfo {year} {2022})},\ \Eprint
  {https://arxiv.org/abs/2107.13084} {arXiv:2107.13084 [hep-ph]} \BibitemShut
  {NoStop}%
\bibitem [{\citenamefont {Hall}\ \emph {et~al.}(2015)\citenamefont {Hall},
  \citenamefont {Kamleh}, \citenamefont {Leinweber}, \citenamefont {Menadue},
  \citenamefont {Owen}, \citenamefont {Thomas},\ and\ \citenamefont
  {Young}}]{Hall:2014uca}%
  \BibitemOpen
  \bibfield  {author} {\bibinfo {author} {\bibfnamefont {J.~M.~M.}\
  \bibnamefont {Hall}}, \bibinfo {author} {\bibfnamefont {W.}~\bibnamefont
  {Kamleh}}, \bibinfo {author} {\bibfnamefont {D.~B.}\ \bibnamefont
  {Leinweber}}, \bibinfo {author} {\bibfnamefont {B.~J.}\ \bibnamefont
  {Menadue}}, \bibinfo {author} {\bibfnamefont {B.~J.}\ \bibnamefont {Owen}},
  \bibinfo {author} {\bibfnamefont {A.~W.}\ \bibnamefont {Thomas}},\ and\
  \bibinfo {author} {\bibfnamefont {R.~D.}\ \bibnamefont {Young}},\ }\bibfield
  {title} {\bibinfo {title} {{Lattice QCD Evidence that the
  \ensuremath{\Lambda}(1405) Resonance is an Antikaon-Nucleon Molecule}},\
  }\href {https://doi.org/10.1103/PhysRevLett.114.132002} {\bibfield  {journal}
  {\bibinfo  {journal} {Phys. Rev. Lett.}\ }\textbf {\bibinfo {volume} {114}},\
  \bibinfo {pages} {132002} (\bibinfo {year} {2015})},\ \Eprint
  {https://arxiv.org/abs/1411.3402} {arXiv:1411.3402 [hep-lat]} \BibitemShut
  {NoStop}%
\bibitem [{\citenamefont {Lang}\ and\ \citenamefont
  {Verduci}(2013)}]{Lang:2012db}%
  \BibitemOpen
  \bibfield  {author} {\bibinfo {author} {\bibfnamefont {C.}~\bibnamefont
  {Lang}}\ and\ \bibinfo {author} {\bibfnamefont {V.}~\bibnamefont {Verduci}},\
  }\bibfield  {title} {\bibinfo {title} {{Scattering in the $\pi N$ negative
  parity channel in lattice QCD}},\ }\href
  {https://doi.org/10.1103/PhysRevD.87.054502} {\bibfield  {journal} {\bibinfo
  {journal} {Phys.Rev.}\ }\textbf {\bibinfo {volume} {D87}},\ \bibinfo {pages}
  {054502} (\bibinfo {year} {2013})},\ \Eprint
  {https://arxiv.org/abs/1212.5055} {arXiv:1212.5055} \BibitemShut {NoStop}%
\bibitem [{\citenamefont {Mohler}\ \emph {et~al.}(2013)\citenamefont {Mohler},
  \citenamefont {Prelovsek},\ and\ \citenamefont {Woloshyn}}]{Mohler:2012na}%
  \BibitemOpen
  \bibfield  {author} {\bibinfo {author} {\bibfnamefont {D.}~\bibnamefont
  {Mohler}}, \bibinfo {author} {\bibfnamefont {S.}~\bibnamefont {Prelovsek}},\
  and\ \bibinfo {author} {\bibfnamefont {R.~M.}\ \bibnamefont {Woloshyn}},\
  }\bibfield  {title} {\bibinfo {title} {{$D \pi$ scattering and $D$ meson
  resonances from lattice QCD}},\ }\href
  {https://doi.org/10.1103/PhysRevD.87.034501} {\bibfield  {journal} {\bibinfo
  {journal} {Phys. Rev.}\ }\textbf {\bibinfo {volume} {D87}},\ \bibinfo {pages}
  {034501} (\bibinfo {year} {2013})},\ \Eprint
  {https://arxiv.org/abs/1208.4059} {arXiv:1208.4059 [hep-lat]} \BibitemShut
  {NoStop}%
\bibitem [{\citenamefont {Wilson}\ \emph {et~al.}(2015)\citenamefont {Wilson},
  \citenamefont {Briceno}, \citenamefont {Dudek}, \citenamefont {Edwards},\
  and\ \citenamefont {Thomas}}]{Wilson:2015dqa}%
  \BibitemOpen
  \bibfield  {author} {\bibinfo {author} {\bibfnamefont {D.~J.}\ \bibnamefont
  {Wilson}}, \bibinfo {author} {\bibfnamefont {R.~A.}\ \bibnamefont {Briceno}},
  \bibinfo {author} {\bibfnamefont {J.~J.}\ \bibnamefont {Dudek}}, \bibinfo
  {author} {\bibfnamefont {R.~G.}\ \bibnamefont {Edwards}},\ and\ \bibinfo
  {author} {\bibfnamefont {C.~E.}\ \bibnamefont {Thomas}},\ }\bibfield  {title}
  {\bibinfo {title} {{Coupled $\pi\pi, K\bar{K}$ scattering in $P$-wave and the
  $\rho$ resonance from lattice QCD}},\ }\href
  {https://doi.org/10.1103/PhysRevD.92.094502} {\bibfield  {journal} {\bibinfo
  {journal} {Phys. Rev.}\ }\textbf {\bibinfo {volume} {D92}},\ \bibinfo {pages}
  {094502} (\bibinfo {year} {2015})},\ \Eprint
  {https://arxiv.org/abs/1507.02599} {arXiv:1507.02599 [hep-ph]} \BibitemShut
  {NoStop}%
\bibitem [{\citenamefont {Fukugita}\ \emph {et~al.}(1995)\citenamefont
  {Fukugita}, \citenamefont {Kuramashi}, \citenamefont {Okawa}, \citenamefont
  {Mino},\ and\ \citenamefont {Ukawa}}]{Fukugita:1994ve}%
  \BibitemOpen
  \bibfield  {author} {\bibinfo {author} {\bibfnamefont {M.}~\bibnamefont
  {Fukugita}}, \bibinfo {author} {\bibfnamefont {Y.}~\bibnamefont {Kuramashi}},
  \bibinfo {author} {\bibfnamefont {M.}~\bibnamefont {Okawa}}, \bibinfo
  {author} {\bibfnamefont {H.}~\bibnamefont {Mino}},\ and\ \bibinfo {author}
  {\bibfnamefont {A.}~\bibnamefont {Ukawa}},\ }\bibfield  {title} {\bibinfo
  {title} {{Hadron scattering lengths in lattice QCD}},\ }\href
  {https://doi.org/10.1103/PhysRevD.52.3003} {\bibfield  {journal} {\bibinfo
  {journal} {Phys. Rev.}\ }\textbf {\bibinfo {volume} {D52}},\ \bibinfo {pages}
  {3003} (\bibinfo {year} {1995})},\ \Eprint
  {https://arxiv.org/abs/hep-lat/9501024} {arXiv:hep-lat/9501024 [hep-lat]}
  \BibitemShut {NoStop}%
\bibitem [{\citenamefont {Bernard}\ and\ \citenamefont
  {Golterman}(1996)}]{Bernard:1995ez}%
  \BibitemOpen
  \bibfield  {author} {\bibinfo {author} {\bibfnamefont {C.~W.}\ \bibnamefont
  {Bernard}}\ and\ \bibinfo {author} {\bibfnamefont {M.~F.~L.}\ \bibnamefont
  {Golterman}},\ }\bibfield  {title} {\bibinfo {title} {{Finite volume two pion
  energies and scattering in the quenched approximation}},\ }\href
  {https://doi.org/10.1103/PhysRevD.53.476} {\bibfield  {journal} {\bibinfo
  {journal} {Phys. Rev. D}\ }\textbf {\bibinfo {volume} {53}},\ \bibinfo
  {pages} {476} (\bibinfo {year} {1996})},\ \Eprint
  {https://arxiv.org/abs/hep-lat/9507004} {arXiv:hep-lat/9507004} \BibitemShut
  {NoStop}%
\bibitem [{\citenamefont {Detmold}\ and\ \citenamefont
  {Nicholson}(2016)}]{Detmold:2015qwf}%
  \BibitemOpen
  \bibfield  {author} {\bibinfo {author} {\bibfnamefont {W.}~\bibnamefont
  {Detmold}}\ and\ \bibinfo {author} {\bibfnamefont {A.}~\bibnamefont
  {Nicholson}},\ }\bibfield  {title} {\bibinfo {title} {{Low energy scattering
  phase shifts for meson-baryon systems}},\ }\href
  {https://doi.org/10.1103/PhysRevD.93.114511} {\bibfield  {journal} {\bibinfo
  {journal} {Phys. Rev.}\ }\textbf {\bibinfo {volume} {D93}},\ \bibinfo {pages}
  {114511} (\bibinfo {year} {2016})},\ \Eprint
  {https://arxiv.org/abs/1511.02275} {arXiv:1511.02275 [hep-lat]} \BibitemShut
  {NoStop}%
\bibitem [{\citenamefont {Torok}\ \emph {et~al.}(2010)\citenamefont {Torok},
  \citenamefont {Beane}, \citenamefont {Detmold}, \citenamefont {Luu},
  \citenamefont {Orginos}, \citenamefont {Parreno}, \citenamefont {Savage},\
  and\ \citenamefont {Walker-Loud}}]{Torok:2009dg}%
  \BibitemOpen
  \bibfield  {author} {\bibinfo {author} {\bibfnamefont {A.}~\bibnamefont
  {Torok}}, \bibinfo {author} {\bibfnamefont {S.~R.}\ \bibnamefont {Beane}},
  \bibinfo {author} {\bibfnamefont {W.}~\bibnamefont {Detmold}}, \bibinfo
  {author} {\bibfnamefont {T.~C.}\ \bibnamefont {Luu}}, \bibinfo {author}
  {\bibfnamefont {K.}~\bibnamefont {Orginos}}, \bibinfo {author} {\bibfnamefont
  {A.}~\bibnamefont {Parreno}}, \bibinfo {author} {\bibfnamefont {M.~J.}\
  \bibnamefont {Savage}},\ and\ \bibinfo {author} {\bibfnamefont
  {A.}~\bibnamefont {Walker-Loud}},\ }\bibfield  {title} {\bibinfo {title}
  {{Meson-Baryon Scattering Lengths from Mixed-Action Lattice QCD}},\ }\href
  {https://doi.org/10.1103/PhysRevD.81.074506} {\bibfield  {journal} {\bibinfo
  {journal} {Phys. Rev.}\ }\textbf {\bibinfo {volume} {D81}},\ \bibinfo {pages}
  {074506} (\bibinfo {year} {2010})},\ \Eprint
  {https://arxiv.org/abs/0907.1913} {arXiv:0907.1913 [hep-lat]} \BibitemShut
  {NoStop}%
\bibitem [{\citenamefont {Meng}\ \emph {et~al.}(2004)\citenamefont {Meng},
  \citenamefont {Miao}, \citenamefont {Du},\ and\ \citenamefont
  {Liu}}]{Meng:2003gm}%
  \BibitemOpen
  \bibfield  {author} {\bibinfo {author} {\bibfnamefont {G.-W.}\ \bibnamefont
  {Meng}}, \bibinfo {author} {\bibfnamefont {C.}~\bibnamefont {Miao}}, \bibinfo
  {author} {\bibfnamefont {X.-N.}\ \bibnamefont {Du}},\ and\ \bibinfo {author}
  {\bibfnamefont {C.}~\bibnamefont {Liu}},\ }\bibfield  {title} {\bibinfo
  {title} {{Lattice study on kaon nucleon scattering length in the $I = 1$
  channel}},\ }\href {https://doi.org/10.1142/S0217751X04019627} {\bibfield
  {journal} {\bibinfo  {journal} {Int. J. Mod. Phys. A}\ }\textbf {\bibinfo
  {volume} {19}},\ \bibinfo {pages} {4401} (\bibinfo {year} {2004})},\ \Eprint
  {https://arxiv.org/abs/hep-lat/0309048} {arXiv:hep-lat/0309048} \BibitemShut
  {NoStop}%
\bibitem [{\citenamefont {Bulava}\ \emph {et~al.}(2023)\citenamefont {Bulava},
  \citenamefont {Hanlon}, \citenamefont {H\"orz}, \citenamefont {Morningstar},
  \citenamefont {Nicholson}, \citenamefont {Romero-L\'opez}, \citenamefont
  {Skinner}, \citenamefont {Vranas},\ and\ \citenamefont
  {Walker-Loud}}]{Bulava:2022vpq}%
  \BibitemOpen
  \bibfield  {author} {\bibinfo {author} {\bibfnamefont {J.}~\bibnamefont
  {Bulava}}, \bibinfo {author} {\bibfnamefont {A.~D.}\ \bibnamefont {Hanlon}},
  \bibinfo {author} {\bibfnamefont {B.}~\bibnamefont {H\"orz}}, \bibinfo
  {author} {\bibfnamefont {C.}~\bibnamefont {Morningstar}}, \bibinfo {author}
  {\bibfnamefont {A.}~\bibnamefont {Nicholson}}, \bibinfo {author}
  {\bibfnamefont {F.}~\bibnamefont {Romero-L\'opez}}, \bibinfo {author}
  {\bibfnamefont {S.}~\bibnamefont {Skinner}}, \bibinfo {author} {\bibfnamefont
  {P.}~\bibnamefont {Vranas}},\ and\ \bibinfo {author} {\bibfnamefont
  {A.}~\bibnamefont {Walker-Loud}},\ }\bibfield  {title} {\bibinfo {title}
  {{Elastic nucleon-pion scattering at $m_\pi=200$~MeV from lattice QCD}},\
  }\href {https://doi.org/10.1016/j.nuclphysb.2023.116105} {\bibfield
  {journal} {\bibinfo  {journal} {Nucl. Phys. B}\ }\textbf {\bibinfo {volume}
  {987}},\ \bibinfo {pages} {116105} (\bibinfo {year} {2023})},\ \Eprint
  {https://arxiv.org/abs/2208.03867} {arXiv:2208.03867 [hep-lat]} \BibitemShut
  {NoStop}%
\bibitem [{\citenamefont {Moir}\ \emph {et~al.}(2016)\citenamefont {Moir},
  \citenamefont {Peardon}, \citenamefont {Ryan}, \citenamefont {Thomas},\ and\
  \citenamefont {Wilson}}]{Moir:2016srx}%
  \BibitemOpen
  \bibfield  {author} {\bibinfo {author} {\bibfnamefont {G.}~\bibnamefont
  {Moir}}, \bibinfo {author} {\bibfnamefont {M.}~\bibnamefont {Peardon}},
  \bibinfo {author} {\bibfnamefont {S.~M.}\ \bibnamefont {Ryan}}, \bibinfo
  {author} {\bibfnamefont {C.~E.}\ \bibnamefont {Thomas}},\ and\ \bibinfo
  {author} {\bibfnamefont {D.~J.}\ \bibnamefont {Wilson}},\ }\bibfield  {title}
  {\bibinfo {title} {{Coupled-Channel $D\pi$, $D\eta$ and $D_{s}\bar{K}$
  Scattering from Lattice QCD}},\ }\href
  {https://doi.org/10.1007/JHEP10(2016)011} {\bibfield  {journal} {\bibinfo
  {journal} {JHEP}\ }\textbf {\bibinfo {volume} {10}},\ \bibinfo {pages}
  {011}},\ \Eprint {https://arxiv.org/abs/1607.07093} {arXiv:1607.07093
  [hep-lat]} \BibitemShut {NoStop}%
\bibitem [{\citenamefont {Briceno}\ \emph {et~al.}(2018)\citenamefont
  {Briceno}, \citenamefont {Dudek}, \citenamefont {Edwards},\ and\
  \citenamefont {Wilson}}]{Briceno:2017qmb}%
  \BibitemOpen
  \bibfield  {author} {\bibinfo {author} {\bibfnamefont {R.~A.}\ \bibnamefont
  {Briceno}}, \bibinfo {author} {\bibfnamefont {J.~J.}\ \bibnamefont {Dudek}},
  \bibinfo {author} {\bibfnamefont {R.~G.}\ \bibnamefont {Edwards}},\ and\
  \bibinfo {author} {\bibfnamefont {D.~J.}\ \bibnamefont {Wilson}},\ }\bibfield
   {title} {\bibinfo {title} {{Isoscalar $\pi\pi, K\overline{K}, \eta\eta$
  scattering and the $\sigma, f_0, f_2$ mesons from QCD}},\ }\href
  {https://doi.org/10.1103/PhysRevD.97.054513} {\bibfield  {journal} {\bibinfo
  {journal} {Phys. Rev.}\ }\textbf {\bibinfo {volume} {D97}},\ \bibinfo {pages}
  {054513} (\bibinfo {year} {2018})},\ \Eprint
  {https://arxiv.org/abs/1708.06667} {arXiv:1708.06667 [hep-lat]} \BibitemShut
  {NoStop}%
\bibitem [{\citenamefont {Woss}\ \emph {et~al.}(2019)\citenamefont {Woss},
  \citenamefont {Thomas}, \citenamefont {Dudek}, \citenamefont {Edwards},\ and\
  \citenamefont {Wilson}}]{Woss:2019hse}%
  \BibitemOpen
  \bibfield  {author} {\bibinfo {author} {\bibfnamefont {A.~J.}\ \bibnamefont
  {Woss}}, \bibinfo {author} {\bibfnamefont {C.~E.}\ \bibnamefont {Thomas}},
  \bibinfo {author} {\bibfnamefont {J.~J.}\ \bibnamefont {Dudek}}, \bibinfo
  {author} {\bibfnamefont {R.~G.}\ \bibnamefont {Edwards}},\ and\ \bibinfo
  {author} {\bibfnamefont {D.~J.}\ \bibnamefont {Wilson}},\ }\bibfield  {title}
  {\bibinfo {title} {{$b_1$ resonance in coupled $\pi\omega$, $\pi\phi$
  scattering from lattice QCD}},\ }\href
  {https://doi.org/10.1103/PhysRevD.100.054506} {\bibfield  {journal} {\bibinfo
   {journal} {Phys. Rev. D}\ }\textbf {\bibinfo {volume} {100}},\ \bibinfo
  {pages} {054506} (\bibinfo {year} {2019})},\ \Eprint
  {https://arxiv.org/abs/1904.04136} {arXiv:1904.04136 [hep-lat]} \BibitemShut
  {NoStop}%
\bibitem [{\citenamefont {Dudek}\ \emph {et~al.}(2014)\citenamefont {Dudek},
  \citenamefont {Edwards}, \citenamefont {Thomas},\ and\ \citenamefont
  {Wilson}}]{Dudek:2014qha}%
  \BibitemOpen
  \bibfield  {author} {\bibinfo {author} {\bibfnamefont {J.~J.}\ \bibnamefont
  {Dudek}}, \bibinfo {author} {\bibfnamefont {R.~G.}\ \bibnamefont {Edwards}},
  \bibinfo {author} {\bibfnamefont {C.~E.}\ \bibnamefont {Thomas}},\ and\
  \bibinfo {author} {\bibfnamefont {D.~J.}\ \bibnamefont {Wilson}} (\bibinfo
  {collaboration} {Hadron Spectrum}),\ }\bibfield  {title} {\bibinfo {title}
  {{Resonances in coupled $\pi K -\eta K$ scattering from quantum
  chromodynamics}},\ }\href {https://doi.org/10.1103/PhysRevLett.113.182001}
  {\bibfield  {journal} {\bibinfo  {journal} {Phys. Rev. Lett.}\ }\textbf
  {\bibinfo {volume} {113}},\ \bibinfo {pages} {182001} (\bibinfo {year}
  {2014})},\ \Eprint {https://arxiv.org/abs/1406.4158} {arXiv:1406.4158
  [hep-ph]} \BibitemShut {NoStop}%
\bibitem [{\citenamefont {Prelovsek}\ \emph {et~al.}(2021)\citenamefont
  {Prelovsek}, \citenamefont {Collins}, \citenamefont {Mohler}, \citenamefont
  {Padmanath},\ and\ \citenamefont {Piemonte}}]{Prelovsek:2020eiw}%
  \BibitemOpen
  \bibfield  {author} {\bibinfo {author} {\bibfnamefont {S.}~\bibnamefont
  {Prelovsek}}, \bibinfo {author} {\bibfnamefont {S.}~\bibnamefont {Collins}},
  \bibinfo {author} {\bibfnamefont {D.}~\bibnamefont {Mohler}}, \bibinfo
  {author} {\bibfnamefont {M.}~\bibnamefont {Padmanath}},\ and\ \bibinfo
  {author} {\bibfnamefont {S.}~\bibnamefont {Piemonte}},\ }\bibfield  {title}
  {\bibinfo {title} {{Charmonium-like resonances with J$^{PC}$ = 0$^{++}$,
  2$^{++}$ in coupled $ \mathrm{D}\overline{\mathrm{D}} $, $
  {\mathrm{D}}_{\mathrm{s}}{\overline{\mathrm{D}}}_{\mathrm{s}} $ scattering on
  the lattice}},\ }\href {https://doi.org/10.1007/JHEP06(2021)035} {\bibfield
  {journal} {\bibinfo  {journal} {JHEP}\ }\textbf {\bibinfo {volume} {06}},\
  \bibinfo {pages} {035}},\ \Eprint {https://arxiv.org/abs/2011.02542}
  {arXiv:2011.02542 [hep-lat]} \BibitemShut {NoStop}%
\bibitem [{\citenamefont {Bulava}\ \emph {et~al.}(2024)\citenamefont {Bulava},
  \citenamefont {Cid-Mora}, \citenamefont {Hanlon}, \citenamefont {H{\"{o}}rz},
  \citenamefont {Mohler}, \citenamefont {Morningstar}, \citenamefont {Moscoso},
  \citenamefont {Nicholson}, \citenamefont {Romero-L{\'{o}}pez}, \citenamefont
  {Skinner},\ and\ \citenamefont {Walker-Loud}}]{Bulava:2023rmn}%
  \BibitemOpen
  \bibfield  {author} {\bibinfo {author} {\bibfnamefont {J.}~\bibnamefont
  {Bulava}}, \bibinfo {author} {\bibfnamefont {B.}~\bibnamefont {Cid-Mora}},
  \bibinfo {author} {\bibfnamefont {A.~D.}\ \bibnamefont {Hanlon}}, \bibinfo
  {author} {\bibfnamefont {B.}~\bibnamefont {H{\"{o}}rz}}, \bibinfo {author}
  {\bibfnamefont {D.}~\bibnamefont {Mohler}}, \bibinfo {author} {\bibfnamefont
  {C.}~\bibnamefont {Morningstar}}, \bibinfo {author} {\bibfnamefont
  {J.}~\bibnamefont {Moscoso}}, \bibinfo {author} {\bibfnamefont
  {A.}~\bibnamefont {Nicholson}}, \bibinfo {author} {\bibfnamefont
  {F.}~\bibnamefont {Romero-L{\'{o}}pez}}, \bibinfo {author} {\bibfnamefont
  {S.}~\bibnamefont {Skinner}},\ and\ \bibinfo {author} {\bibfnamefont
  {A.}~\bibnamefont {Walker-Loud}},\ }\bibfield  {title} {\bibinfo {title}
  {{Two-pole nature of the $\Lambda(1405)$ from lattice QCD}},\ }\href
  {https://doi.org/10.1103/PhysRevLett.132.051901} {\bibfield  {journal}
  {\bibinfo  {journal} {Phys. Rev. Lett.}\ }\textbf {\bibinfo {volume} {132}},\
  \bibinfo {pages} {051901} (\bibinfo {year} {2024})},\ \Eprint
  {https://arxiv.org/abs/2307.10413} {arXiv:2307.10413 [hep-lat]} \BibitemShut
  {NoStop}%
\bibitem [{\citenamefont {Bruno}\ \emph {et~al.}(2015)\citenamefont {Bruno}
  \emph {et~al.}}]{Bruno:2014jqa}%
  \BibitemOpen
  \bibfield  {author} {\bibinfo {author} {\bibfnamefont {M.}~\bibnamefont
  {Bruno}} \emph {et~al.},\ }\bibfield  {title} {\bibinfo {title} {{Simulation
  of QCD with N$_{f} =$ 2 $+$ 1 flavors of non-perturbatively improved Wilson
  fermions}},\ }\href {https://doi.org/10.1007/JHEP02(2015)043} {\bibfield
  {journal} {\bibinfo  {journal} {JHEP}\ }\textbf {\bibinfo {volume} {02}},\
  \bibinfo {pages} {043}},\ \Eprint {https://arxiv.org/abs/1411.3982}
  {arXiv:1411.3982 [hep-lat]} \BibitemShut {NoStop}%
\bibitem [{\citenamefont {Luscher}\ and\ \citenamefont
  {Weisz}(1985)}]{Luscher:1984xn}%
  \BibitemOpen
  \bibfield  {author} {\bibinfo {author} {\bibfnamefont {M.}~\bibnamefont
  {Luscher}}\ and\ \bibinfo {author} {\bibfnamefont {P.}~\bibnamefont
  {Weisz}},\ }\bibfield  {title} {\bibinfo {title} {{On-Shell Improved Lattice
  Gauge Theories}},\ }\href {https://doi.org/10.1007/BF01206178} {\bibfield
  {journal} {\bibinfo  {journal} {Commun. Math. Phys.}\ }\textbf {\bibinfo
  {volume} {97}},\ \bibinfo {pages} {59} (\bibinfo {year} {1985})},\ \bibinfo
  {note} {[Erratum: Commun. Math. Phys.98,433(1985)]}\BibitemShut {NoStop}%
\bibitem [{\citenamefont {Bulava}\ and\ \citenamefont
  {Schaefer}(2013)}]{Bulava:2013cta}%
  \BibitemOpen
  \bibfield  {author} {\bibinfo {author} {\bibfnamefont {J.}~\bibnamefont
  {Bulava}}\ and\ \bibinfo {author} {\bibfnamefont {S.}~\bibnamefont
  {Schaefer}},\ }\bibfield  {title} {\bibinfo {title} {{Improvement of
  $N_{\mathrm{f}}=3$ lattice QCD with Wilson fermions and tree-level improved
  gauge action}},\ }\href {https://doi.org/10.1016/j.nuclphysb.2013.05.019}
  {\bibfield  {journal} {\bibinfo  {journal} {Nucl. Phys.}\ }\textbf {\bibinfo
  {volume} {B874}},\ \bibinfo {pages} {188} (\bibinfo {year} {2013})},\ \Eprint
  {https://arxiv.org/abs/1304.7093} {arXiv:1304.7093 [hep-lat]} \BibitemShut
  {NoStop}%
\bibitem [{\citenamefont {{L\"{u}scher}}\ and\ \citenamefont
  {Schaefer}(2011)}]{Luscher:2011kk}%
  \BibitemOpen
  \bibfield  {author} {\bibinfo {author} {\bibfnamefont {M.}~\bibnamefont
  {{L\"{u}scher}}}\ and\ \bibinfo {author} {\bibfnamefont {S.}~\bibnamefont
  {Schaefer}},\ }\bibfield  {title} {\bibinfo {title} {{Lattice QCD without
  topology barriers}},\ }\href {https://doi.org/10.1007/JHEP07(2011)036}
  {\bibfield  {journal} {\bibinfo  {journal} {JHEP}\ }\textbf {\bibinfo
  {volume} {07}},\ \bibinfo {pages} {036}},\ \Eprint
  {https://arxiv.org/abs/1105.4749} {arXiv:1105.4749 [hep-lat]} \BibitemShut
  {NoStop}%
\bibitem [{\citenamefont {Clark}\ and\ \citenamefont
  {Kennedy}(2007)}]{Clark:2006fx}%
  \BibitemOpen
  \bibfield  {author} {\bibinfo {author} {\bibfnamefont {M.~A.}\ \bibnamefont
  {Clark}}\ and\ \bibinfo {author} {\bibfnamefont {A.~D.}\ \bibnamefont
  {Kennedy}},\ }\bibfield  {title} {\bibinfo {title} {{Accelerating dynamical
  fermion computations using the rational hybrid Monte Carlo (RHMC) algorithm
  with multiple pseudofermion fields}},\ }\href
  {https://doi.org/10.1103/PhysRevLett.98.051601} {\bibfield  {journal}
  {\bibinfo  {journal} {Phys. Rev. Lett.}\ }\textbf {\bibinfo {volume} {98}},\
  \bibinfo {pages} {051601} (\bibinfo {year} {2007})},\ \Eprint
  {https://arxiv.org/abs/hep-lat/0608015} {arXiv:hep-lat/0608015 [hep-lat]}
  \BibitemShut {NoStop}%
\bibitem [{\citenamefont {Mohler}\ and\ \citenamefont
  {Schaefer}(2020)}]{Mohler:2020txx}%
  \BibitemOpen
  \bibfield  {author} {\bibinfo {author} {\bibfnamefont {D.}~\bibnamefont
  {Mohler}}\ and\ \bibinfo {author} {\bibfnamefont {S.}~\bibnamefont
  {Schaefer}},\ }\bibfield  {title} {\bibinfo {title} {{Remarks on
  strange-quark simulations with Wilson fermions}},\ }\href
  {https://doi.org/10.1103/PhysRevD.102.074506} {\bibfield  {journal} {\bibinfo
   {journal} {Phys. Rev. D}\ }\textbf {\bibinfo {volume} {102}},\ \bibinfo
  {pages} {074506} (\bibinfo {year} {2020})},\ \Eprint
  {https://arxiv.org/abs/2003.13359} {arXiv:2003.13359 [hep-lat]} \BibitemShut
  {NoStop}%
\bibitem [{\citenamefont {{L\"{u}scher}}\ and\ \citenamefont
  {Palombi}(2008)}]{Luscher:2008tw}%
  \BibitemOpen
  \bibfield  {author} {\bibinfo {author} {\bibfnamefont {M.}~\bibnamefont
  {{L\"{u}scher}}}\ and\ \bibinfo {author} {\bibfnamefont {F.}~\bibnamefont
  {Palombi}},\ }\bibfield  {title} {\bibinfo {title} {{Fluctuations and
  reweighting of the quark determinant on large lattices}},\ }\bibfield
  {booktitle} {\emph {\bibinfo {booktitle} {{Proceedings, 26th International
  Symposium on Lattice field theory (Lattice 2008): Williamsburg, USA, July
  14-19, 2008}}},\ }\href@noop {} {\bibfield  {journal} {\bibinfo  {journal}
  {PoS}\ }\textbf {\bibinfo {volume} {LATTICE2008}},\ \bibinfo {pages} {049}
  (\bibinfo {year} {2008})},\ \Eprint {https://arxiv.org/abs/0810.0946}
  {arXiv:0810.0946 [hep-lat]} \BibitemShut {NoStop}%
\bibitem [{\citenamefont {Kuberski}(2023)}]{Kuberski:2023zky}%
  \BibitemOpen
  \bibfield  {author} {\bibinfo {author} {\bibfnamefont {S.}~\bibnamefont
  {Kuberski}},\ }\bibfield  {title} {\bibinfo {title} {{Low-mode deflation for
  twisted-mass and RHMC reweighting in lattice QCD}},\ }\href@noop {} {\
  (\bibinfo {year} {2023})},\ \Eprint {https://arxiv.org/abs/2306.02385}
  {arXiv:2306.02385 [hep-lat]} \BibitemShut {NoStop}%
\bibitem [{\citenamefont {Bruno}\ \emph {et~al.}(2017)\citenamefont {Bruno},
  \citenamefont {Korzec},\ and\ \citenamefont {Schaefer}}]{Bruno:2016plf}%
  \BibitemOpen
  \bibfield  {author} {\bibinfo {author} {\bibfnamefont {M.}~\bibnamefont
  {Bruno}}, \bibinfo {author} {\bibfnamefont {T.}~\bibnamefont {Korzec}},\ and\
  \bibinfo {author} {\bibfnamefont {S.}~\bibnamefont {Schaefer}},\ }\bibfield
  {title} {\bibinfo {title} {{Setting the scale for the CLS $2 + 1$ flavor
  ensembles}},\ }\href {https://doi.org/10.1103/PhysRevD.95.074504} {\bibfield
  {journal} {\bibinfo  {journal} {Phys. Rev.}\ }\textbf {\bibinfo {volume}
  {D95}},\ \bibinfo {pages} {074504} (\bibinfo {year} {2017})},\ \Eprint
  {https://arxiv.org/abs/1608.08900} {arXiv:1608.08900 [hep-lat]} \BibitemShut
  {NoStop}%
\bibitem [{\citenamefont {Strassberger}\ \emph {et~al.}(2022)\citenamefont
  {Strassberger} \emph {et~al.}}]{Strassberger:2021tsu}%
  \BibitemOpen
  \bibfield  {author} {\bibinfo {author} {\bibfnamefont {B.}~\bibnamefont
  {Strassberger}} \emph {et~al.},\ }\bibfield  {title} {\bibinfo {title}
  {{Scale setting for CLS 2+1 simulations}},\ }\href
  {https://doi.org/10.22323/1.396.0135} {\bibfield  {journal} {\bibinfo
  {journal} {PoS}\ }\textbf {\bibinfo {volume} {LATTICE2021}},\ \bibinfo
  {pages} {135} (\bibinfo {year} {2022})},\ \Eprint
  {https://arxiv.org/abs/2112.06696} {arXiv:2112.06696 [hep-lat]} \BibitemShut
  {NoStop}%
\bibitem [{\citenamefont {Bali}\ \emph {et~al.}(2023)\citenamefont {Bali},
  \citenamefont {Collins}, \citenamefont {Georg}, \citenamefont {Jenkins},
  \citenamefont {Korcyl}, \citenamefont {Sch\"afer}, \citenamefont {Scholz},
  \citenamefont {Simeth}, \citenamefont {S\"oldner},\ and\ \citenamefont
  {Weish\"aupl}}]{RQCD:2022xux}%
  \BibitemOpen
  \bibfield  {author} {\bibinfo {author} {\bibfnamefont {G.~S.}\ \bibnamefont
  {Bali}}, \bibinfo {author} {\bibfnamefont {S.}~\bibnamefont {Collins}},
  \bibinfo {author} {\bibfnamefont {P.}~\bibnamefont {Georg}}, \bibinfo
  {author} {\bibfnamefont {D.}~\bibnamefont {Jenkins}}, \bibinfo {author}
  {\bibfnamefont {P.}~\bibnamefont {Korcyl}}, \bibinfo {author} {\bibfnamefont
  {A.}~\bibnamefont {Sch\"afer}}, \bibinfo {author} {\bibfnamefont {E.~E.}\
  \bibnamefont {Scholz}}, \bibinfo {author} {\bibfnamefont {J.}~\bibnamefont
  {Simeth}}, \bibinfo {author} {\bibfnamefont {W.}~\bibnamefont {S\"oldner}},\
  and\ \bibinfo {author} {\bibfnamefont {S.}~\bibnamefont {Weish\"aupl}}
  (\bibinfo {collaboration} {RQCD}),\ }\bibfield  {title} {\bibinfo {title}
  {{Scale setting and the light baryon spectrum in N$_{f}$ = 2 + 1 QCD with
  Wilson fermions}},\ }\href {https://doi.org/10.1007/JHEP05(2023)035}
  {\bibfield  {journal} {\bibinfo  {journal} {JHEP}\ }\textbf {\bibinfo
  {volume} {05}},\ \bibinfo {pages} {035}},\ \Eprint
  {https://arxiv.org/abs/2211.03744} {arXiv:2211.03744 [hep-lat]} \BibitemShut
  {NoStop}%
\bibitem [{\citenamefont {Morningstar}\ \emph {et~al.}(2013)\citenamefont
  {Morningstar}, \citenamefont {Bulava}, \citenamefont {Fahy}, \citenamefont
  {Foley}, \citenamefont {Jhang} \emph {et~al.}}]{Morningstar:2013bda}%
  \BibitemOpen
  \bibfield  {author} {\bibinfo {author} {\bibfnamefont {C.}~\bibnamefont
  {Morningstar}}, \bibinfo {author} {\bibfnamefont {J.}~\bibnamefont {Bulava}},
  \bibinfo {author} {\bibfnamefont {B.}~\bibnamefont {Fahy}}, \bibinfo {author}
  {\bibfnamefont {J.}~\bibnamefont {Foley}}, \bibinfo {author} {\bibfnamefont
  {Y.}~\bibnamefont {Jhang}}, \emph {et~al.},\ }\bibfield  {title} {\bibinfo
  {title} {{Extended hadron and two-hadron operators of definite momentum for
  spectrum calculations in lattice QCD}},\ }\href
  {https://doi.org/10.1103/PhysRevD.88.014511} {\bibfield  {journal} {\bibinfo
  {journal} {Phys.Rev.}\ }\textbf {\bibinfo {volume} {D88}},\ \bibinfo {pages}
  {014511} (\bibinfo {year} {2013})},\ \Eprint
  {https://arxiv.org/abs/1303.6816} {arXiv:1303.6816 [hep-lat]} \BibitemShut
  {NoStop}%
\bibitem [{\citenamefont {Morningstar}\ \emph {et~al.}(2011)\citenamefont
  {Morningstar}, \citenamefont {Bulava}, \citenamefont {Foley}, \citenamefont
  {Juge}, \citenamefont {Lenkner}, \citenamefont {Peardon},\ and\ \citenamefont
  {Wong}}]{Morningstar:2011ka}%
  \BibitemOpen
  \bibfield  {author} {\bibinfo {author} {\bibfnamefont {C.}~\bibnamefont
  {Morningstar}}, \bibinfo {author} {\bibfnamefont {J.}~\bibnamefont {Bulava}},
  \bibinfo {author} {\bibfnamefont {J.}~\bibnamefont {Foley}}, \bibinfo
  {author} {\bibfnamefont {K.~J.}\ \bibnamefont {Juge}}, \bibinfo {author}
  {\bibfnamefont {D.}~\bibnamefont {Lenkner}}, \bibinfo {author} {\bibfnamefont
  {M.}~\bibnamefont {Peardon}},\ and\ \bibinfo {author} {\bibfnamefont {C.~H.}\
  \bibnamefont {Wong}},\ }\bibfield  {title} {\bibinfo {title} {{Improved
  stochastic estimation of quark propagation with Laplacian Heaviside smearing
  in lattice QCD}},\ }\href {https://doi.org/10.1103/PhysRevD.83.114505}
  {\bibfield  {journal} {\bibinfo  {journal} {Phys. Rev.}\ }\textbf {\bibinfo
  {volume} {D83}},\ \bibinfo {pages} {114505} (\bibinfo {year} {2011})},\
  \Eprint {https://arxiv.org/abs/1104.3870} {arXiv:1104.3870 [hep-lat]}
  \BibitemShut {NoStop}%
\bibitem [{\citenamefont {Peardon}\ \emph {et~al.}(2009)\citenamefont
  {Peardon}, \citenamefont {Bulava}, \citenamefont {Foley}, \citenamefont
  {Morningstar}, \citenamefont {Dudek}, \citenamefont {Edwards}, \citenamefont
  {Joo}, \citenamefont {Lin}, \citenamefont {Richards},\ and\ \citenamefont
  {Juge}}]{Peardon:2009gh}%
  \BibitemOpen
  \bibfield  {author} {\bibinfo {author} {\bibfnamefont {M.}~\bibnamefont
  {Peardon}}, \bibinfo {author} {\bibfnamefont {J.}~\bibnamefont {Bulava}},
  \bibinfo {author} {\bibfnamefont {J.}~\bibnamefont {Foley}}, \bibinfo
  {author} {\bibfnamefont {C.}~\bibnamefont {Morningstar}}, \bibinfo {author}
  {\bibfnamefont {J.}~\bibnamefont {Dudek}}, \bibinfo {author} {\bibfnamefont
  {R.~G.}\ \bibnamefont {Edwards}}, \bibinfo {author} {\bibfnamefont
  {B.}~\bibnamefont {Joo}}, \bibinfo {author} {\bibfnamefont {H.-W.}\
  \bibnamefont {Lin}}, \bibinfo {author} {\bibfnamefont {D.~G.}\ \bibnamefont
  {Richards}},\ and\ \bibinfo {author} {\bibfnamefont {K.~J.}\ \bibnamefont
  {Juge}} (\bibinfo {collaboration} {Hadron Spectrum}),\ }\bibfield  {title}
  {\bibinfo {title} {{A Novel quark-field creation operator construction for
  hadronic physics in lattice QCD}},\ }\href
  {https://doi.org/10.1103/PhysRevD.80.054506} {\bibfield  {journal} {\bibinfo
  {journal} {Phys. Rev.}\ }\textbf {\bibinfo {volume} {D80}},\ \bibinfo {pages}
  {054506} (\bibinfo {year} {2009})},\ \Eprint
  {https://arxiv.org/abs/0905.2160} {arXiv:0905.2160 [hep-lat]} \BibitemShut
  {NoStop}%
\bibitem [{\citenamefont {Morningstar}\ and\ \citenamefont
  {Peardon}(2004)}]{Morningstar:2003gk}%
  \BibitemOpen
  \bibfield  {author} {\bibinfo {author} {\bibfnamefont {C.}~\bibnamefont
  {Morningstar}}\ and\ \bibinfo {author} {\bibfnamefont {M.~J.}\ \bibnamefont
  {Peardon}},\ }\bibfield  {title} {\bibinfo {title} {{Analytic smearing of
  SU(3) link variables in lattice QCD}},\ }\href
  {https://doi.org/10.1103/PhysRevD.69.054501} {\bibfield  {journal} {\bibinfo
  {journal} {Phys. Rev.}\ }\textbf {\bibinfo {volume} {D69}},\ \bibinfo {pages}
  {054501} (\bibinfo {year} {2004})},\ \Eprint
  {https://arxiv.org/abs/hep-lat/0311018} {arXiv:hep-lat/0311018 [hep-lat]}
  \BibitemShut {NoStop}%
\bibitem [{\citenamefont {H\"orz}\ and\ \citenamefont
  {Hanlon}(2019)}]{Horz:2019rrn}%
  \BibitemOpen
  \bibfield  {author} {\bibinfo {author} {\bibfnamefont {B.}~\bibnamefont
  {H\"orz}}\ and\ \bibinfo {author} {\bibfnamefont {A.}~\bibnamefont
  {Hanlon}},\ }\bibfield  {title} {\bibinfo {title} {{Two- and three-pion
  finite-volume spectra at maximal isospin from lattice QCD}},\ }\href
  {https://doi.org/10.1103/PhysRevLett.123.142002} {\bibfield  {journal}
  {\bibinfo  {journal} {Phys. Rev. Lett.}\ }\textbf {\bibinfo {volume} {123}},\
  \bibinfo {pages} {142002} (\bibinfo {year} {2019})},\ \Eprint
  {https://arxiv.org/abs/1905.04277} {arXiv:1905.04277 [hep-lat]} \BibitemShut
  {NoStop}%
\bibitem [{\citenamefont {Bali}\ \emph {et~al.}(2021)\citenamefont {Bali},
  \citenamefont {Braun}, \citenamefont {Collins}, \citenamefont {Sch\"afer},\
  and\ \citenamefont {Simeth}}]{Bali:2021qem}%
  \BibitemOpen
  \bibfield  {author} {\bibinfo {author} {\bibfnamefont {G.~S.}\ \bibnamefont
  {Bali}}, \bibinfo {author} {\bibfnamefont {V.}~\bibnamefont {Braun}},
  \bibinfo {author} {\bibfnamefont {S.}~\bibnamefont {Collins}}, \bibinfo
  {author} {\bibfnamefont {A.}~\bibnamefont {Sch\"afer}},\ and\ \bibinfo
  {author} {\bibfnamefont {J.}~\bibnamefont {Simeth}} (\bibinfo {collaboration}
  {RQCD}),\ }\bibfield  {title} {\bibinfo {title} {{Masses and decay constants
  of the $\eta$ and $\eta'$ mesons from lattice QCD}},\ }\href
  {https://doi.org/10.1007/JHEP08(2021)137} {\bibfield  {journal} {\bibinfo
  {journal} {JHEP}\ }\textbf {\bibinfo {volume} {08}},\ \bibinfo {pages}
  {137}},\ \Eprint {https://arxiv.org/abs/2106.05398} {arXiv:2106.05398
  [hep-lat]} \BibitemShut {NoStop}%
\bibitem [{\citenamefont {C\`e}\ \emph {et~al.}(2022)\citenamefont {C\`e} \emph
  {et~al.}}]{Ce:2022kxy}%
  \BibitemOpen
  \bibfield  {author} {\bibinfo {author} {\bibfnamefont {M.}~\bibnamefont
  {C\`e}} \emph {et~al.},\ }\bibfield  {title} {\bibinfo {title} {{Window
  observable for the hadronic vacuum polarization contribution to the muon g-2
  from lattice QCD}},\ }\href {https://doi.org/10.1103/PhysRevD.106.114502}
  {\bibfield  {journal} {\bibinfo  {journal} {Phys. Rev. D}\ }\textbf {\bibinfo
  {volume} {106}},\ \bibinfo {pages} {114502} (\bibinfo {year} {2022})},\
  \Eprint {https://arxiv.org/abs/2206.06582} {arXiv:2206.06582 [hep-lat]}
  \BibitemShut {NoStop}%
\bibitem [{\citenamefont {Michael}\ and\ \citenamefont
  {Teasdale}(1983)}]{Michael:1982gb}%
  \BibitemOpen
  \bibfield  {author} {\bibinfo {author} {\bibfnamefont {C.}~\bibnamefont
  {Michael}}\ and\ \bibinfo {author} {\bibfnamefont {I.}~\bibnamefont
  {Teasdale}},\ }\bibfield  {title} {\bibinfo {title} {{Extracting Glueball
  Masses From Lattice {QCD}}},\ }\href
  {https://doi.org/10.1016/0550-3213(83)90674-0} {\bibfield  {journal}
  {\bibinfo  {journal} {Nucl. Phys.}\ }\textbf {\bibinfo {volume} {B215}},\
  \bibinfo {pages} {433} (\bibinfo {year} {1983})}\BibitemShut {NoStop}%
\bibitem [{\citenamefont {Luscher}\ and\ \citenamefont
  {Wolff}(1990)}]{Luscher:1990ck}%
  \BibitemOpen
  \bibfield  {author} {\bibinfo {author} {\bibfnamefont {M.}~\bibnamefont
  {Luscher}}\ and\ \bibinfo {author} {\bibfnamefont {U.}~\bibnamefont
  {Wolff}},\ }\bibfield  {title} {\bibinfo {title} {{How to Calculate the
  Elastic Scattering Matrix in Two-dimensional Quantum Field Theories by
  Numerical Simulation}},\ }\href
  {https://doi.org/10.1016/0550-3213(90)90540-T} {\bibfield  {journal}
  {\bibinfo  {journal} {Nucl. Phys.}\ }\textbf {\bibinfo {volume} {B339}},\
  \bibinfo {pages} {222} (\bibinfo {year} {1990})}\BibitemShut {NoStop}%
\bibitem [{\citenamefont {Blossier}\ \emph {et~al.}(2009)\citenamefont
  {Blossier}, \citenamefont {Della~Morte}, \citenamefont {von Hippel},
  \citenamefont {Mendes},\ and\ \citenamefont {Sommer}}]{Blossier:2009kd}%
  \BibitemOpen
  \bibfield  {author} {\bibinfo {author} {\bibfnamefont {B.}~\bibnamefont
  {Blossier}}, \bibinfo {author} {\bibfnamefont {M.}~\bibnamefont
  {Della~Morte}}, \bibinfo {author} {\bibfnamefont {G.}~\bibnamefont {von
  Hippel}}, \bibinfo {author} {\bibfnamefont {T.}~\bibnamefont {Mendes}},\ and\
  \bibinfo {author} {\bibfnamefont {R.}~\bibnamefont {Sommer}},\ }\bibfield
  {title} {\bibinfo {title} {{On the generalized eigenvalue method for energies
  and matrix elements in lattice field theory}},\ }\href
  {https://doi.org/10.1088/1126-6708/2009/04/094} {\bibfield  {journal}
  {\bibinfo  {journal} {JHEP}\ }\textbf {\bibinfo {volume} {04}},\ \bibinfo
  {pages} {094}},\ \Eprint {https://arxiv.org/abs/0902.1265} {arXiv:0902.1265
  [hep-lat]} \BibitemShut {NoStop}%
\bibitem [{\citenamefont {Aoki}\ \emph {et~al.}(2007)\citenamefont {Aoki} \emph
  {et~al.}}]{CP-PACS:2007wro}%
  \BibitemOpen
  \bibfield  {author} {\bibinfo {author} {\bibfnamefont {S.}~\bibnamefont
  {Aoki}} \emph {et~al.} (\bibinfo {collaboration} {CP-PACS}),\ }\bibfield
  {title} {\bibinfo {title} {{Lattice QCD Calculation of the rho Meson Decay
  Width}},\ }\href {https://doi.org/10.1103/PhysRevD.76.094506} {\bibfield
  {journal} {\bibinfo  {journal} {Phys. Rev. D}\ }\textbf {\bibinfo {volume}
  {76}},\ \bibinfo {pages} {094506} (\bibinfo {year} {2007})},\ \Eprint
  {https://arxiv.org/abs/0708.3705} {arXiv:0708.3705 [hep-lat]} \BibitemShut
  {NoStop}%
\bibitem [{\citenamefont {{L\"{u}scher}}(1991)}]{Luscher:1990ux}%
  \BibitemOpen
  \bibfield  {author} {\bibinfo {author} {\bibfnamefont {M.}~\bibnamefont
  {{L\"{u}scher}}},\ }\bibfield  {title} {\bibinfo {title} {{Two particle
  states on a torus and their relation to the scattering matrix}},\ }\href
  {https://doi.org/10.1016/0550-3213(91)90366-6} {\bibfield  {journal}
  {\bibinfo  {journal} {Nucl. Phys.}\ }\textbf {\bibinfo {volume} {B354}},\
  \bibinfo {pages} {531} (\bibinfo {year} {1991})}\BibitemShut {NoStop}%
\bibitem [{\citenamefont {Rummukainen}\ and\ \citenamefont
  {Gottlieb}(1995)}]{Rummukainen:1995vs}%
  \BibitemOpen
  \bibfield  {author} {\bibinfo {author} {\bibfnamefont {K.}~\bibnamefont
  {Rummukainen}}\ and\ \bibinfo {author} {\bibfnamefont {S.~A.}\ \bibnamefont
  {Gottlieb}},\ }\bibfield  {title} {\bibinfo {title} {{Resonance scattering
  phase shifts on a nonrest frame lattice}},\ }\href
  {https://doi.org/10.1016/0550-3213(95)00313-H} {\bibfield  {journal}
  {\bibinfo  {journal} {Nucl. Phys.}\ }\textbf {\bibinfo {volume} {B450}},\
  \bibinfo {pages} {397} (\bibinfo {year} {1995})},\ \Eprint
  {https://arxiv.org/abs/hep-lat/9503028} {arXiv:hep-lat/9503028 [hep-lat]}
  \BibitemShut {NoStop}%
\bibitem [{\citenamefont {Kim}\ \emph {et~al.}(2005)\citenamefont {Kim},
  \citenamefont {Sachrajda},\ and\ \citenamefont {Sharpe}}]{Kim:2005gf}%
  \BibitemOpen
  \bibfield  {author} {\bibinfo {author} {\bibfnamefont {C.~H.}\ \bibnamefont
  {Kim}}, \bibinfo {author} {\bibfnamefont {C.~T.}\ \bibnamefont {Sachrajda}},\
  and\ \bibinfo {author} {\bibfnamefont {S.~R.}\ \bibnamefont {Sharpe}},\
  }\bibfield  {title} {\bibinfo {title} {{Finite-volume effects for two-hadron
  states in moving frames}},\ }\href
  {https://doi.org/10.1016/j.nuclphysb.2005.08.029} {\bibfield  {journal}
  {\bibinfo  {journal} {Nucl. Phys.}\ }\textbf {\bibinfo {volume} {B727}},\
  \bibinfo {pages} {218} (\bibinfo {year} {2005})},\ \Eprint
  {https://arxiv.org/abs/hep-lat/0507006} {arXiv:hep-lat/0507006 [hep-lat]}
  \BibitemShut {NoStop}%
\bibitem [{\citenamefont {He}\ \emph {et~al.}(2005)\citenamefont {He},
  \citenamefont {Feng},\ and\ \citenamefont {Liu}}]{He:2005ey}%
  \BibitemOpen
  \bibfield  {author} {\bibinfo {author} {\bibfnamefont {S.}~\bibnamefont
  {He}}, \bibinfo {author} {\bibfnamefont {X.}~\bibnamefont {Feng}},\ and\
  \bibinfo {author} {\bibfnamefont {C.}~\bibnamefont {Liu}},\ }\bibfield
  {title} {\bibinfo {title} {{Two particle states and the S-matrix elements in
  multi-channel scattering}},\ }\href
  {https://doi.org/10.1088/1126-6708/2005/07/011} {\bibfield  {journal}
  {\bibinfo  {journal} {JHEP}\ }\textbf {\bibinfo {volume} {07}},\ \bibinfo
  {pages} {011}},\ \Eprint {https://arxiv.org/abs/hep-lat/0504019}
  {arXiv:hep-lat/0504019 [hep-lat]} \BibitemShut {NoStop}%
\bibitem [{\citenamefont {Bernard}\ \emph {et~al.}(2011)\citenamefont
  {Bernard}, \citenamefont {Lage}, \citenamefont {Meissner},\ and\
  \citenamefont {Rusetsky}}]{Bernard:2010fp}%
  \BibitemOpen
  \bibfield  {author} {\bibinfo {author} {\bibfnamefont {V.}~\bibnamefont
  {Bernard}}, \bibinfo {author} {\bibfnamefont {M.}~\bibnamefont {Lage}},
  \bibinfo {author} {\bibfnamefont {U.~G.}\ \bibnamefont {Meissner}},\ and\
  \bibinfo {author} {\bibfnamefont {A.}~\bibnamefont {Rusetsky}},\ }\bibfield
  {title} {\bibinfo {title} {{Scalar mesons in a finite volume}},\ }\href
  {https://doi.org/10.1007/JHEP01(2011)019} {\bibfield  {journal} {\bibinfo
  {journal} {JHEP}\ }\textbf {\bibinfo {volume} {01}},\ \bibinfo {pages}
  {019}},\ \Eprint {https://arxiv.org/abs/1010.6018} {arXiv:1010.6018
  [hep-lat]} \BibitemShut {NoStop}%
\bibitem [{\citenamefont {{G\"{o}ckeler}}\ \emph {et~al.}(2012)\citenamefont
  {{G\"{o}ckeler}}, \citenamefont {Horsley}, \citenamefont {Lage},
  \citenamefont {{Mei{\ss}ner}}, \citenamefont {Rakow}, \citenamefont
  {Rusetsky}, \citenamefont {Schierholz},\ and\ \citenamefont
  {Zanotti}}]{Gockeler:2012yj}%
  \BibitemOpen
  \bibfield  {author} {\bibinfo {author} {\bibfnamefont {M.}~\bibnamefont
  {{G\"{o}ckeler}}}, \bibinfo {author} {\bibfnamefont {R.}~\bibnamefont
  {Horsley}}, \bibinfo {author} {\bibfnamefont {M.}~\bibnamefont {Lage}},
  \bibinfo {author} {\bibfnamefont {U.~G.}\ \bibnamefont {{Mei{\ss}ner}}},
  \bibinfo {author} {\bibfnamefont {P.~E.~L.}\ \bibnamefont {Rakow}}, \bibinfo
  {author} {\bibfnamefont {A.}~\bibnamefont {Rusetsky}}, \bibinfo {author}
  {\bibfnamefont {G.}~\bibnamefont {Schierholz}},\ and\ \bibinfo {author}
  {\bibfnamefont {J.~M.}\ \bibnamefont {Zanotti}},\ }\bibfield  {title}
  {\bibinfo {title} {{Scattering phases for meson and baryon resonances on
  general moving-frame lattices}},\ }\href
  {https://doi.org/10.1103/PhysRevD.86.094513} {\bibfield  {journal} {\bibinfo
  {journal} {Phys. Rev.}\ }\textbf {\bibinfo {volume} {D86}},\ \bibinfo {pages}
  {094513} (\bibinfo {year} {2012})},\ \Eprint
  {https://arxiv.org/abs/1206.4141} {arXiv:1206.4141 [hep-lat]} \BibitemShut
  {NoStop}%
\bibitem [{\citenamefont {Briceno}\ and\ \citenamefont
  {Davoudi}(2013)}]{Briceno:2012yi}%
  \BibitemOpen
  \bibfield  {author} {\bibinfo {author} {\bibfnamefont {R.~A.}\ \bibnamefont
  {Briceno}}\ and\ \bibinfo {author} {\bibfnamefont {Z.}~\bibnamefont
  {Davoudi}},\ }\bibfield  {title} {\bibinfo {title} {{Moving multichannel
  systems in a finite volume with application to proton-proton fusion}},\
  }\href {https://doi.org/10.1103/PhysRevD.88.094507} {\bibfield  {journal}
  {\bibinfo  {journal} {Phys. Rev.}\ }\textbf {\bibinfo {volume} {D88}},\
  \bibinfo {pages} {094507} (\bibinfo {year} {2013})},\ \Eprint
  {https://arxiv.org/abs/1204.1110} {arXiv:1204.1110 [hep-lat]} \BibitemShut
  {NoStop}%
\bibitem [{\citenamefont {Briceno}(2014)}]{Briceno:2014oea}%
  \BibitemOpen
  \bibfield  {author} {\bibinfo {author} {\bibfnamefont {R.~A.}\ \bibnamefont
  {Briceno}},\ }\bibfield  {title} {\bibinfo {title} {{Two-particle
  multichannel systems in a finite volume with arbitrary spin}},\ }\href
  {https://doi.org/10.1103/PhysRevD.89.074507} {\bibfield  {journal} {\bibinfo
  {journal} {Phys. Rev.}\ }\textbf {\bibinfo {volume} {D89}},\ \bibinfo {pages}
  {074507} (\bibinfo {year} {2014})},\ \Eprint
  {https://arxiv.org/abs/1401.3312} {arXiv:1401.3312 [hep-lat]} \BibitemShut
  {NoStop}%
\bibitem [{\citenamefont {Morningstar}\ \emph {et~al.}(2017)\citenamefont
  {Morningstar}, \citenamefont {Bulava}, \citenamefont {Singha}, \citenamefont
  {Brett}, \citenamefont {Fallica}, \citenamefont {Hanlon},\ and\ \citenamefont
  {{H\"{o}rz}}}]{Morningstar:2017spu}%
  \BibitemOpen
  \bibfield  {author} {\bibinfo {author} {\bibfnamefont {C.}~\bibnamefont
  {Morningstar}}, \bibinfo {author} {\bibfnamefont {J.}~\bibnamefont {Bulava}},
  \bibinfo {author} {\bibfnamefont {B.}~\bibnamefont {Singha}}, \bibinfo
  {author} {\bibfnamefont {R.}~\bibnamefont {Brett}}, \bibinfo {author}
  {\bibfnamefont {J.}~\bibnamefont {Fallica}}, \bibinfo {author} {\bibfnamefont
  {A.}~\bibnamefont {Hanlon}},\ and\ \bibinfo {author} {\bibfnamefont
  {B.}~\bibnamefont {{H\"{o}rz}}},\ }\bibfield  {title} {\bibinfo {title}
  {{Estimating the two-particle $K$-matrix for multiple partial waves and decay
  channels from finite-volume energies}},\ }\href
  {https://doi.org/10.1016/j.nuclphysb.2017.09.014} {\bibfield  {journal}
  {\bibinfo  {journal} {Nucl. Phys.}\ }\textbf {\bibinfo {volume} {B924}},\
  \bibinfo {pages} {477} (\bibinfo {year} {2017})},\ \Eprint
  {https://arxiv.org/abs/1707.05817} {arXiv:1707.05817 [hep-lat]} \BibitemShut
  {NoStop}%
\bibitem [{\citenamefont {{G. K\"all\'en}}(1964)}]{KallenBook}%
  \BibitemOpen
  \bibfield  {author} {\bibinfo {author} {\bibnamefont {{G. K\"all\'en}}},\
  }\href@noop {} {\emph {\bibinfo {title} {Elementary Particle Physics}}}\
  (\bibinfo  {publisher} {Addison-Wesley},\ \bibinfo {year} {1964})\BibitemShut
  {NoStop}%
\bibitem [{\citenamefont {Blatt}\ and\ \citenamefont
  {Biedenharn}(1952)}]{Blatt:zz1952a}%
  \BibitemOpen
  \bibfield  {author} {\bibinfo {author} {\bibfnamefont {J.~M.}\ \bibnamefont
  {Blatt}}\ and\ \bibinfo {author} {\bibfnamefont {L.~C.}\ \bibnamefont
  {Biedenharn}},\ }\bibfield  {title} {\bibinfo {title} {{Neutron-Proton
  Scattering with Spin-Orbit Coupling. 1. General Expressions}},\ }\href
  {https://doi.org/10.1103/PhysRev.86.399} {\bibfield  {journal} {\bibinfo
  {journal} {Phys. Rev.}\ }\textbf {\bibinfo {volume} {86}},\ \bibinfo {pages}
  {399} (\bibinfo {year} {1952})}\BibitemShut {NoStop}%
\bibitem [{\citenamefont {Guo}\ \emph {et~al.}(2013)\citenamefont {Guo},
  \citenamefont {Dudek}, \citenamefont {Edwards},\ and\ \citenamefont
  {Szczepaniak}}]{Guo:2012hv}%
  \BibitemOpen
  \bibfield  {author} {\bibinfo {author} {\bibfnamefont {P.}~\bibnamefont
  {Guo}}, \bibinfo {author} {\bibfnamefont {J.}~\bibnamefont {Dudek}}, \bibinfo
  {author} {\bibfnamefont {R.}~\bibnamefont {Edwards}},\ and\ \bibinfo {author}
  {\bibfnamefont {A.~P.}\ \bibnamefont {Szczepaniak}},\ }\bibfield  {title}
  {\bibinfo {title} {{Coupled-channel scattering on a torus}},\ }\href
  {https://doi.org/10.1103/PhysRevD.88.014501} {\bibfield  {journal} {\bibinfo
  {journal} {Phys. Rev. D}\ }\textbf {\bibinfo {volume} {88}},\ \bibinfo
  {pages} {014501} (\bibinfo {year} {2013})},\ \Eprint
  {https://arxiv.org/abs/1211.0929} {arXiv:1211.0929 [hep-lat]} \BibitemShut
  {NoStop}%
\bibitem [{\citenamefont {Draper}\ \emph
  {et~al.}(2023{\natexlab{a}})\citenamefont {Draper}, \citenamefont {Hanlon},
  \citenamefont {H\"orz}, \citenamefont {Morningstar}, \citenamefont
  {Romero-L\'opez},\ and\ \citenamefont {Sharpe}}]{Draper:2023boj}%
  \BibitemOpen
  \bibfield  {author} {\bibinfo {author} {\bibfnamefont {Z.~T.}\ \bibnamefont
  {Draper}}, \bibinfo {author} {\bibfnamefont {A.~D.}\ \bibnamefont {Hanlon}},
  \bibinfo {author} {\bibfnamefont {B.}~\bibnamefont {H\"orz}}, \bibinfo
  {author} {\bibfnamefont {C.}~\bibnamefont {Morningstar}}, \bibinfo {author}
  {\bibfnamefont {F.}~\bibnamefont {Romero-L\'opez}},\ and\ \bibinfo {author}
  {\bibfnamefont {S.~R.}\ \bibnamefont {Sharpe}},\ }\bibfield  {title}
  {\bibinfo {title} {{Interactions of \ensuremath{\pi}K,
  \ensuremath{\pi}\ensuremath{\pi}K and KK\ensuremath{\pi} systems at maximal
  isospin from lattice QCD}},\ }\href {https://doi.org/10.1007/JHEP05(2023)137}
  {\bibfield  {journal} {\bibinfo  {journal} {JHEP}\ }\textbf {\bibinfo
  {volume} {05}},\ \bibinfo {pages} {137}},\ \Eprint
  {https://arxiv.org/abs/2302.13587} {arXiv:2302.13587 [hep-lat]} \BibitemShut
  {NoStop}%
\bibitem [{\citenamefont {Jay}\ and\ \citenamefont {Neil}(2021)}]{Jay:2020jkz}%
  \BibitemOpen
  \bibfield  {author} {\bibinfo {author} {\bibfnamefont {W.~I.}\ \bibnamefont
  {Jay}}\ and\ \bibinfo {author} {\bibfnamefont {E.~T.}\ \bibnamefont {Neil}},\
  }\bibfield  {title} {\bibinfo {title} {{Bayesian model averaging for analysis
  of lattice field theory results}},\ }\href
  {https://doi.org/10.1103/PhysRevD.103.114502} {\bibfield  {journal} {\bibinfo
   {journal} {Phys. Rev. D}\ }\textbf {\bibinfo {volume} {103}},\ \bibinfo
  {pages} {114502} (\bibinfo {year} {2021})},\ \Eprint
  {https://arxiv.org/abs/2008.01069} {arXiv:2008.01069 [stat.ME]} \BibitemShut
  {NoStop}%
\bibitem [{\citenamefont {Hunter}(2007)}]{Hunter:2007}%
  \BibitemOpen
  \bibfield  {author} {\bibinfo {author} {\bibfnamefont {J.~D.}\ \bibnamefont
  {Hunter}},\ }\bibfield  {title} {\bibinfo {title} {Matplotlib: A 2d graphics
  environment},\ }\href {https://doi.org/10.1109/MCSE.2007.55} {\bibfield
  {journal} {\bibinfo  {journal} {Computing in Science \& Engineering}\
  }\textbf {\bibinfo {volume} {9}},\ \bibinfo {pages} {90} (\bibinfo {year}
  {2007})}\BibitemShut {NoStop}%
\bibitem [{\citenamefont {Green}\ \emph {et~al.}(2021)\citenamefont {Green},
  \citenamefont {Hanlon}, \citenamefont {Junnarkar},\ and\ \citenamefont
  {Wittig}}]{Green:2021qol}%
  \BibitemOpen
  \bibfield  {author} {\bibinfo {author} {\bibfnamefont {J.~R.}\ \bibnamefont
  {Green}}, \bibinfo {author} {\bibfnamefont {A.~D.}\ \bibnamefont {Hanlon}},
  \bibinfo {author} {\bibfnamefont {P.~M.}\ \bibnamefont {Junnarkar}},\ and\
  \bibinfo {author} {\bibfnamefont {H.}~\bibnamefont {Wittig}},\ }\bibfield
  {title} {\bibinfo {title} {{Weakly Bound H Dibaryon from
  SU(3)-Flavor-Symmetric QCD}},\ }\href
  {https://doi.org/10.1103/PhysRevLett.127.242003} {\bibfield  {journal}
  {\bibinfo  {journal} {Phys. Rev. Lett.}\ }\textbf {\bibinfo {volume} {127}},\
  \bibinfo {pages} {242003} (\bibinfo {year} {2021})},\ \Eprint
  {https://arxiv.org/abs/2103.01054} {arXiv:2103.01054 [hep-lat]} \BibitemShut
  {NoStop}%
\bibitem [{\citenamefont {Draper}\ \emph
  {et~al.}(2023{\natexlab{b}})\citenamefont {Draper}, \citenamefont {Hansen},
  \citenamefont {Romero-L\'opez},\ and\ \citenamefont
  {Sharpe}}]{Draper:2023xvu}%
  \BibitemOpen
  \bibfield  {author} {\bibinfo {author} {\bibfnamefont {Z.~T.}\ \bibnamefont
  {Draper}}, \bibinfo {author} {\bibfnamefont {M.~T.}\ \bibnamefont {Hansen}},
  \bibinfo {author} {\bibfnamefont {F.}~\bibnamefont {Romero-L\'opez}},\ and\
  \bibinfo {author} {\bibfnamefont {S.~R.}\ \bibnamefont {Sharpe}},\ }\bibfield
   {title} {\bibinfo {title} {{Three relativistic neutrons in a finite
  volume}},\ }\href@noop {} {\  (\bibinfo {year} {2023}{\natexlab{b}})},\
  \Eprint {https://arxiv.org/abs/2303.10219} {arXiv:2303.10219 [hep-lat]}
  \BibitemShut {NoStop}%
\bibitem [{\citenamefont {Stanzione}\ \emph {et~al.}(2020)\citenamefont
  {Stanzione}, \citenamefont {West}, \citenamefont {Evans}, \citenamefont
  {Minyard}, \citenamefont {Ghattas},\ and\ \citenamefont {Panda}}]{frontera}%
  \BibitemOpen
  \bibfield  {author} {\bibinfo {author} {\bibfnamefont {D.}~\bibnamefont
  {Stanzione}}, \bibinfo {author} {\bibfnamefont {J.}~\bibnamefont {West}},
  \bibinfo {author} {\bibfnamefont {R.}~\bibnamefont {Evans}}, \bibinfo
  {author} {\bibfnamefont {T.}~\bibnamefont {Minyard}}, \bibinfo {author}
  {\bibfnamefont {O.}~\bibnamefont {Ghattas}},\ and\ \bibinfo {author}
  {\bibfnamefont {D.}~\bibnamefont {Panda}},\ }\bibfield  {title} {\bibinfo
  {title} {Frontera: The evolution of leadership computing at the national
  science foundation},\ }in\ \href@noop {} {\emph {\bibinfo {booktitle}
  {Proceedings of Practice and Experience in Advanced Research Computing (PEARC
  '20)}}}\ (\bibinfo {year} {2020})\BibitemShut {NoStop}%
\bibitem [{\citenamefont {Harris}\ \emph {et~al.}(2020)\citenamefont {Harris},
  \citenamefont {Millman}, \citenamefont {Van Der~Walt}, \citenamefont
  {Gommers}, \citenamefont {Virtanen}, \citenamefont {Cournapeau},
  \citenamefont {Wieser}, \citenamefont {Taylor}, \citenamefont {Berg},
  \citenamefont {Smith} \emph {et~al.}}]{harris2020array}%
  \BibitemOpen
  \bibfield  {author} {\bibinfo {author} {\bibfnamefont {C.~R.}\ \bibnamefont
  {Harris}}, \bibinfo {author} {\bibfnamefont {K.~J.}\ \bibnamefont {Millman}},
  \bibinfo {author} {\bibfnamefont {S.~J.}\ \bibnamefont {Van Der~Walt}},
  \bibinfo {author} {\bibfnamefont {R.}~\bibnamefont {Gommers}}, \bibinfo
  {author} {\bibfnamefont {P.}~\bibnamefont {Virtanen}}, \bibinfo {author}
  {\bibfnamefont {D.}~\bibnamefont {Cournapeau}}, \bibinfo {author}
  {\bibfnamefont {E.}~\bibnamefont {Wieser}}, \bibinfo {author} {\bibfnamefont
  {J.}~\bibnamefont {Taylor}}, \bibinfo {author} {\bibfnamefont
  {S.}~\bibnamefont {Berg}}, \bibinfo {author} {\bibfnamefont {N.~J.}\
  \bibnamefont {Smith}}, \emph {et~al.},\ }\bibfield  {title} {\bibinfo {title}
  {Array programming with numpy},\ }\href@noop {} {\bibfield  {journal}
  {\bibinfo  {journal} {Nature}\ }\textbf {\bibinfo {volume} {585}},\ \bibinfo
  {pages} {357} (\bibinfo {year} {2020})}\BibitemShut {NoStop}%
\bibitem [{\citenamefont {Edwards}\ and\ \citenamefont
  {Joo}(2005)}]{Edwards:2004sx}%
  \BibitemOpen
  \bibfield  {author} {\bibinfo {author} {\bibfnamefont {R.~G.}\ \bibnamefont
  {Edwards}}\ and\ \bibinfo {author} {\bibfnamefont {B.}~\bibnamefont {Joo}}
  (\bibinfo {collaboration} {SciDAC}),\ }\bibfield  {title} {\bibinfo {title}
  {{The Chroma software system for lattice QCD}},\ }\href@noop {} {\bibfield
  {journal} {\bibinfo  {journal} {Nucl. Phys. Proc. Suppl.}\ }\textbf {\bibinfo
  {volume} {140}},\ \bibinfo {pages} {832} (\bibinfo {year}
  {2005})}\BibitemShut {NoStop}%
\bibitem [{zen()}]{zenododata}%
  \BibitemOpen
  \href@noop {} {}\bibinfo {howpublished}
  {\url{https://zenodo.org/records/10425662}}\BibitemShut {NoStop}%
\end{thebibliography}%

\end{document}